\documentclass[3p]{elsarticle}

\usepackage{lineno,hyperref}
\modulolinenumbers[5]

\journal{Physica E: Low-dimensional Systems and Nanostructures}
\usepackage{graphicx}
\usepackage{amsmath,amsthm,amssymb}
\usepackage{amsbsy}
\usepackage{bm}
\usepackage{gt16}

\usepackage{numcompress}\biboptions{authoryear}

\begin{document}

\begin{frontmatter}

\title{
Effective gauge field theory of spintronics}

\author{Gen Tatara}
\address{RIKEN Center for Emergent Matter Science (CEMS), 
2-1 Hirosawa, Wako, Saitama 351-0198, Japan}


\begin{abstract}
The aim of  this paper  is to present a comprehensive theory of spintronics phenomena based on the concept of effective gauge field, the spin gauge field.
An effective gauge field generally arises when we change a basis to describe system and describes low energy properties of the system.
In the case of ferromagnetic metals we consider, it arises from structures of localized spin (magnetization) and couples to spin current of conduction electron.
The first half of the paper is devoted to quantum mechanical arguments and phenomenology.
We show that the spin gauge field has adiabatic and nonadiabatic (off-diagonal) components, consisting an SU(2) gauge field.
The adiabatic component gives rise to spin Berry's phase, topological Hall effect and spin motive force, while nonadiabatic components are essential for spin-transfer torque and spin pumping effects by inducing nonequilibrium spin accumulation.
In the latter part of the paper, field theoretic approaches are described. Dynamics of localized spins in the presence of applied spin-polarized current is studied in a microscopic viewpoint, and current-driven domain wall motion is discussed.
Recent developments on interface spin-orbit interaction are also mentioned.

\end{abstract}


\end{frontmatter}


\newcommand{\alphatil}{\widetilde{\alpha}}
\newcommand{\alphavhat}{\hat{\alphav}}
\newcommand{\betav}{{\bm \beta}}
\newcommand{\kappaE}{\kappa_{\rm E}}
\newcommand{\kappaIE}{\kappa_{\rm IE}}
\newcommand{\IE}{{\rm IE}}
\newcommand{\mev}{{\mv}}
\newcommand{\omegaF}{\omega_{\rm F}}
\newcommand{\omegaR}{\omega_{\rm R}}
\newcommand{\omegaP}{\omega_{\rm p}}

\tableofcontents

\section{Introduction}
Electromagnetism is absolutely essential for the present technologies.
Electromagnetism is described by the two field, electric field, $\Ev$, and magnetic field, $\Bv$.
They satisfy four equations called the Maxwell's equations, 
\begin{align}
\nabla\cdot\Bv&=0 \nnr
\nabla\times\Ev&=-\delp{\Bv}{t},\label{faraday}
\end{align} 
and
\begin{align}
\nabla\cdot\Ev&=\frac{\rho}{\epsilon_0} \nnr
\nabla\times\Bv&= \muz\jv+\epsilon_0 \muz \delp{\Ev}{t}, \label{matter}
\end{align} 
where where $\rho$ and $\jv$ are density of charge and current, respectively and $\epsilon_0$ and 
$\muz$ are dielectric constant and magnetic permeability of vacuum, respectively.
The first two equations (\ref{faraday}) allows us to write the two fields by a scalar and vector potential, $\phi$ and $\Av$, respectively as
\begin{align}
 \Bv&=\nabla\times \Av \nnr
 \Ev&=-\nabla\Phi-\dot{\Av}. \label{potentials}
\end{align}
The six components of vectors $\Ev$ and $\Bv$ are therefore described by the four components of $\Phi$ and $\Av$.
The equations for $\Ev$ and $\Bv$ are similar, but not completely symmetric, because they represent different features of $\Av$ and $\Phi$.
The fields $\Phi$ (scalar potential) and $\Av$ (vector potential) are called (electromagnetic) gauge field.
In terms of the gauge field, the four equations reduces to even simpler two equations if we introduce a relativistic notation (see textbooks such as Ref. \citep{Ryder96}).

Electromagnetic effects on charged particles are represented conveniently in terms of the gauge field. 
The electric force and the Lorentz force acting on free electrons with charge $e$ and mass $m$ is represented by the electron Hamiltonian 
\begin{align}
 H=\frac{1}{2m}(\pv-e\Av)^2+e\Phi,\label{AphicouplingQM}
\end{align}
where $\pv$ is momentum.
The coupling obtained by replacing $\pv$ in the kinetic energy by $\pv-e\Av$ is called the minimal coupling.

\subsection{Symmetry and conservation law}
Gauge fields arise from symmetries.  
The symmetry for the electromagnetism is the invariance under local phase transformations, called U(1) symmetry, and it ensures the conservation of electric charge.
A gauge field couples to a current that corresponds to the conservation 
law. In the case of electromagnetic field, it is charge current.

We demonstrate this fact using field representation for clearness.
Let us denote the field and its conjugate by $\psi$ and $\psi^\dagger$, and denote the Lagrangian density by ${\cal L}(\psi^\dagger,\psi)$. 
The Lagrangian density contains field derivatives only to the linear order with respect to each field  $\psi$ and $\psi^\dagger$.
The equation the field satisfies is given by the condition of least action (the time-integral of the Lagrangian), ${\cal S}\equiv \int d^4x  {\cal L}(\psi^\dagger(\rv,t),\psi(\rv,t))$, where 
$\int d^4x\equiv \int dt \intr$.
Namely, for any small variation $\delta\psi^\dagger(\rv,t)$ and $\delta \psi(\rv,t)$ of the fields, the action remains the same (stationary condition) 
($\frac{\delta {\cal L}}{\delta \psi^\dagger}$ is a functional derivative);
\begin{align}
 \delta {\cal S}
 &=\int d^4x  \lt[  \delta \psi^\dagger \frac{\delta {\cal L}}{\delta \psi^\dagger}
+\frac{\delta {\cal L}}{\delta {\psi}}\delta {\psi} 
+\delta \partial_\mu \psi^\dagger \frac{\delta {\cal L}}{\delta \partial_\mu \psi^\dagger}
+\frac{\delta {\cal L}}{\delta \partial_\mu {\psi}}\delta \partial_\mu {\psi} \rt]  \nnr
&=
\int d^4x  \lt[  \delta \psi^\dagger 
\lt[ \frac{\delta {\cal L}}{\delta \psi^\dagger} - \frac{\partial}{\partial x_\mu} \frac{\delta {\cal L}}{\delta \partial_\mu \psi^\dagger} \rt]
+ \lt[\frac{\delta {\cal L}}{\delta {\psi}}
- \frac{\partial}{\partial x_\mu} \frac{\delta {\cal L}}{\delta \partial_\mu {\psi}} \rt] \delta {\psi} 
+\frac{\partial}{\partial x_\mu} 
\lt( \delta \psi^\dagger  \frac{\delta {\cal L}}{\delta \partial_\mu \psi^\dagger}
+ \frac{\delta {\cal L}}{\delta \partial_\mu {\psi}} \delta {\psi}   \rt)
\rt],\label{deltaS}
 \end{align}
where we used $\delta \partial_\mu {\psi}=\partial_\mu \delta {\psi}$ and  $x_\mu=\rv,t$, $\partial_\mu \equiv \frac{\partial}{\partial x_\mu}$.
The last total derivative term of the right-hand side vanishes, and we obtain field equation of motions,
\begin{align}
\frac{\delta {\cal L}}{\delta \psi^\dagger} - \frac{\partial}{\partial x_\mu} \frac{\delta {\cal L}}{\delta \partial_\mu \psi^\dagger} &=0, 
&
\frac{\delta {\cal L}}{\delta {\psi}}
- \frac{\partial}{\partial x_\mu} \frac{\delta {\cal L}}{\delta \partial_\mu {\psi}}
=0.
\end{align}

Equation (\ref{deltaS}) is used to find a conservation law, as known as the Noether's theorem \citep{Noether18}.
Suppose that the Lagrangian density is invariant under a certain transformation and that the 
variation $\delta {\psi} $ and $\delta \psi^\dagger $ are those for the invariant transformation.
As a result of equation of motion, Eq. (\ref{deltaS}) (without integrals) then indicates that 
\begin{align}
\frac{\partial}{\partial x_\mu} J_\mu=0, \label{noether}
\end{align}
where 
\begin{align}
J_\mu\equiv 
\lt( \delta \psi^\dagger  \frac{\delta {\cal L}}{\delta \partial_\mu \psi^\dagger}
+ \frac{\delta {\cal L}}{\delta \partial_\mu {\psi}} \delta {\psi}   \rt). \label{noethercurrent}
\end{align} 
Namely, there is a conserved current $J_\mu$ associated with the symmetric transformation 
$\delta {\psi} $ and $\delta \psi^\dagger $. 
We note that the expression (\ref{noethercurrent}) is a conserved current for internal degrees of freedom.
Original Noether's current is general one including for example the one for translation in time and space.

An example of the conserved current is the electric charge and current.
Physical quantities of electron like density are invariant by phase transformation,
\begin{align}
 \psi(\rv,t) &\ra e^{-i\epsilon}\psi(\rv,t), &
 \psi^\dagger(\rv,t) &\ra e^{i\epsilon}\psi^\dagger(\rv,t),
\end{align}
where $\epsilon$ is a real constant independent of position and time.
For small $\epsilon$, we have $\delta {\psi}=i\epsilon \psi $ and $\delta \psi^\dagger =-i\epsilon \psi^\dagger$.
For the case of free electron, the conserved current, Eq. (\ref{noethercurrent}), for the phase transformation is 
(multiplying by $e/\epsilon$)
\begin{align}
J_t &= e \psi^\dagger{\psi}, 
& 
J_i = -i\frac{\hbar e}{2m}\psi^\dagger\nablalr_i{\psi},
\label{chargecurrent}
\end{align} 
which are electric charge and current.
Therefore conservation of electric charge is a result of invariance under a global (i.e., independent of position and time) phase transformation.

A gauge field arises if we impose a stronger requirement that the system if invariant under local transformation. In the case of phase transformation, it is to require the invariance under 
\begin{align}
 \psi(\rv,t) &\ra e^{-i\epsilon(\rv,t)}\psi(\rv,t), &
 \psi^\dagger(\rv,t) &\ra e^{i\epsilon(\rv,t)}\psi^\dagger(\rv,t),
\end{align}
for phase factor depending on the space time.
Derivative of field then becomes 
$\partial_\mu \psi \ra e^{i\epsilon(\rv,t)}[\partial_\mu-i(\partial_\mu\epsilon)]\psi$, and the Lagrangian as it is is modified.
The Lagrangian is kept invariant, if we introduce another field $\Av$ coupled to the field derivative as 
$\partial_\mu \psi \ra D_\mu\psi$, where $D_\mu \equiv \nabla_\mu+iA_\mu$ is a covariant derivative.
The Lagragian is invariant if we define the field $A_\mu$  to be transformed as 
$A_\mu\ra A_\mu+\partial_\mu\epsilon$. 
This field $A_\mu$ is a gauge field, which defines relative relations among local coordinates (how to define the origin of the phase in the case of phase transformation).

The nature appears to possess symmetries under local transformations, gauge symmetries.  
In condensed matter, various symmetries other than U(1) symmetry of charge 
exist approximately for low energy phenomena.
Such gauge fields are called the effective gauge fields.
The concept of gauge field is highly useful for describing 
low energy transport effects in condensed matter.

\begin{figure}[tb]
  \begin{center}
 \includegraphics[width=0.7\hsize]{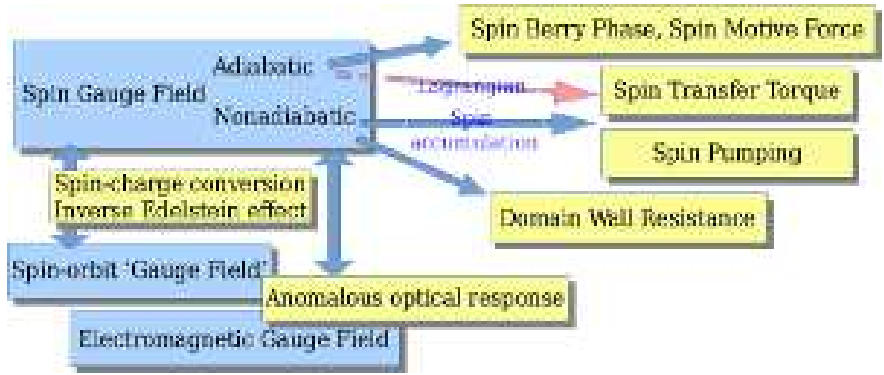}
  \end{center}
\caption{Scope of the paper.
\label{FIGscheme}}
\end{figure}

The objective of this paper is to demonstrate that spintronics effects in ferromagnetic metals are beautifully described in the framework of electromagnetism by introducing an effective spin gauge field that couples to electron spin current (Fig. \ref{FIGscheme}).
The effective gauge field has three components corresponding to three components of spin, forming an SU(2) gauge field. 
Its adiabatic component is a U(1) gauge field having the same mathematical structure as charge electromagnetism.
Spin Berry's phase, spin motive force, spin transfer effect and spin pumping effects are discussed in detail, and roles of the adiabatic and nonadiabatic components in inducing these effects are clarified.
Various spin-charge conversion effects arise when spin-orbit interaction, approximated as another gauge field, is introduced. 
Coupling of the spin gauge field to  electromagnetism results in anomalous optical properties. 

\section{Unitary transformation and gauge field}
An effective gauge field is a general concept that arises naturally when we 
diagonalize quantum systems.
Let us consider an electron with mass $m$ described by a Hamiltonian $H$, given by a sum of free 
electron kinetic energy and a potential $V$ as 
$H=-\frac{\hbar^2\nabla^2}{2m}+V$.
An eigenstate $\ket{\psi(t)}$ satisfies the Schr\"odinger's 
equation
\begin{align}
i\hbar\delpo{t}\ket{\psi}=H\ket{\psi} .
\label{Seqpsi}
\end{align}
Usually the state $\ket{\psi}$ cannot be fully solved, and so we may like to 
define the state using a unitary transformation $U$ from 
a state we are familiar with (like the states for non-interacting or uniform potential cases), $\ket{\phi}$, as 
\begin{align}
 \ket{\psi}\equiv U \ket{\phi} . \label{psiphirelation}
\end{align}
A typical example is the case of slowly changing potential as function of time 
or space. One should then solve for uniform potential to obtain $\ket{\phi}$ 
and include the temporal and/or spatial dependence in terms of the matrix $U$.
The Sch\"odinger's equation for state $\ket{\phi}$ reads 
\begin{align}
i\hbar\lt( \frac{\partial}{\partial t}+ U^{-1}\delpo{t} U\rt) \ket{\phi}= 
\lt(-\frac{\hbar^2}{2m}\lt(\nabla + U^{-1}\nabla U\rt)^2+ \tilde{V} \rt)\ket{\phi},
\label{UinvUeq}
\end{align}
where $\tilde{V}\equiv U^{-1} V U$.
We see that now derivatives are replaced by a 'covariant' one,
\begin{align}
D_\mu\equiv  \partial_\mu \pm \frac{i}{\hbar}A_\mu,
\end{align}
where $\mu=t,x,y,z$ and sign $+$ and $-$ corresponds to $\mu=t$ and $\mu=x,y,z$, respectively.
(The sign convention is chosen in accordance with Lorentz invariance.) 
In this paper, greek letters suffix denotes space and time and roman letters are used for spatial suffix.
The quantity
\begin{align}
 A_\mu\equiv \mp i\hbar U^{-1}\partial_\mu U,\label{Adefgeneral}
\end{align}
describes a modification of derivative is an effective gauge field.
It couples to the matter in the same manner as the electromagnetic gauge field (Eq. (\ref{AphicouplingQM})). 
The spatial component may be called an effective vector potential, as it modifies the kinetic energy, and time-component is an effective scalar potential.
Effective gauge field couples to a current corresponding to the unitary transformation $U$ via the minimal coupling.
Explicit example for the spin case is described in Sec. \ref{SEC:adiabatic}.

\subsection{Berry's phase}
In the presence of slowly changing potential, time-development of quantum system is represented by solely by a phase factor called the Berry's phase \citep{Berry87}.
Let us briefly mention the effect.
The condition assumed is so called the adiabatic condition, 
\begin{align}
 \hat{H}(t)\ket{\psi}=E(t)\ket{\psi},
 \label{adiabaticcondition0}
\end{align}
where $\hat{H}(t)$ is the Hamiltonian operator with slowly varying potential and $E(t)$ is a real number. 
The adiabatic condition means that the state $\ket{\psi}$, defined with respect to eigenstates of $H(t)$, remains to be the eigenstate of the Hamiltonian at each instance.
This condition puts a constraint which allows only a phase as dynamic variable.

The phase arises because of a change of the frame (basis to describe quantum states) as a result of a time-development of the Hamiltonian. 
What is essential for the phase appearance is that we do not notice a slow change of the frame, and observe the system based on the initial frame. 
A change of frame is described by a unitary transformation.
We denote the state $\ket{\psi}$ as the one in the 'correct' basis defined with respect to the Hamiltonian $H(t)$ at each time $t$, and the state $\ket{\phi}$ as the state of the observer, i.e., the one represented using the basis defined at $t=0$.
They are related by a unitary matrix $U$ as in Eq.(\ref{psiphirelation}).
For the observer's state $\ket{\phi}$, the Schr\"odinger equation 
(\ref{UinvUeq}) reads
\begin{align}
 i(\hbar\partial_t+iA_t)\ket{\phi}=\widetilde{H}\ket{\phi},\label{Seqphiadiabatic}
\end{align}
where $A_t$ is a gauge field defined in Eq. (\ref{Adefgeneral}) and  $\widetilde{H}\equiv U^{-1}HU$.
Because of the adiabatic condition (\ref{adiabaticcondition0}), we have $\widetilde{H}\ket{\phi}=E(t)\ket{\phi}$, and thus  
the equation reduces to
\begin{align}
 \partial_t\ket{\phi}=-\frac{i}{\hbar}(E(t)+A_t)\ket{\phi}.
 \label{Seqphiadiabatic2}
\end{align}
The term $E(t)$ on the right-hand side describes the standard time-development with energy $E(t)$, while the term $A_t$ describes the effects of variation of the Hamiltonian.
In general $A_t$ is a matrix including off-diagonal components that causes transition to different states.
In the adiabatic limit, the off-diagonal components are neglected, because they give rise to rapidly oscillating term like $e^{-it\Delta/\hbar }$ at long time ($t$), where $\Delta $ is excitation energy for transitions. (In the case of spin, $\Delta=\spol$ is energy of spin splitting.) 
Thus $A_t$ can be treated as a constant in the adiabatic limit, 
and  Eq. (\ref{Seqphiadiabatic2}) is integrated to obtain
\begin{align}
\ket{\phi(t)}=e^{i\gamma(t)} e^{-\frac{i}{\hbar}\int^t_0dt'E(t')}\ket{\phi(0)},\label{phigamma}
\end{align}
where 
\begin{align}
\gamma(t)\equiv \frac{1}{\hbar} \int^t_0dt'A_t(t'),
\end{align}
is the Berry's phase arising from the change of the frame, while the second phase factor of Eq. (\ref{phigamma}) is the ordinary dynamic phase.
The gauge field describing the Berry's phase is written explicitly as 
\begin{align}
A_t(t)= -i\hbar \average{\phi(0)|U^{-1}(t)\partial_t U(t)|\phi(0)}.
\end{align}
It is written using a derivative of the state as (neglecting higher orders of time derivative)  
\begin{align}
A_t(t)= -i\hbar\average{\psi(t)|\partial_t |\psi(t)}.\label{BerryphaseQM}
\end{align}
The Berry's phase is therefore a result of an effective gauge field $A_t$ arising from a unitary transformation (\ref{Adefgeneral}).

\section{Localized spin \label{SEC:spin}}
In this section,  theoretical treatments of ferromagnetism are briefly summarized.

\subsection{Spin dynamics}
Magnetism is collective property arising from an ensemble of many localized spins.
Each spin, $\hat{\Sv}=(\hat{S}_x,\hat{S}_y,\hat{S}_z)$ 
\footnote{In this section, quantum operators are denoted by $\hat{\ }$. }, is a quantum object governed by commutation relation 
\begin{align}
   [\hat S_i,\hat S_j] &= \hat S_i,\hat S_j-\hat S_j,\hat S_i
   =i\hbar \epsilon_{ijk} \hat S_k,
 \label{spincomm}
\end{align}
where $i,j,k=x,y,z$, $\hbar$ is Planck constant divided by $2\pi$ \footnote{In most part of this paper, $\hbar$ is set to unity.}
and $\epsilon_{ijk}$ is a totally antisymmetric tensor that satisfies 
$\epsilon_{ijk}=\epsilon_{jki}=\epsilon_{kij}$,  $\epsilon_{jik}=-\epsilon_{ijk}$ and $\epsilon_{xyz}=1$.
Summation over repeated index is assumed (Einstein's convention but for spatial index, $x,y,z$.).
Spin is an angular momentum and thus create magnetic moment 
\begin{align}
\hat \mv=\frac{e\hbar}{m}\hat\Sv,    \label{mandS}  
\end{align}
which couples to an magnetic field by the interaction 
\begin{align}
   \hat{H}_S &= - \Bv\cdot\hat\mv = -\hbar \gamma \Bv\cdot\hat\Sv, \label{HBquantum}
\end{align}
where $\gamma\equiv \frac{e}{m}(<0)$ is gyromagnetic ratio.
Because the electron charge $e$ is negative, magnetization and spin points opposite, and spin tends to point antiparallel to the magnetic field (Fig. \ref{FIGprecess}).
The dynamics of spin is described by the Heisenberg equation
\begin{align}
  \frac{\partial}{\partial t} \hat S_i & = \frac{i}{\hbar}[\hat{H}_S,\hat S_i], 
\end{align}
which reads
\begin{align}
  \frac{\partial\hat\Sv}{\partial t}  & = -\gamma \Bv\times\hat\Sv.
  \label{singlespineq}
\end{align}
This is a quantum mechanical equation, but interestingly, this form equivalent to the one describing torque on classical objects, applies to macroscopic magnetization.
\begin{figure}[tb]
\begin{minipage}{0.48\hsize}
  \begin{center}
  \end{center}
\caption{Magnetization $\Mv$ precesses anticlockwise around $\Bv$, while  $\Sv$, pointing opposite to $\Mv$, precesses clockwise around direction $-\Bv$.
\label{FIGprecess}}
\end{minipage}
\hfill
\begin{minipage}{0.48\hsize}
  \begin{center}
  \end{center}
\caption{LLG equation describes the damping of spin, which tends to point the spin along $-\Bv$ direction.
\label{FIGdecay}}
\end{minipage}
\end{figure}

The equation of motion of spin (\ref{singlespineq}) is derived from a Lagrangian
\begin{align}
L_S= \hbar S\dot{\phi}(\cos\theta-1)-H_S 
,
\label{Lspin}
\end{align}
where $\theta,\phi$ are polar angles of $\Sv$, namely 
$\Sv\equiv S\nv$,  
\begin{align}
\nv\equiv(\sin\theta\cos\phi,\sin\theta\sin\phi,\cos\theta),
\end{align}
and $\dot{\cal O}=\frac{\partial}{\partial t}{\cal O}$ denotes time derivative of field ${\cal O}$.
In the Lagrangian representation, variables $\theta$, $\phi$ and $\nv$ are classical variables and quantum nature is embedded in the first time-derivative term  of Eq. (\ref{Lspin}). 
The equation of motion derived from the Lagrangian is 
\begin{align}
\frac{d}{dt} \frac{\delta L}{\delta \dot{\theta}}-\frac{\delta L}{\delta \theta}&=0,
& \frac{d}{dt} \frac{\delta L}{\delta \dot{\phi}}-\frac{\delta L}{\delta \phi}=0,
\end{align}
 namely,
\begin{align}
 \hbar S \sin\theta \dot{\theta} &= \frac{\delta H_S}{\delta \phi} \nnr
 -\hbar S \sin\theta \dot{\phi} &= \frac{\delta H_S}{\delta \theta}.
 \label{thetaphieq}
\end{align}
Using 
\begin{align}
 \frac{\delta H_S}{\delta \theta} &= \cos\theta\lt(\cos\phi  \frac{\delta H_S}{\delta n_x}+\sin\phi  \frac{\delta H_S}{\delta n_y}\rt)
  -\sin\theta \frac{\delta H_S}{\delta n_z} =\evth\cdot  \frac{\delta H_S}{\delta \nv} =-\hbar\gamma S \evth\cdot \Bv \nnr
 \frac{\delta H_S}{\delta \phi} &= \sin\theta\lt(-\sin\phi  \frac{\delta H_S}{\delta n_x}+\cos\phi  \frac{\delta H_S}{\delta n_y}\rt)
 = \sin\theta\evph\cdot  \frac{\delta H_S}{\delta \nv} =-\hbar\gamma S \sin\theta\evph\cdot \Bv ,
\end{align}
we see that the equations (\ref{thetaphieq}) leads to  
\begin{align}
 \dot{\nv}=  \dot{\theta}\evth+\sin\theta \dot{\phi} \evph
 =\gamma[ \evph(\evth\cdot\Bv) - \evth(\evph\cdot\Bv) ]
 =\gamma \nv\times\Bv,
\end{align}
which is  Eq. (\ref{singlespineq}).


\subsection{Spin relaxation}
In ferromagnets, large number of localized spins moves coherently, and this case is described by replacing quantum variable $\Svhat$ by a classical vector $\Sv$ whose magnitude is proportional to the number of coherent spins.
The magnetization $\Mv$, commonly used to describe macroscopic magnetism, is related to it as 
\begin{align}
 \Mv 
 = \muz\frac{\hbar \gamma}{a^3}\Sv,
\end{align}
assuming that each localized spin contribute independently to the magnetization, 
where $a$ is the lattic constant.
In this paper, we discuss in terms of the localized spin.
The fundamental equation of motion for classical spin $\Sv$ is the same as 
the quantum one, Eq. (\ref{singlespineq}), with $\Bv$ the total magnetic field. 
In solids, there are various microscopic sources for $\Bv$, such as conduction electron in metals and lattice vibration (phonons).
The total magnetic field acting on each localized spin is therefore not simply written as $\Bv=\muz\Hv+\Mv$, the sum of external magnetic field $\Hv$ and macroscopic magnetization. 
Instead, magnetization is taken account of by considering microscopic exchange interaction (and dipole interactions).
In fact, the Hamiltonian with an exchange interaction $J_0$ and an external magnetic field $\Hv$,
\begin{align}
 H=-J_0\sum_{ij}\Sv_i\cdot\Sv_j -\hbar\gamma \muz \sum_{i}\Hv \cdot\Sv_i,  \label{HJHlattice}
\end{align}
leads, under a mean field approximation, to 
\begin{align}
 H_{\rm mf}= -\hbar\gamma \sum_{i} \Sv_i\cdot\Bv ,\label{HJBlattice}
\end{align}
with the total magnetic field of 
$\Bv\equiv \muz\Hv+\frac{3J_0 }{\hbar\gamma}\average{\Sv}$, where 
$\average{\Sv}$ is average of localized spin.
Therefore magnetization in this case is $\Mv=\frac{3J_0 }{\hbar\gamma}\average{\Sv}$.
Effects from uncontrollable magnetic interaction such as spin flip scattering by magnetic impurities or phonons lead to a relaxation (damping) of localized spins.
Damping is essential in magnetism, as we are familiar with magnetic moment pointing along an applied magnetic field (Fig. \ref{FIGdecay}), which does not occur in the absence of damping.

There is a long history how to incorporate damping in equation of motion (\ref{singlespineq}). 
One way proposed by Gilbert is to modify it to be
\begin{align}
  \frac{\partial\Sv}{\partial t}  & = - \gamma \Bv\times\Sv -\frac{\alpha}{S} \lt(\Sv\times\frac{\partial\Sv}{\partial t} \rt),
  \label{LLG}
\end{align}
where the coefficient $\alpha$ is dimensionless, positive and is called the Gilbert damping constant.
The equation is called the Landau-Lifshitz-Gilbert (LLG) equation.
When $-\Bv$ is along $z$ axis, small amplitude oscillation of $\nv$ around $z$ axis obtained from Eq. (\ref{LLG}) is 
\begin{align}
 \phi & = \gamma |B| t \nnr
 \theta & = \theta_0 e^{\alpha \gamma |B| t},
\end{align}
indicating that the equation describes a relaxation process to the equilibrium direction with the period of precession $\frac{2\pi}{|\gamma B|}$ and the decay time of $\frac{1}{\alpha |\gamma B|}$.

There are other ways to introduce damping, like the one called the Landau-Lifshitz equation, 
\begin{align}
  \frac{\partial\Sv}{\partial t}  & = -\gamma \Bv\times\Sv -\alpha\frac{\gamma}{S}[\Sv\times(\Sv\times \Bv)].
  \label{LL}
\end{align}
Those equations including damping implicitly assume weak damping, and Eqs. (\ref{LL}) and (\ref{LLG}) are equivalent if quantities of the order of $\alpha^2$ are neglected.
From microscopic viewpoint, LLG equation treating damping by introducing time-derivative is natural, as such a damping term is derived systematically by a gradient expansion \citep{KTS06}, as we shall mention later (Sec. \ref{SECLLGwithelectron}).

To treat spin damping in Lagrangian formulation, we use the Rayleigh's method, and introduce a dissipation function, 
\begin{align}
 W_S \equiv \sumr \frac{\hbar \alpha}{2S} \dot{\Sv}^2.
\end{align}
The modified equation of motion,
\begin{align}
 \frac{d}{dt}\frac{\delta L_S}{\delta \dot{q}} -\frac{\delta L_S}{\delta {q}} 
 = -\frac{\delta W_S}{\delta \dot{q}} ,
 \label{Rayleigheq}
\end{align}
where $q=\theta,\phi$,
turns out to be the LLG equation.

Damping leads to an energy dissipation as confirmed by calculating the time-derivative of the Hamiltonian using the LLG equation (\ref{LLG});  
\begin{align}
  \frac{dH_{B}}{dt}=
-{\alpha}{S}\lt(\frac{d\nv}{dt}\rt)^2 +O(\alpha^2).
\end{align}

\subsection{Domain wall}
\begin{figure}[tbh]
  \begin{center}
  \includegraphics[width=0.3\linewidth]{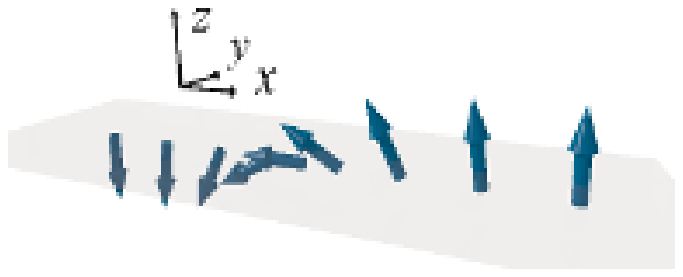}
  \includegraphics[width=0.3\linewidth]{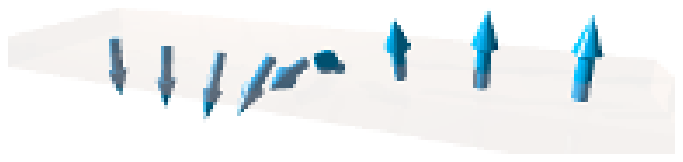}
  \includegraphics[width=0.3\linewidth]{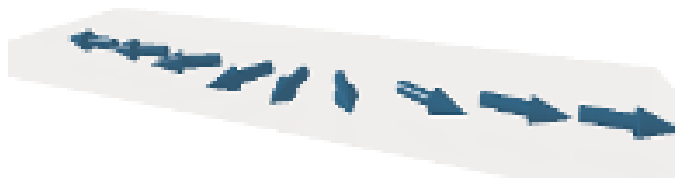}
  \end{center}
\caption{Domain wall structures. Choosing the wall direction as the $x$ axis, left figure corresponds to a N\'eel wall with Eq. (\ref{DWsolrest1}) and $\phi=0$, the middle figure is a Bloch wall with $\phi=\frac{\pi}{2}$.
Right figure is another type of N'eel wall realized in wires.
\label{FIGDWs}}
\end{figure}

Domain wall is a spatial structure between magnetic domains having different localized spin directions. In the wall, localized spins rotates as  function of spatial coordinate (Fig. \ref{FIGDWs}) . The thickness of the wall, $\lambda$, is determined by the competition between the exchange energy and magnetic anisotropy energy, and is typically 10-100 nm.
We consider an one-dimensional and rigid wall, neglecting deformation.
We consider first a system with only an easy axis magnetic anisotropy energy, which is necessary for creation of a wall.
For discussing dynamics, we shall later include a hard-axis anisotropy energy.
Choosing the easy axis along the $z$ direction, the anisotropy energy is represented by the Hamiltonian
\begin{align}
 H_{K} &\equiv -\frac{KS^2}{2}\sumr  \cos^2\theta.
\end{align}
where $K$ is the easy-axis anisotropy energy ($K>0$). 
Including the ferromagnetic exchange coupling, the Hamiltonian in the continuum expression reads 
\begin{align}
 H=\sumr \lt[\frac{JS^2}{2} [(\nabla\theta )^2+\sin^2\theta (\nabla\phi )^2 ]+\frac{KS^2}{2}\sin^2\theta \rt].\label{HJK}
\end{align}
The total energy is minimized by the conditions
\begin{align}
 \lambda^2\nabla^2\theta-\sin\theta\cos\theta(1+\lambda^2(\nabla\phi)^2)&=0 \nnr
\nabla(\sin^2\theta\nabla\phi)&=0, \label{thetaeq1}
\end{align}
where
\begin{align}
\lambda=\sqrt{\frac{J}{K}},
\end{align} 
turns out to be the thickness of the wall.
A static domain wall solution is obtained as 
\begin{align}
 \cos\theta&=\pm \tanh\frac{x-X}{\lambda} \nnr  
 \sin\theta&=\frac{1}{\cosh\frac{x-X}{\lambda}},
 \label{DWsolrest1}
\end{align}
and $\phi$ is any constant.
We chose the wall direction along the $x$ axis, but the choice is mathematically arbitrary as far as the spin space and coordinate space are decoupled,  i.e., if spin-orbit interaction is neglected.
Value of $\phi$ is also arbitrary in the present system without hard-axis anisotropy.
Historically, a wall with $\phi=0$ in Eq. (\ref{DWsolrest1}), where the localized spins in the wall has a component perpendicular to the wall plane (the $yz$-plane), is called the N\'eel wall, while the case of $\phi=\frac{\pi}{2}$ with localized spins rotating in the wall plane is called the Bloch wall.
In wires, anisotropy axis tends to be along the wire direction (here $x$) to reduce the magnetostatic energy, and another type of N\'eel wall is realized.

\subsection{Domain wall dynamics \label{SECDWB}}
\newcommand{\Hwb}{H_{\rm w}}
We describe here the wall dynamics when a magnetic field is applied along the easy axis. 
It may appear that dynamics wall is described simply by replacing the wall center coordinate $X$ in Eq. (\ref{DWsolrest1}) by a time-dependent variable, $X(t)$. This is not, however, sufficient, and we need to introduce $\phi$ as another dynamical variable, $\phi(t)$ \citep{Slonczewski72,TKS_PR08}.
A hard-axis anisotropy energy, therefore, plays an essential role in the wall dynamics. 
We introduce it choosing the hard axis as the $y$ axis. 
Anisotropy energies we consider are thus 
\begin{align}
 H_{K} &\equiv \frac{KS^2}{2}\sumr  \sin^2\theta(1+\kappa\sin^2\phi),
\end{align}
where $\kappa\equiv \frac{K_\perp}{K}$ with $K_\perp$ being the hard-axis anisotropy energy.
The external magnetic field $H$ along $z$ axis is represented by the Hamiltonian 
\begin{align}
 H_H &= \sumr \muz \hbar \gamma H (S_z(X(t))-S_z(X=0)),
\end{align}
where we subtracted an irrelevant constant.
We consider a rigid wall, namely, the wall structure does not change when dynamic, which requires that $K \gg K_\perp$.
The low energy dynamics of the wall is thus described by the wall profile of  
\begin{align}
 n_z(x,t)=\tanh\frac{x-X(t)}{\lambda}, \;\;\;  n_\pm(x,t)\equiv n_x\pm in_y=\frac{e^{\pm i\phi(t)}}{\cosh\frac{x-X(t)}{\lambda}},
 \label{DWsol}
\end{align}
where two dynamics variables, $X(t)$ and $\phi(t)$ are called the collective coordinates.
Rewriting the spin Lagrangian using Eq. (\ref{DWsol}), we obtain 
\begin{align}
 L&=\frac{\hbar \Nw S}{\lambda} 
 \lt[ -\dot{\phi} X-\frac{\Kp \lambda S}{2\hbar}\sin^2\phi +\muz\gamma H X \rt] ,
 \label{LDW}
\end{align}
where we used $\int dx \frac{1}{\cosh^2 (x/\lambda)}=2\lambda$ and $\Nw\equiv \frac{2A\lambda}{a^3}$ is the number of localized spins in the wall, with $A$ being the cross sectional area of the system.
The dissipation function is written using collective coordinates as  
\begin{align}
 W_S&=\alpha \frac{\hbar \Nw S}{2} \lt[ \frac{\dot{X}^2}{\lambda^2}+\dot{\phi}^2 \rt] .\label{WDW}
\end{align}
The equations of motion obtained from Eqs. (\ref{LDW}) (\ref{WDW}) read 
\begin{align}
 \dot{X}-\alpha\lambda\dot{\phi} &= \vc \sin 2\phi \nonumber\\ 
 \dot{\phi}+\alpha\frac{\dot{X}}{\lambda} &= \muz \gamma H,
 \label{DWeq}
\end{align}
where 
\begin{align}
\vc\equiv \frac{\Kp \lambda S}{2\hbar}.                                             \end{align}

These equations (neglecting dissipation) are Hamilton's equation for position and canonical momentum ($P$),
\begin{align}
 \dot{X} &= \frac{\delta H}{\delta P}, & 
 \dot{P} = - \frac{\delta H}{\delta X}. \label{Hamiltoneq}
\end{align}
This means that the canonical momentum of domain wall is $\phi$ and not simply proportional to $\dot{X}$ like an particle.
This fact is obviously seen in the Lagrangian (\ref{LDW}), where the first term describes the canonical relation between variables.
A domain wall therefore has an intriguing property that angle $\phi$ needs to be finite to have a translational motion.
%
\begin{figure}[tbh]
  \begin{center}
  \includegraphics[width=0.25\linewidth]{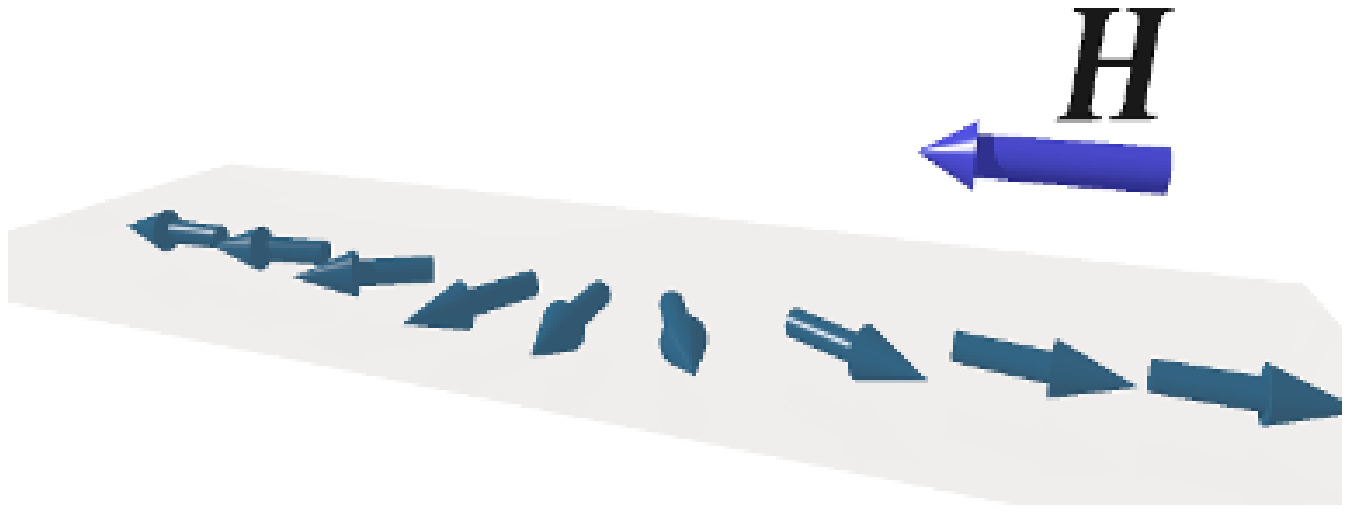}\\
  \includegraphics[width=0.25\linewidth]{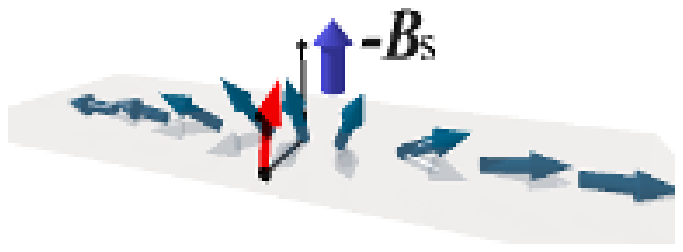}\\
  \includegraphics[width=0.25\linewidth]{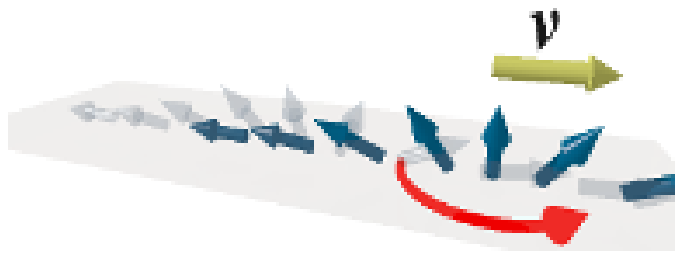}
  \end{center}
\caption{
Mechanism of domain wall motion when an external magnetic field $\Hv$ is applied along the easy axis (top figure).
The magnetic field induces a precession of localized spins and out-of-plane component (middle figure).
This results in an out-of-plane magnetic field $\Bv_{\rm s}$, which induces a precession within the plane, which is equivalent to a translational motion of the wall (bottom figure).  
\label{FIGDWmotion}}
\end{figure}
This feature can be understood based on the spin dynamics (Fig. \ref{FIGDWmotion}).
As spin dynamics is always a precession around a magnetic field, localized spins in the wall tends to be out-of the wall plane when magnetic field is applied along the easy axis, namely, $\phi$ necessarily develops.
The out-of the plane component then induces another precession within the wall plane, and this is equivalent to a translational motion of the whole wall.
This complex behavior is expressed theoretically by a single term, $\dot{\phi}X$, in the Lagrangian.

\begin{figure}[tbh]
  \begin{center}
  \includegraphics[width=0.4\linewidth]{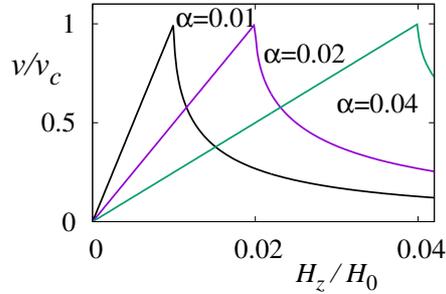}
  \end{center}
\caption{
Left: Domain wall averaged speed for $\alpha=0.01, 0.02, 0.04$ under an easy axis external magnetic field $H$.
Speed and magnetic field are normalized by $v_{\rm c}$ and $H_0\equiv \frac{\vc}{\muz\gyro \lambda}$, respectively.
Right: High-field regime for $\alpha=0.1$, where linear dependence on the field is seen. 
\label{FIG:dwvb}}
\end{figure}

The solution for Eq. (\ref{DWeq}) shows different behavior for two regimes of magnetic field, $H<H_{\rm w}$ and $H\geq H_{\rm w}$, where ($\gamma<0$)
\begin{align}
H_{\rm w}\equiv -\muz 
\frac{\alpha \vc}{\lambda\gamma}.
\end{align}
In the weak field regime, wall has a constant speed 
\begin{align}
\dot{X} = \frac{\muz \lambda \gamma H}{\alpha},
\end{align}
and the angle $\phi$ is determined by the speed as $\dot{X}= \vc \sin2\phi$.
This means that the torque necessary for wall motion is supplied by tilting the wall plane by the finite angle $\phi$.
When $B>B_{\rm w}$, the static tilt of the wall cannot support the wall motion, resulting in an oscillating motion of $X$ and $\phi$
(Walker's breakdown).
The solution in this case is
\begin{align}
\sin2\phi &= \frac{H}{\Hwb}
 +\frac{1-\lt( \frac{H}{\Hwb}\rt)^2}
   { \frac{H}{\Hwb}+\sin 2\omega t}  \\
\omega &= \frac{\vc}{\lambda} \frac{\alpha}{1+\alpha^2} 
  \sqrt{\lt( \frac{H}{\Hwb}\rt)^2-1}.  
\end{align}
The wall speed is 
\begin{align}
\dot{X} &= \vc
\lt( \frac{H}{\Hwb}
 +\frac{1}{1+\alpha^2} \frac{1-\lt( \frac{H}{\Hwb}\rt)^2}
   { \frac{H}{\Hwb}+\sin 2\omega t}  \rt)  ,
\end{align}
and its time-average is
\begin{align}
\average{\dot{X}} &= \vc
\lt( \frac{H}{\Hwb}
 - \frac{1}{1+\alpha^2}
\sqrt{\lt( \frac{H}{\Hwb}\rt)^2-1} \rt).
\end{align}
Average speed is plotted in Fig. \ref{FIG:dwvb}.
In the limit of high field, $H \gg \Hwb$, 
$\average{\dot{X}} \ra 
-\muz{\lambda\gamma H}\frac{\alpha}{1+\alpha^2}
$.

\section{Adiabatic spin gauge field in ferromagnetic metal  \label{SEC:SEMF}}
In this section, we include conduction electron to describe a ferromagnetic metal  and study transport properties.
The ferromagnetic metal is modeled by a simple Hamiltonian of a free electron with an $sd$ exchange interaction with localized spin, $\Sv(\rv,t)=S\nv(\rv,t)$;
\begin{align}
 H=\frac{\pv^2}{2m} -\spol \nv\cdot \sigmav,
\end{align}
where $\nv$ is a unit vector along $\Sv$ and $\spol\equiv \Jsd S$ is the spin energy splitting ($\Jsd$ is the $sd$ exchange constant) and $\sigmav\equiv (\sigma_x,\sigma_y,\sigma_z)$ is the vector of Pauli matrix representing the electron spin operator.
In most $3d$ metallic ferromagnets, $sd$ exchange interaction is strongest energy scale for spin dynamics. 
(Compared to Fermi energy $\ef$, the ratio is $\spol/\ef\gtrsim0.1$ in most cases \citep{Kittel96}.) 
The expectation value of electron spin, $\average{\sigmav}$, is therefore locked to the direction $\nv$ almost perfectly.
This limit is called the adiabatic limit.

\subsection{Phase from spin texture \label{SEC:adiabatic}}

\begin{figure}[tbh]
\begin{minipage}{0.48\hsize}
 \begin{center}
  \end{center}
\caption{  In the presence of localized spin structure, conduction electron spin feels a position-dependent effective magnetic field.  
\label{FIGhop}}
\end{minipage}
\hfill
\begin{minipage}{0.48\hsize}
 \begin{center}
  \includegraphics[height=7\baselineskip]{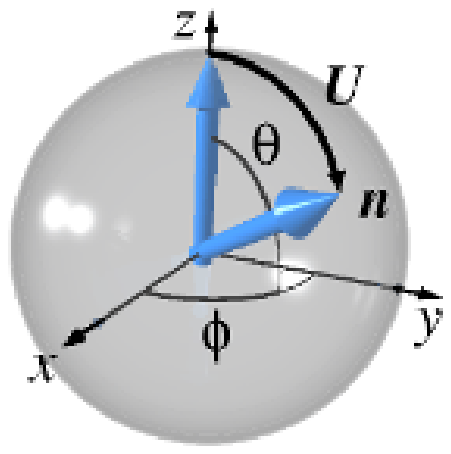}
  \end{center}
\caption{ Unitary transformation $U$ is defined to connect up spin state $\ket{\uparrow}$ and a state pointing $\nv$.
\label{FIGU}}
\end{minipage}
\end{figure}

Transport of conduction electrons in the adiabatic limit is 
theoretically studied by calculating the quantum mechanical phase attached to 
the wave function of electron spin. 
Let us consider a conduction electron hopping from a site $\rv$ to a 
different site at $\rv'$ (Fig. \ref{FIGhop}).
The localized spin direction at those sites are $\nv(\rv)\equiv\nv$ and 
$\nv(\rv')\equiv\nv'$, respectively, and the electron spin's wave function at the 
two sites are  
\begin{align}
|\nv\rangle 
&=\cos\frac{\theta}{2}|\!\uparrow\rangle+\sin\frac{\theta}{2}e^{i\phi}
|\!\downarrow\rangle \nnr 
|\nv'\rangle 
&=\cos\frac{\theta'}{2}|\!\uparrow\rangle+\sin\frac{\theta'}{2}e^{i\phi'}
|\!\downarrow\rangle,
\end{align}
where $|\!\uparrow\rangle$ and $|\!\downarrow\rangle$ are spin up and downs states, respectively and 
$\theta$, $\phi$ and $\theta'$, $\phi'$  are the polar angle of $\nv(\rv)$ 
and $\nv(\rv')$, respectively (Fig. \ref{FIGU}).
The wave functions are concisely written by use of matrices, $U(\rv)$, which rotates the spin up state $|\!\uparrow\rangle$ to $|\nv\rangle$, as  $|\nv\rangle = U(\rv)|\!\uparrow\rangle$ 
(Fig. \ref{FIGU}).
The rotation is a combination of rotation of angle $\theta$ around $+y$ axis followed by a rotation of $\phi$ around $+z$ axis, represented by a matrix $e^{-\frac{i}{2}\phi\sigma_z} e^{-\frac{i}{2}\theta\sigma_y} $.
We add irrelevant phase factors and define the rotation matrix as
\begin{align}   
U &\equiv  e^{\frac{i\pi}{2}} 
    e^{-\frac{i}{2}\phi\sigma_z} e^{-\frac{i}{2}\theta\sigma_y} e^{-\frac{i}{2}(\pi-\phi)\sigma_z} 
    =
  \lt(\begin{array}{cc} \cos\frac{\theta}{2} & e^{-i\phi}\sin\frac{\theta}{2} \\
        e^{i\phi}\sin\frac{\theta}{2} & -\cos\frac{\theta}{2} \end{array} \rt) 
=\mv\cdot\sigmav,
\label{Udef}
\end{align}
where
\begin{align}
 \mv\equiv \lt(\sin\frac{\theta}{2}\cos\phi,\sin\frac{\theta}{2}\sin\phi,\cos\frac{\theta}{2}\rt).
 \label{mvdef}
\end{align}
The overlap of the electron wave functions at the two sites is thus 
$\langle \nv'|\nv\rangle= \langle \uparrow\!|U(\rv')^{-1} 
U(\rv)|\!\uparrow\rangle$.
When localized spin texture is slowly varying, we can expand
the matrix product with respect to $\av\equiv \rv'-\rv$ as  
$U(\rv')^{-1}U(\rv)=1-U(\rv)^{-1}(\av\cdot\nabla)U(\rv)+O(a^2)$ to obtain 
\begin{align}
\langle \nv'|\nv\rangle \simeq 1-\langle 
\uparrow\!|U(\rv)^{-1}(\av\cdot\nabla)U(\rv)|\!\uparrow\rangle
= e^{i\varphi}+O(a^2),
\label{phasedef}
\end{align}
where 
\begin{align}
\varphi\equiv i\av\cdot\langle\uparrow|U(\rv)^{-1}\nabla 
U(\rv)|\uparrow\rangle\equiv \frac{1}{\hbar}\av\cdot\Asv.\label{varphidef}
\end{align}
Since $(U^{-1}\nabla U)^\dagger =-U^{-1}\nabla U$, $\varphi$ is real.
A vector $\Asv$ here plays a role of a gauge field, similarly to that of the 
electromagnetism, 
and it is  called (adiabatic) spin gauge field. 
By use of Eq. (\ref{Udef}), this gauge field reads (including the factor of $\frac{1}{2}$ 
representing the magnitude of electron spin)
\begin{align}
\Asv\equiv i\hbar \langle\uparrow|U(\rv)^{-1}\nabla 
U(\rv)|\uparrow\rangle =-\frac{\hbar}{2}(1-\cos\theta)\nabla\phi.
\label{Asdef}
\end{align}
For a transport between general points connected by a path $C$, the phase is written as an integral  along $C$ as 
\begin{align}
\varphi=\frac{1}{\hbar} \int_C d\rv\cdot \Asv.\label{phaseAsv}
\end{align}

\subsection{Spin electromagnetic field}
Existence of path-dependent phase means that there is an effective magnetic 
field, $\Bsv$, defined when the contour $C$ is a closed path.
In fact, the contour integral is written by use of the Stokes theorem as a surface integral as 
\begin{align}
\varphi=\frac{e}{\hbar} \int_S d\Sv\cdot\Bsv,
\end{align}
 where
\begin{align}
\Bsv\equiv \nabla\times\Asv,\label{Bsdef}
\end{align}
represents a curvature or an effective magnetic field.
This phase $\varphi$, arising from strong $sd$ interaction, couples to electron 
spin, and is called the spin Berry's phase.  

Time-derivative of phase is equivalent to a voltage, and thus we have an effective 
electric field, too. 
Applying the argument of Eq.(\ref{phasedef}) to the case where spin direction is changing with time, the phase factor attached during $t=0$ to $t=t$ on the electron wave function is $e^{i\varphi(t)}$ with 
\begin{align}
\varphi(t) =\frac{1}{\hbar} \int^t_0 dt A_{{\rm s},t},\label{phaseAst}
\end{align}
where 
\begin{align}
A_{{\rm s},t}\equiv \frac{\hbar}{2}(1-\cos\theta)\partial_t\phi,
\label{Astdef}
\end{align}
is a scalar potential arising from spin dynamics.
The sum of the two contributions to the phase, Eqs. (\ref{phaseAsv})(\ref{phaseAst}) leads to the time-derivative of the total phase as 
\begin{align}
\dot{\varphi}=-\frac{1}{\hbar} \int_Cd\rv\cdot\Esv,
\end{align}
 where 
 \begin{align}
\Esv\equiv-\dot\Asv-\nabla A_{{\rm s},t},\label{Esdef}
\end{align}
is the effective electric field.
The definitions of the fields (\ref{Bsdef})(\ref{Esdef}) are the same as the electromagnetic field of Eq. (\ref{potentials}).
The two fields $\Esv$ and $\Bsv$  coupling to the electron spin are called spin electromagnetic 
fields ($\Asv$ is spin gauge field). 

In terms of vector $\nv$ the effective fields read 
\begin{align}
{\Ev}_{{\rm s},i}&
             = \frac{\hbar}{2} \nv \cdot (\dot{\nv} \times \nabla_i \nv)     
                         \nnr
{\Bv}_{{\rm s},i}&= -\frac{\hbar}{4}{\sum}_{jk}\epsilon_{ijk} \nv \cdot 
(\nabla_j \nv \times \nabla_k \nv).
\label{EsBsdef}
\end{align}
In terms of polar coordinates, the magnetic component reads 
\begin{align}
{\Bv}_{{\rm s},i}&= -\frac{\hbar}{2}{\sum}_{jk}\epsilon_{ijk} \sin\theta (\nabla_j\theta)(\nabla_k\phi),
\label{Bsdef2}
\end{align}
indicating that it has a geometrical meaning of the area defined by the magnetization structure
($\sin\theta\delta\theta\delta\phi$ is the area element for small angle variation $\delta\theta$ and $\delta\phi$). 
The classical effect of the spin electromagnetic field for electron is given by the same as the  conventional electromagnetism ('charge $\frac{1}{2}$ is included in definition of $\Esv$ and $\Bsv$);
\begin{align}
 m\ddot{\rv}&= \pm \Esv \pm \dot{\rv}\times\Bsv,\label{EeqEsBs}
\end{align}
where the sing $\pm$ denotes spin direction.
This is obvious from the minimal coupling form, which we shall argue later in Eqs. (\ref{HAgeneral}) (\ref{gaugeH}).

\begin{figure}[bt]
\begin{center}
\includegraphics[width=0.25\textwidth]{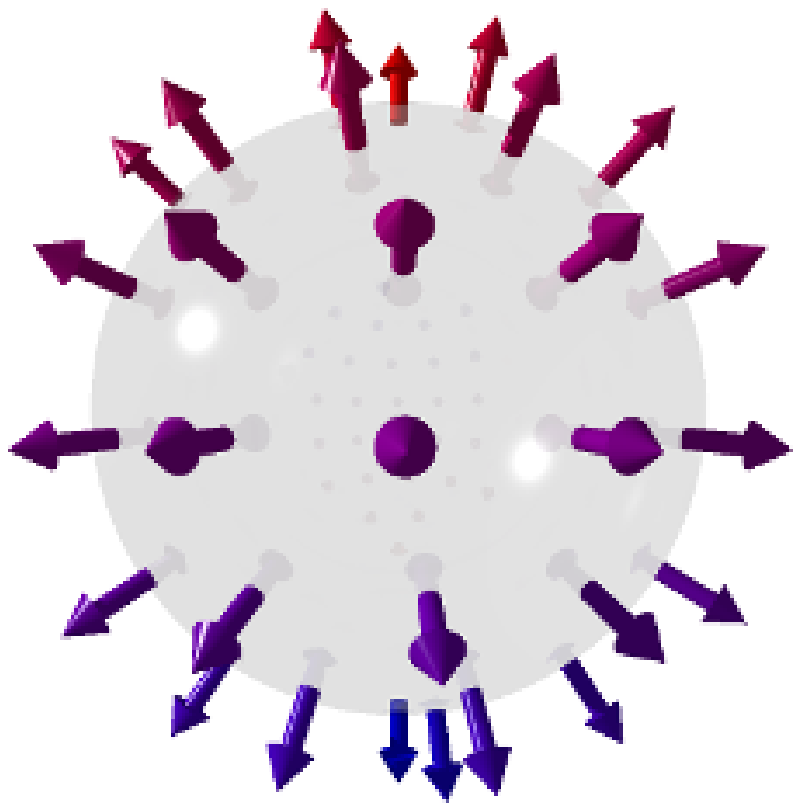}
\includegraphics[width=0.25\textwidth]{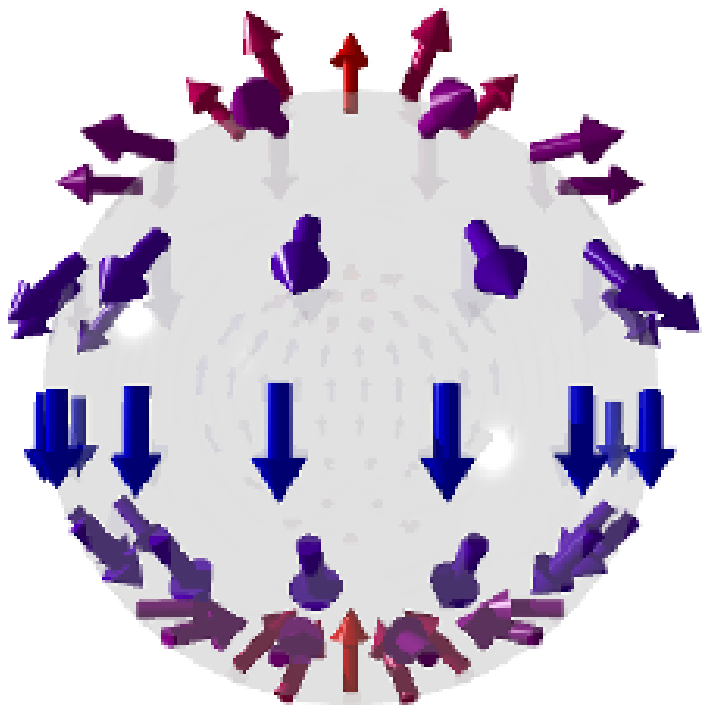}
\end{center}
\caption{
Magnetization structures, $\nv(\rv)$, of a hedgehog monopole having a monopole 
charge of $Q_{\rm m}=1$ and the one with $Q_{\rm m}=2$ .
At the center, $\nv(\rv)$ has a singularity and this gives rise to a finite 
monopole charge. 
}
\label{FIGHH}
\end{figure}
%

\subsection{Topological monopole}
As a trivial consequence of the definition, they appear to satisfy the Faraday's law and condition of no monopole,
\begin{align}
\nabla\times\Esv+\dot{\Bsv}&=0, &
\nabla\cdot\Bsv=0,
\end{align}
because spin vector with fixed 
length has only two independent variables, and therefore 
${\sum}_{ijk}\epsilon_{ijk} (\nabla_i\nv) \cdot (\nabla_j \nv \times \nabla_k 
\nv)=0$.
They are correct as a local equation. The nature, however, sometimes exhibit surprising possibilities that we may not imagine straightforwardly.
In the present case, those equations may be broken globally due to a topological reason.
Let us define a spin magnetic charge (monopole charge) as 
\begin{align}
\nabla\cdot\Bsv\equiv -\rho_{\rm m},
\end{align}
which appears to vanish locally, and show that its volume integral, $Q_{\rm m}\equiv \intr 
\rho_{\rm m}$, can be finite. 
In fact, using the Gauss's law we can write 
\begin{align}
Q_{\rm m} =\int_{r=\infty} d\Sv\cdot \Bsv = \frac{h}{4\pi }\int d\Omega,
\end{align}
where $\int_{r=\infty} d\Sv$ denotes integral over surface at spatial infinity and 
the last integral, $\int d\Omega\equiv \int\sin\theta d\theta d\phi$, is over the spin direction at the infinity.
It thus follows that 
\begin{align}
Q_{\rm m} ={h}\times\textrm{integer}    \label{MPquantization}                                                       \end{align}
 since 
$\frac{1}{4\pi}\int d{\Omega}$ is a winding number, an integer, of a 
mapping from a sphere in the coordinate space to a sphere in spin space. 
If the mapping is topologically non-trivial, the 
monopole charge becomes finite. 
Typical nontrivial structures of $\nv$ are shown in Fig.  \ref{FIGHH}.
The  singular structure with a single monopole charge is called the hedgehog 
monopole from its shape.
In a local picture, such topological monopole arises because the spin configurations having monopole always contain at least one singular point where the derivative $\partial_\mu \nv$ diverges. 
In the case of as symmetric hedgehog monopole, singularity is at the center of monopole. 
Such singularities cannot be removed by continuous deformation of spin configuration, and is therefore topologically stable in a continuum.
One should notice, however, that the topological stability is not exact in solids; since localized spins are on a discrete lattice, singularities can be annihilated or created with a finite excitation energy. This fact reduces mathematical beautifulness, but is essential for applications, since 'topological' objects like domain wall or vortex can be created externally.

Similarly, the Faraday's law reads
\begin{align}
 (\nabla\times\Esv)_i+\dot{\Bsv}_i=-\frac{\hbar}{4}\sum_{ijk}\epsilon_{ijk} 
\dot\nv \cdot (\nabla_j \nv \times \nabla_k \nv) \equiv -\jv_{\rm m},
\end{align}
 which vanishes locally but is finite when integrated, allowing a topological monopole current $\jv_{\rm m}$ to be finite.

The gauge field has a constraint arising from the requirement 
that a gauge field covering the whole space without singularity be constructed by patching together locally-defined gauge fields.
In fact, a definition (for spatial component)
\begin{align}
{A}_{{\rm s},i}^{\rm N}= -\frac{g}{4\pi}(1-\cos\theta)\nabla_i \phi ,\label{ANorth}
\end{align}
where superscript N denotes north and $g$ is the monopole charge,
is not well-defined at $\theta=\pi$ (south pole).
A gauge field that is regular at the south pole is defined as
\begin{align}
{A}_{{\rm s},i}^{\rm S}= \frac{g}{4\pi}(1+\cos\theta)\nabla_i \phi .\label{ASouth}
\end{align}
This field has a singularity at the north pole ($\theta=0$), but represents the same effective magnetic field ($\nabla\times {\Av}_{\rm s}^{\rm N}=\nabla\times {\Av}_{\rm s}^{\rm S}$ except at poles). 
To cover the whole space by either of the gauge fields,  we have necessarily a singularity. 
The singularity is the Dirac string.
Instead of playing with singular gauge field, we can cover the whole space by patching two gauge fields, ${A}_{{\rm s},i}^{\rm S}$ and ${A}_{{\rm s},i}^{\rm N}$.
They are related by a gauge transformation
\begin{align}
{A}_{{\rm s},i}^{\rm N}={A}_{{\rm s},i}^{\rm S} +i{\hbar} \Theta^{-1}\nabla_i \Theta,
\end{align}
where 
\begin{align}
\Theta \equiv e^{-i\frac{g}{h}\phi}
\end{align}
is a gauge transform function.
This function must be single-valued, i.e., be invariant under $\phi\ra\phi+2\pi$.
Thus, a condition 
\begin{align}
g=2\pi\hbar n,\label{diracquantization2}
\end{align}
where $n$ is an integer is imposed (Dirac's quantization condition).
This condition imposing the magnitude of spin to be $\frac{n}{2}$ is quantization of spin.

The other two Maxwell's equations describing $\nabla\cdot\Esv$ and 
$\nabla\times\Bsv$ are derived by evaluating the induced spin density and spin 
current based on linear response theory \citep{Takeuchi_LT12,Tatara12}.
It may appear surprising that the complete Maxwell's equations for electromagnetism is derived by discussion of spin-polarized electron. 
It is, however, just natural, because electromagnetism arises from a conservation law of charge, which corresponds to spin angular momentum in our adiabatic context.
Therefore, electromagnetism is automatically derived if we carry out a correct calculation that keeps the conservation law.

\begin{figure}[tb]
  \begin{center}
    \includegraphics[width=0.33\hsize]{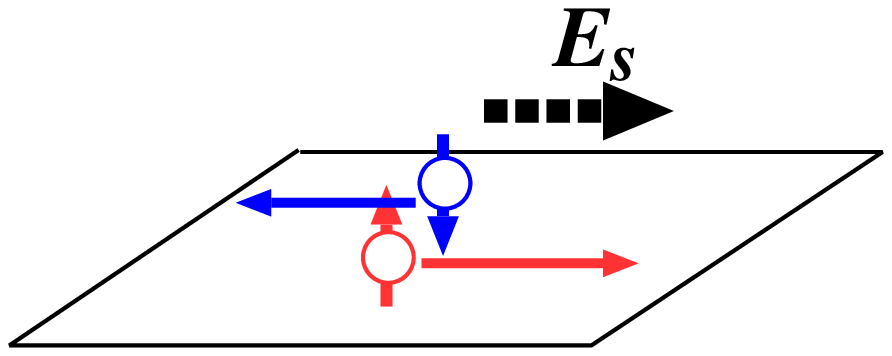}
    \includegraphics[width=0.33\hsize]{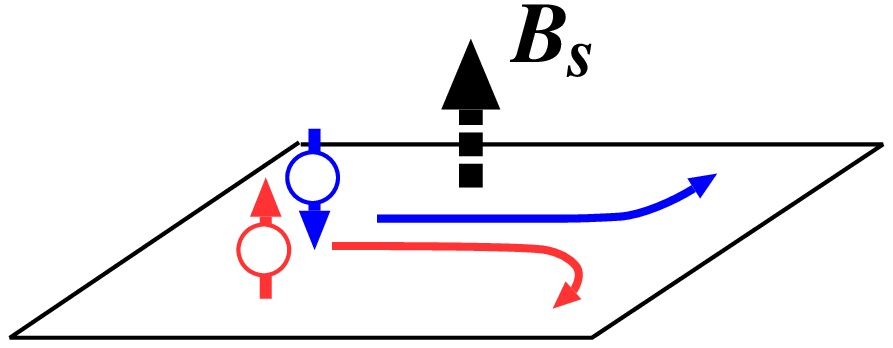}
  \end{center}
\caption{Spin electric field $\bm{E}_{\rm s}$ and spin magnetic field 
$\bm{B}_{\rm s}$ act oppositely for electrons with opposite spin, 
generating spin current.   
\label{FIGEsBs}}
\end{figure}

The electromagnetism of spin gauge field was discussed by G. Volovik \citep{Volovik87}, and SU(2) hedgehog monopole was argued. The mechanism for emergence of spin gauge field and monopole is identical to the one pointed out by G. t'Hooft and A. M. Polyakov \citep{tHooft74,Polyakov74} for the case of larger non-Abelian group, introduced for explaining the emergence of electromagnetism as a result of a symmetry breaking of a grand unified theory (GUT). 

The idea of spin gauge field (spin vector potential) was introduced to describe Heisenberg models by P. Chandra et al. \citep{Chandra90}.
SU(2) gauge description of spin and charge transport was discussed in the Boltzmann equation approach in Ref. \citep{Raimondi12}.
Effective gauge field (artificial gauge field) plays important roles also in cold atom systems \citep{Phuc15,Aidelsburger17}.

\subsection{Detection of spin electromagnetic fields}
Although spin magnetic field is often called a 'fictitious' magnetic field, spin magnetic fields are real fields detectable in transport measurements.
They couple to the spin polarization of the electrons according to Eq. (\ref{EeqEsBs}) (Fig. \ref{FIGEsBs}), and so they are measurable by spin current measurements.
Fortunately, in ferromagnetic conductors, conventional electric measurements are sufficient for detection,  
because spin current $\js$ is always 
accompanied with electric current $j$ as $\js=Pj$, where $P$ is the spin polarization.
The electric component $\Esv$ is therefore directly observable as a voltage generation 
from magnetization dynamics. In experiments, voltage signals of $\mu$V order have been 
observed for the motion of domain walls and vortices \citep{Yang09,Tanabe12}.
The spin magnetic field causes an anomalous Hall effect of spin, i.e., the spin 
Hall effect or the topological Hall effect.
The spin electric field arises if magnetization structure carrying spin magnetic 
field becomes dynamical due to the Lorentz force from $\Bsv$ according to 
$\Esv=\vv\times\Bsv$, where $\vv$ denotes the electron spin's velocity.
The topological Hall effect due to skyrmion lattice turned out to induce Hall 
resistivity of 4n$\Omega$cm \citep{Neubauer09,Schulz12}.
Although those signals are not large, existence of spin electromagnetic fields 
is thus confirmed experimentally.
It was recently shown theoretically that spin magnetic field couples to helicity 
of circularly polarized light (topological inverse Faraday effect) 
\citep{Taguchi12}, and an optical detection is thus possible.

\section{Minimal coupling of spin gauge field\label{SECminimamcoupling}}
So far we have considered adiabatic component of effective gauge field, starting from 
a phase factor attached to electron spin.
As we see from its construction, the gauge field has originally three components 
corresponding to spin space, and thus is an SU(2) gauge field.
It reduced to an effective U(1) gauge field when an expectation value was taken in Eq. (\ref{varphidef}). 
Here we discuss the effect of the effective gauge field taking account of its  
SU(2) nature.

The effective gauge field arising from a unitary transformation $U$, defined in Eq. (\ref{Udef}), is (negative and positive signs correspond to $\mu=t$ and $\mu=x,y,z$, respectively)
\begin{align}
 {\cal A}_{{\rm s},\mu} &= \mp i\hbar U^{-1}\partial_\mu U,
\end{align}
and is expressed by use of Pauli matrices as
\begin{align}
 {\cal A}_{{\rm s},\mu} &=  {\cal A}_{{\rm s},\mu}^\alpha \sigma_\alpha \equiv \Ascalv{\mu}\cdot\sigmav.\label{Ascomp}
\end{align}
The three spin components of vector $ \Ascalv{\mu}$ are
\begin{equation}
\Ascalv{\mu}= \pm \frac{\hbar}{2}
\vecth{
-\partial_\mu \theta \sin \phi -\sin\theta \cos\phi \partial_\mu \phi }{
\partial_\mu \theta \cos \phi -\sin\theta \sin\phi \partial_\mu \phi }{
  (1-\cos\theta)\partial_\mu \phi  }.
\label{Aexpression}
\end{equation}
It can be represented as 
\begin{align}
\Ascalv{\mu}= \pm\frac{\hbar}{2}\nv\times\partial_\mu\nv -{A}_{{\rm s},\mu} \nv,\label{ADMAz}
\end{align}
where $A_{{\rm s},\mu}\equiv {\cal A}_{{\rm s},\mu}^z $ is the adiabatic spin gauge field we discussed in Sec. \ref{SEC:SEMF}.

Being a gauge field for electron spin, the spin gauge field couples to the spin 
current via the minimal coupling.
To the first order, the coupling reads (see Sec. \ref{SEC:fieldlagrangian} for derivation) 
\begin{align}
 H_{\cal A}= -\intr\lt[ j_{{\rm s},i}^\alpha {\cal A}_{{\rm s},i}^\alpha -s^\alpha {\cal A}_{{\rm s},t}^\alpha \rt].
 \label{HAgeneral}
\end{align}
Here $j_{{\rm s},i}^\alpha$ and $s^\alpha$ denote spin current and spin density in the frame after unitary transformation, i.e., in the rotated frame, respectively. (The effective gauge field is a quantity defined in the rotated frame.) 
Let us write the rotated frame spin current and density as 
\begin{align}
 j_{{\rm s},i}^\alpha&=j_{{\rm s},i} \hat{z}_\alpha +(j_{{\rm s},i}^\perp)^\alpha,
& s^\alpha =s \hat{z}_\alpha +(s^\perp)^\alpha,
\end{align}
where $(j_{{\rm s},i}^\perp)^\alpha\equiv j_{{\rm s},i}^\alpha-\hat{z}_\alpha j_{{\rm s},i}$ and
$(s^\perp)^\alpha\equiv s^\alpha-\hat{z}_\alpha s$ represent nonadiabatic components, with 
$ j_{{\rm s},i}\equiv  j_{{\rm s},i}^z$ and $s\equiv s^z$.
The gauge coupling then reads 
\begin{align}
 H_{\cal A}= -\intr\lt[j_{{\rm s},i} {A}_{{\rm s},i}  -s {A}_{{\rm s},t}  +  (j_{{\rm s},i}^\perp)^\alpha ({\cal A}_{{\rm s},i}^\perp)^\alpha -(s^\perp)^\alpha ({\cal A}_{{\rm s},t}^\perp)^\alpha \rt],
 \label{HAgeneral2}
\end{align}
where  $({\cal A}_{{\rm s},\mu}^\perp)^\alpha\equiv {\cal A}_{{\rm s},\mu}^\alpha-\hat{z}_\alpha A_{{\rm s},\mu}$ is the nonadiabatic gauge field.

\begin{figure}[tb]
  \begin{center}
    \includegraphics[width=0.3\hsize]{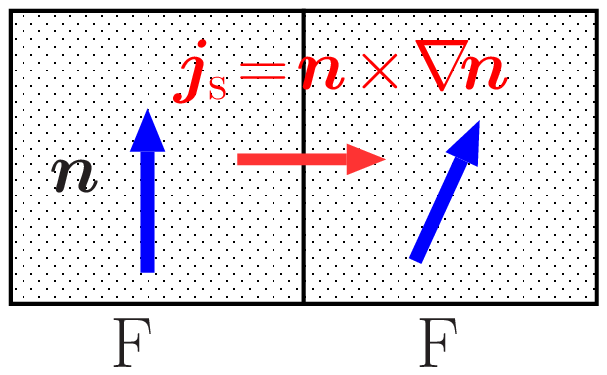}
    \includegraphics[width=0.3\hsize]{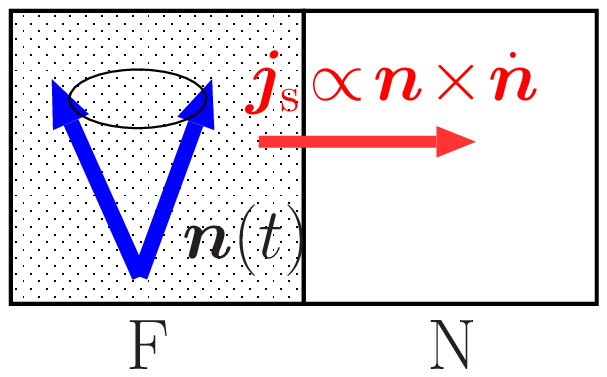}
  \end{center}
\caption{Spin current ($\js$) generation from magnetization. Left figure represetns an equilibrium spin current due to magnetization structure, while right figure describes the spin pumping effect in a FN junction.
In spin pumping, spin accumulation is generated by effective chemical potential $\nv\times\dot{\nv}$ at the interface, resulting in a spin current proportional to  $\nv\times\dot{\nv}$ in N.
The  effects are  described by the minimal coupling between the spin current and spin gauge field arising from the magnetization texture or dynamics.  
\label{FIGFFFN}}
\end{figure}

We now argue that the gauge coupling directly indicates the following several important effects.
\begin{enumerate}
 \item Effects on electron transport
 \begin{enumerate}
   \item Adiabatic spin electromagnetic field 
   \item Spin current  generation  
 \end{enumerate}
 \item Effects on magnetism
 \begin{enumerate}
   \item Spin-transfer effect 
   \item Antisymmetric exchange (DM) interaction   
 \end{enumerate}
\end{enumerate}
The adiabatic spin gauge field was already explained 
 in Sec. \ref{SEC:SEMF}. Let us briefly explain other effects.
\paragraph{Spin current generation} 
Application of spin gauge field, $\Ascalv{\mu}^\alpha$, leads to generation of spin 
current and density.
Of particular interest is the first term of Eq. (\ref{ADMAz}).
Its spatial component induces equilibrium spin current 
proportional to $\nv\times\nabla_i\nv$ (Fig. \ref{FIGFFFN}). This spin current  represents a torque acting between non collinear localized spins \citep{TKS_PR08}. 
If in a junction of two ferromagnets with localized spin, $\Sv_i$ ($i=1,2$), the current reduces to a discrete form of $j_{{\rm s},i}^\alpha\propto (\Sv_1\times\Sv_2)^\alpha(\ev_{12})_i$ proportional to vector chirality ($\ev_{12}$ represents the vector connecting site 1 to site 2).
The time-component $\Ascalv{t}$ couples to spin density, and forms a spin accumulation; It is an effective chemical potential for electron spin.
In a junction of ferromagnet and normal metal, the accumulation at the interface leads to a spin current generation into the normal metal proportional to $\nv\times\dot{\nv}$  (Fig. \ref{FIGFFFN}), i.e.,  spin pumping effect occurs \citep{Silsbee79,Tserkovnyak02,TataraSP17}.
This effect is explained in detail in Sec. \ref{SEC:SP}. 

\paragraph{Spin-transfer torque}
The opposite effects of the interaction Eq. (\ref{HAgeneral2})  is the electrons' effects on magnetization. 
In the adiabatic limit,  Eq. (\ref{HAgeneral2}) reduces to 
\begin{align}
 H_A^z= -\intr \lt[\Asv\cdot \jsv-sA_{{\rm s},t}\rt].
 \label{HAz}
\end{align}
Noting that the expression for ${A}_{{\rm s},t}$ coincides with that of spin 
Berry's phase term of the spin Lagrangian, $L_{\rm B}$, 
the spin Lagrangian taking account of Eq. (\ref{HAz}) reads 
\begin{align}
 L_{\rm B}-H_{A}^z  &= \intr \lt[ -\frac{2}{a^3} A_{{\rm s},0}S-sA_{{\rm s},t} + \Asv\cdot \jsv\rt] \nnr
 &= \sumr \lt[ \hbar \bar{S}(\cos\theta-1) \lt( \frac{\partial}{\partial 
t}+\vv_{\rm s}\cdot\nabla \rt) \phi\rt] ,
 \label{LsAs}
\end{align}
where $\bar{S}\equiv S+\frac{sa^3}{2}$ is the magnitude of the total spin and 
\begin{align}
\vv_{\rm s}\equiv \frac{a^3}{2\bar{S}}\jsv.\label{vstt}
\end{align}
This Galilean invariant form indicates that any spin 
structure under spin current flows along the current with velocity $\vv_{\rm 
s}$  in the adiabatic limit. 
This is the spin-transfer effect, which can be applied to drive magnetization 
structure by injecting electric current.
(As the total spin measured in experiments always contains the adiabatic component of electron spin, the localized spin magnitude $S$ (like the on in Sec. \ref{SEC:spin}) should be regarded as $\bar{S}$, although we use notation $S$ for the total spin for simplicity.)

\paragraph{  Dzyaloshinskii-Moriya (DM) interaction}
In contrast, nonadiabatic spin current, $(j_{{\rm s},i}^\perp)^\alpha$, induces antisymmetric exchange interaction, the  Dzyaloshinskii-Moriya (DM) interaction.
This is seen from an identity 
\begin{align}
R_{\alpha\beta}({\cal A}_{{\rm s},i}^\perp)^\beta=-\frac{\hbar}{2}(\nv\times\nabla_i\nv),
\end{align}
where $R_{\alpha\beta}\equiv 2m_\alpha m_\beta-\delta_{\alpha\beta}$ is a rotation matrix, $\mv$ being defined in Eq. (\ref{mvdef}).
In fact, using this identity, the spatial nonadiabatic terms of Eq. (\ref{HAgeneral2}) turns out  to be 
the DM interaction, 
\begin{align}
 H_A^\perp&\equiv -\intr  (j_{{\rm s},i}^\perp)^\alpha ({\cal A}_{{\rm s},i}^\perp)^\alpha  
 = \sumr D_{i}^\alpha (\nv\times \nabla_i\nv)_\alpha,
\end{align}
where 
\begin{align}
 D_i^\alpha= -\frac{\hbar a^3}{2}R_{\alpha\beta}(j^{\perp}_{{\rm s},i})^\beta. \label{Djs}
\end{align}
We have therefore an interesting identity that DM coefficient is the 
magnitude of nonadiabatic spin current in the laboratory frame, 
$(j^{\perp,{\rm (L)}}_{{\rm s},i})^\beta\equiv R_{\alpha\beta}(j^{\perp}_{{\rm s},i})^\beta$ \citep{Kikuchi16}.
(Here the spin current density is defined without electric charge $e$ and spin magnitude $\frac{1}{2}$).
This simple formula tells us a microscopic mechanism for emergence of DM interaction, namely,  inversion symmetry breaking gives rise to a finite intrinsic spin current, and DM interaction arises as a result of 'Doppler shift' \citep{Kikuchi16}.
The form (\ref{Djs}) is unique in the sense that the DM coefficient is not described by a correlation function like most physical parameters like exchange interaction. 
The formula thus enables us numerical evaluation with less computing time than previous formula.
For strong spin-orbit interaction, deviation from Eq. (\ref{Djs}) is theoretically predicted \citep{FreimuthDM17}.

The Doppler shift picture becomes clear if we regard the DM interaction as a modification of ferromagnetic exchange interaction due to spin current. 
In fact, in a moving frame, a spatial derivative is replaced by a covariant form 
\citep{KimChirality13,Kikuchi16}
\begin{align}
 \mathfrak D_i n_\alpha &= \nabla_i n_\alpha +\eta\epsilon_{\alpha\beta\gamma}(j_{{\rm s},i}^{\rm (L)})^\beta n_\gamma, \label{covariant}
\end{align}
where $\eta$ is a coefficient. 
Similar Doppler shift for a vector in a moving medium has been known
in the case of the velocity vector of sound wave \citep{LandauLifshitz-FluidMechanics}.
The magnetic exchange energy induced by electron, proportional to $(\nabla\nv)^2$ in the rest frame, is then modified to be  
$(\mathfrak D_i\nv)^2=(\nabla\nv)^2+2\eta \sum_i\jv_{{\rm s},i}^{\rm (L)}\cdot(\nv\times\nabla_i\nv)+O(\eta^2)$, resulting in a DM interaction

It has been known that in the presence of DM interaction spin waves around uniform ferromagnetic state show  nonreciprocal propagation as a result of Doppler shift \citep{Kataoka87}, as confirmed in recent experiments  \citep{Iguchi15,Seki16}. 
This effect is natural from our theory, because DM interaction itself is a sign of internal flow of spin.

The spin current that determines the DM interaction by Eq.  (\ref{Djs}) can be the equilibrium one or the non-equilibrium  one such as the one injected externally. 
Our formula (\ref{Djs}) thus indicates an interesting possibility to modulate DM interaction by injecting spin current.
For a spin current density of $e\js=10^{12}$A/m$^2$, the modulation is estimated to be 
$\delta D=\frac{\hbar a^3}{2}j_{\rm s}=2.6\times10^{-33}$ Jm$=0.16$ meV\AA\ for $a=2$\AA.
This value is an order of magnitude smaller than the one in natural strongly chiral materials such as MnFeGe. 
In this sense, intrinsic spin current induced by atomic spin-orbit interaction is larger than what we can do.
Nevertheless, external control of DM interaction by current application would be useful for weakly chiral magnets.
Voltage  control of  DM interaction  has been experimentally demonstrated  recently \citep{Nawaoka15}.

\subsection{Perturbative picture of spin gauge field}
We discussed emergence of spin gauge field so far in the adiabatic limit. 
The concept of spin gauge field exists also in the weak $sd$ exchange interaction regime.
In fact, adiabatic condition justifies gradient expansions and adiabatic limit can be realized even in the weak $sd$ coupling case if the gradient is small enough.
(See Eq. (\ref{adiabaticcondition}).) 
In this subsection, we present a perturbative picture of spin gauge field in the weak $sd$ limit.

\begin{figure}[tb]
  \begin{center}
    \includegraphics[width=0.3\hsize]{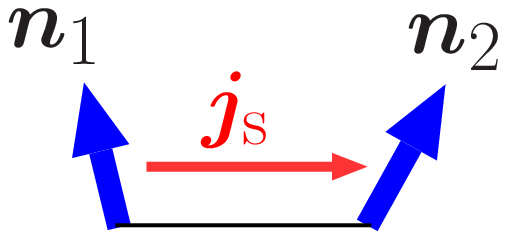} \hspace{0.05\hsize}
    \includegraphics[width=0.3\hsize]{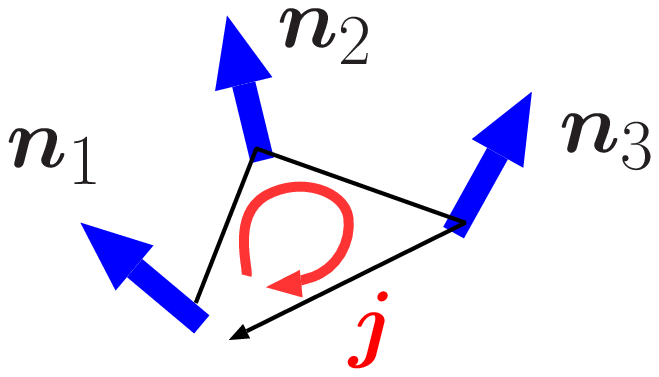}
  \end{center}
\caption{ Equilibrium currents generated by localized spin structures, represented by $\nv_i$. 
Non collinear spin structure (vector spin chirality) induces a spin current with polarization proportional to the vector chirality, $\jsv\propto \nv_1\times\nv_2$, while  
non coplanar spin structure (scalar spin chirality) leads to a charge current $j\propto \nv_1\cdot(\nv_2\times\nv_3)$ as a result of breaking of time-reversal symmetry.   
\label{FIGjjs}}
\end{figure}

The $sd$ exchange interaction with a localized spin $\Sv$ is 
$H_{sd}=-\Jsd\Sv\cdot\sigmav$.
Let us consider interaction with two localized spins 
$\Sv_1$ and $\Sv_2$.
As a second order contribution, the electron spin wave function acquires a phase proportional to 
\begin{align}
 {\cal V}_2 &\equiv (\Jsd)^2 (\Sv_1\cdot\sigmav)(\Sv_2\cdot\sigmav)
 = (\Jsd)^2[(\Sv_1\cdot\Sv_2)+i(\Sv_1\times\Sv_2)\cdot\sigmav] .\label{A2}
\end{align}
The first term on the right-hand side describes the amplitude of charge part, in other words, magnetoresistance effect. 
The second term containing Pauli matrix indicates that spin current (and/or density) are induced as a result of non collinear localized spin as (Fig. \ref{FIGjjs})
\begin{align}
 j_{{\rm s},i}^\alpha &\propto (\Jsd)^2 (\Sv_1\times\Sv_2)^\alpha(\ev_{12})_i, \label{Js2}
\end{align}
where $\ev_{12}$ is a unit vector representing relative spatial position of $\Sv_1$ and $\Sv_2$.
(Here spin current is in the laboratory frame, as rotating frame description is not valid in the perturbative regime.) 
Let us consider a junction of two ferromagnets with localized spins 
$\Sv_1$ and $\Sv_2$ (Fig. \ref{FIGFFFN}).
The spin current (\ref{Js2}) in this case flows between the two ferromagnetic layers.
It is an equilibrium current that arises even in the static spin configuration, and is a kind of persistent or super current if spin relaxation effect is neglected.
Spin current indicates that dynamics is induced as a result of angular momentum change.
The equations of motion for the two localized spins read (neglecting external magnetic field)
\begin{align}
 \dot{\Sv}_1 &= -\frac{\alpha}{S}(\Sv_1\times\dot{\Sv}_1) +c(\Sv_1\times \Sv_2) \nnr
 \dot{\Sv}_2 &= -\frac{\alpha}{S}(\Sv_2\times\dot{\Sv}_2)-c(\Sv_1\times \Sv_2), \label{S1S2eq}
\end{align}
where $c$ is a constant.
Equation (\ref{S1S2eq}) indicates that the two ferromagnets tends to align  
$\Sv_1$ and $\Sv_2$ parallel or anti parallel. 
This is natural because localized spin $\Sv_2$ acts as an effective magnetic field for localized spin $\Sv_1$ and vice versa.
The equilibrium spin current of Eq. (\ref{Js2}) therefore represents the torque mediated by the conduction electron between the two localized spins.
For smooth localized spins, the expression reduces to a continuum form of 
\begin{align}
 j_{{\rm s},i}^\alpha &\propto (\Jsd)^2  (\Sv\times\nabla_i \Sv)^\alpha. \label{Js2cont}
\end{align}

If we regard the two localized spins as the one at different time, Eq. (\ref{Js2}) is a dynamic spin current, namely, we have a spin pumping effect. 
In the slow change of localized spins, the current is proportional to 
\begin{align}
 (\Jsd)^2 (\Sv\times\dot{\Sv}),  \label{A2t}
\end{align}
which is a perturbative picture of spin pumping effect \citep{TataraSP17}.
 
We saw that the second order contribution of the $sd$ exchange interaction is governed by a vector chirality of spins, $ (\Sv_1\times\Sv_2)$ for two spins.
This quantity is the non-adiabatic component of the spin gauge field in the adiabatic limit, as seen in Eq. (\ref{ADMAz}).
The spin gauge field, therefore, arises from the non-commutative algebra of spins.

We can extend the discussion to the third order.
The charge part of the third order amplitude is 
\begin{align}
\tr[{\cal V}_3] &= (\Jsd)^3\tr[(\Sv_1\cdot\sigmav)(\Sv_2\cdot\sigmav)(\Sv_3\cdot\sigmav)]
 = 2i (\Jsd)^3 \Sv_1\cdot(\Sv_2\times\Sv_3), \label{scalarchirality}
\end{align}
namely, proportional to the scalar chirality $\Sv_1\cdot(\Sv_2\times\Sv_3)$ of the three spins.
This indicates that a spontaneous charge current is induced by the scalar chirality of localized spins as a result of broken time-reversal symmetry  (Fig. \ref{FIGjjs}) \citep{Loss92,TK03}. 
This effect is in fact the spin Berry's phase effect as seen by noticing that the continuum limit of the scalar chirality is 
$\Sv\cdot(\partial_i\Sv\times\partial_j\Sv)$ ($i$ and $j$ are direction of relative spin positions), which agrees with the spin magnetic field, Eq. (\ref{EsBsdef}).
The persistent current represented by Eq. (\ref{scalarchirality}) is described as the Amp'ere's law, $\nabla\times\Bsv=\jv$ \citep{Takeuchi_LT12}.

Spin chirality persistent charge current gives rise to an anomalous Hall effect \citep{TK02,TK03}.
It was predicted that spin chirality also affects optical response to circularly polarized light (topological inverse Faraday effect) \citep{Taguchi12}.
Direct observation of persistent current was carried out recently for the case of neutron \citep{Tatarskiy16}.

The spin current (\ref{Js2cont}) is an equilibrium one and cannot be 'converted' into a charge current by use of the inverse spin Hall effect, as was mentioned based on a microscopic analysis  \citep{Takeuchi10}; As for magnetically induced spin current, the inverse spin Hall effect acts only for non equilibrium one, where dynamics is involved. 
In the case of junction of two ferromagnets (Fig.\ref{FIGFFFN}), inverse spin Hall signal shall arise when the magnetizations start to precess following Eq. (\ref{S1S2eq}).
The excess magnetic energy the initial state had  is dissipated as joule heat associated with the charge current.

Equilibrium spin current has been pointed out to induce electric polarization in insulators \citep{Katsura05}. 
This magnetoelectric effect due to magnetic inhomogeneity (spin vector chirality) was predicted earlier in Ref. \citep{Baryakhtar83}.

\subsection{Momentum space monopole}
In Sec. \ref{SEC:SEMF} and in the previous subsection, we discussed spin gauge field in the real space picture.
On the other hand, it has been known that the Berry's curvature in the momentum space plays essential roles in transport phenomena such as anomalous Hall effect \citep{Thouless82,Nagaosa10}.
In clean frustrated magnets anomalous Hall conductivity has been shown to arise from monopoles in the momentum space as a result of a non-coplanar spin structure.
In this momentum picture, role of real space spin magnetic field $\Bsv$ is not clear.
In  contrast, chirality-induced anomalous Hall conductivity in disordered metals was shown to be governed by real space chirality \citep{TK02,Nakazawa14}. 
These features are understood as follows \citep{OTN04}.
In the clean limit, electrons form bands defined including effects of $sd$ exchange interaction, $\Jsd$. The effect of localized spin structures such as chirality are contained in each bands as monopole density. 
In the disordered limit, $\Jsd\taue/\hbar\ll1$ ($\taue$ is elastic lifetime of electron), in contrast,	 bands smeared by energy scale of $\hbar/\taue$ no longer keep the information of spin structure; Instead, the real space  spin structure affects the electron hopping amplitude and transport.

\section{Spin-transfer effect : Phenomenology \label{SECDWtransmit}}
\begin{figure}[tb]
\begin{center}
\includegraphics[width=0.4\hsize]{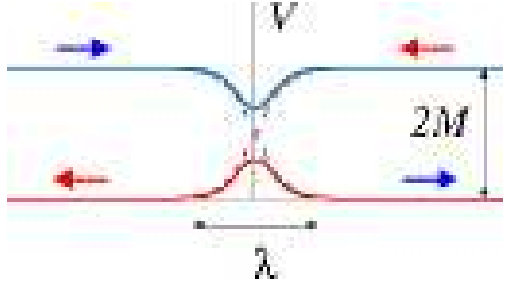}
\end{center}
\caption{Potential energy $V(z)$ arising from $sd$ exchange interaction for conduction electron with spin $\ra$ and $\la$.
The energy gaps is $2\spol=2S\Jsd$.
Dotted lines are the cases 
neglecting spin flip inside the wall, while solid lines are with spin flip. 
\label{FIGDWpotentialenergy}}
\end{figure}

As we have seen in Sec. \ref{SECminimamcoupling}, spin transfer-effect is a direct consequence of minimal coupling between spin current and adiabatic spin gauge field representing magnetization structure.
Here let us present a phenomenological theory for the effect for the case of transmission through a domain wall based on quantum mechanics.
The issue here is how the angular momentum is transfered between conduction electron and localized spin via the $sd$ exchange interaction.
The thickness of the wall, $\lambda$, in typical ferromagnets is $\lambda=10-100$nm, and is much larger 
than the typical length scale of electron, the Fermi wavelength, $1/\kf$, which is atomic scale in metals.
The wall is therefore a slowly varying spin structure for conduction electron.
We choose the $z$ axis along the direction of localized spins' change, and magnetic easy axis for localized spins is chosen 
as along $z$ axis.
\footnote{The mutual direction between the localized spin and direction of spin change is irrelevant in the case without spin-orbit interaction.}
At $z=\infty$ the localized spin is $S_z=S$, and is $S_z=-S$ at $z=-\infty$. We describe electron states with spin along $+z$ and $-z$ direction by $\ra$ and $\la$, respectively.
Because of the domain wall, the electron is in a potential barrier, 
\begin{align}
V_\rightarrow(z)=-\Jsd S_z(z),  V_\leftarrow(z)=\Jsd S_z(z).
\end{align}
Namely, for $\leftarrow$ electron, the potential in the left regime is low, while that in the right region is high (dotted lines in Fig. \ref{FIGDWpotentialenergy}).
\footnote{We choose the sign of $sd$ exchange interaction as positive, but the sign does not matter for the spin-transfer effect.}
Considering the domain wall centered at $x=0$ having profile of 
\begin{align}
 S_z(z)&=S\tanh\frac{z}{\lambda}, \;\;\;  
S_x(z)=\frac{S}{\cosh\frac{z}{\lambda}},\;\;\;  S_y=0 ,
 \label{DWsolrest}
\end{align}
conduction electron's Schr\"odinger equation with energy $E$ reads 
\begin{align}
 \lt[-\frac{\hbar^2}{2m}\frac{d^2}{dz^2}-\Jsd S
  \lt(\sigma_z \tanh\frac{z}{\lambda}+\sigma_x 
\frac{1}{\cosh\frac{z}{\lambda}}\rt) \rt]\Psi=E\Psi,
 \label{elecSeqinDW}
\end{align}
$\Psi(z)=(\Psi_\rightarrow(z),\Psi_\leftarrow(z))$ begin the two-component wave 
function.
If the spin direction of the conduction electron is fixed along the $z$ axis, 
the potential barrier represented by the term proportional to $\sigma_z$ leads 
to reflection of electron, but in reality, the electron spin can rotate inside 
the wall as a result of the term proportional to $\sigma_x$ in Eq. 
(\ref{elecSeqinDW}).
The mixing of $\leftarrow$ and $\rightarrow$ electron leads to the smooth 
potential barrier plotted as solid lines in Fig. \ref{FIGDWpotentialenergy}.

\begin{figure}[tb]
\begin{center}
\includegraphics[width=0.3\hsize]{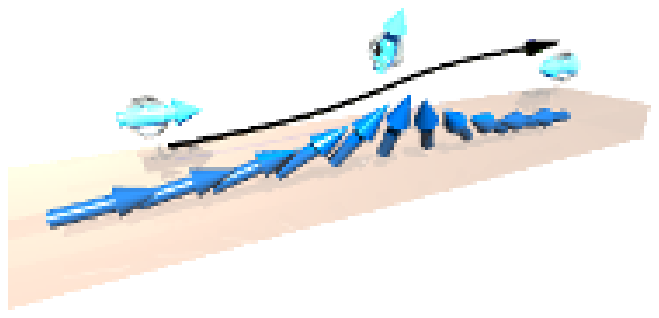}
\hspace{0.1\hsize}
\includegraphics[width=0.3\hsize]{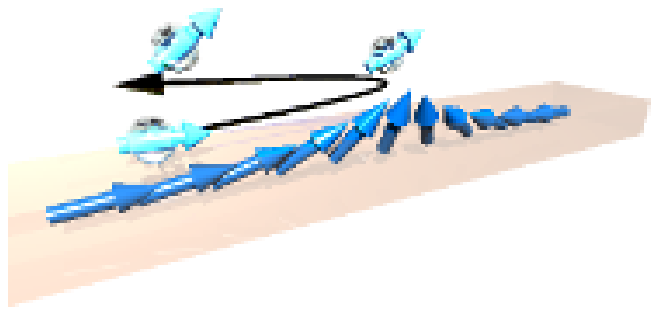}
\caption{ 
Conduction electron incident on a domain wall may go through the wall or get reflected. 
The former process occurs for a thick wall (adiabatic limit) and electron spin is rotated, resulting in a spin transfer effect. The latter process is an nonadiabatic effect, and leads to a force on a domain wall and electric resistance.
\label{FIGDWelec}}
\end{center}
\end{figure}
Let us consider an incident $\leftarrow$ electron from the left.
If the electron is slow, the electron spin can keep the lowest energy state by 
gradually rotating its direction inside the wall.
This is the adiabatic limit.
As there is no potential barrier for the electron in this limit, no reflection 
arises from the domain wall, resulting in a vanishing resistance  (Fig. 
\ref{FIGDWelec}(a))
In contrast, if the electron is fast, the electron spin cannot follow the 
rotation of the localized spin, resulting in a reflection and finite resistance 
(Fig. \ref{FIGDWelec}(b)).
The condition for slow and fast is determined by the relation between the time 
for the electron to pass the wall and the time for electron spin rotation. 
The former is $\lambda/\vf$ for electron with Fermi velocity $\vf(=\hbar\kf/m)$ 
(spin-dependence of the Fermi wave vector is neglected  and $m$ is the electron 
mass).
The latter time is $\hbar/\Jsd S$, as the electron spin is rotated by the $sd$ 
exchange interaction in the wall.
Therefore, if 
\begin{align}
\frac{\lambda}{\vf}\gg \frac{\hbar}{\Jsd S} , \label{adiabaticcondition}
\end{align}
is satisfied, the electron is in the adiabatic limit \citep{Waintal04}.
The condition of adiabatic limit here is the case of clean metal (long mean free 
path);
 In dirty metals, it is modified \citep{Stern92,TKS_PR08}.

The transmission of electron through a domain wall was calculated by G. G. 
Cabrera and L. M. Falicov \citep{Cabrera74}, and its physical aspects were 
discussed by L. Berger \citep{Berger78,Berger86}.
Linear response formulation and scattering approach were presented in Refs. 
\citep{TF97,GT00,GT01}.

\begin{figure}[tb]
\begin{center}
\includegraphics[width=0.3\hsize]{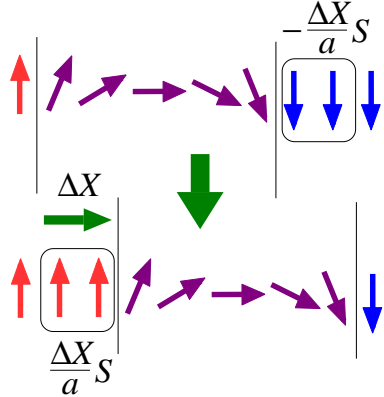}
\end{center}
\caption{ The shift of the domain wall by a distance $\Delta X$ results in a 
change of the spin of the localized spins $\frac{\Delta X}{a}S-\lt(-\frac{\Delta 
X}{a}S\rt)=2S\frac{\Delta X}{a}$.
The angular momentum change is therefore $\hbar$ if $\Delta X=\frac{a}{2S}$. 
\label{FIGdwdisplacement}}
\end{figure}

As we have seen above, in the adiabatic limit, the electron spin gets rotated 
after passing through the wall (Fig. \ref{FIGDWelec}(a)).
The change of spin angular momentum, $2\times\frac{\hbar}{2}=\hbar$, must be 
absorbed by the localized spins.
(Angular momentum dissipation as a result of spin relaxation is slow compared to 
the exchange of the angular momentum via the $sd$ exchange interaction.)
To absorb the spin change of $\hbar$, the domain wall must shift to the right, 
resulting in an increase of the spins $\leftarrow$.
We consider for simplicity the case of cubic lattice with lattice constant $a$.
The distance of the wall shift $\Delta X$  necessary to absorb the electron's 
spin angular momentum of $\hbar$ is then $[\hbar/(2\hbar S)]a$ (Fig. 
\ref{FIGdwdisplacement}).
When we apply a spin-polarized current through the wall with the spin current density $j_{\rm 
s}$
(spin current density is defined to have the unit of 1/(m$^2$s) and without spin magnetitude of $\frac{1}{2}$), 
the rate of the change of spin angular momentum of conduction electron per unit time 
and unit area is 
$\hbar j_{\rm s}$.
As the number of the localized spins in the unit area is $1/a^2$, the wall must 
keep moving a distance of 
$j_{\rm s}(a^3/2S)$ per unit time.
Namely, when a spin current density is applied, the wall moves with the speed 
\begin{align} 
v_{\rm s}\equiv \frac{a^3}{2S} j_{\rm s} \nonumber
\end{align} 
which agrees with the speed we obtained in Eq. (\ref{vstt}). 
It should be noted that a simple Lagrangian argument of Eq. (\ref{LsAs}), even without physical argument, is sufficient to draw the conclusion.

The effect was pointed out by L. Berger \citep{Berger86} in 1986, and is now 
called the spin-transfer effect after the papers by J. Slonczewski 
\citep{Slonczewski96}.

From the above considerations  in the adiabatic limit, we found that a 
domain wall is driven by  spin-polarized current, while the electrons do not get 
reflected and no resistance arises from the wall.
These two facts naively seem inconsistent, but are direct consequence of the 
fact that  a domain wall is a composite structure having both linear momentum 
and angular momentum.
The adiabatic limit is the limit where angular momentum is transfered between 
the electron and the wall, while no linear momentum is transfered.

\section{Spin pumping effect \label{SEC:SP}}
Spin pumping effect is a method to generate spin current in a junction of a ferromagnet (F) and a normal metal (N) (Fig. \ref{FIGFFFN}) by exciting magnetization precession by applying an oscillating magnetic field.
The generated spin current density has two independent components, proportional to $\dot{\nv}$ and $\nv\times\dot{\nv}$, where $\nv$ is a unit vector describing the direction of localized spin, and thus is represented phenomenologically as 
\begin{equation}
{\jv}_s = \frac{1}{4\pi} \left( A_{\rm r} \nv \times \dot{\nv}+ A_{\rm i} \dot{\nv} \right),\label{Jsphenom}
\end{equation}
where $A_{\rm r} $ and $A_{\rm i}$ are phenomenological constants having unit of $1/$m$^2$.
(Spin current here is obviously in the laboratory frame, as it is the one in the normal metal.)
Spin pumping effect was theoretically formulated by Tserkovnyak et al. \citep{Tserkovnyak02}
 by use of scattering matrix approach.
This approach, widely applied in mesoscopic physics, describes transport phenomena in terms of transmission and reflection amplitudes (scattering matrix), and provides quantum mechanical pictures of the phenomena without calculating explicitly the amplitudes \citep{Moskalets12}. 
Tserkovnyak et al. applied the scattering matrix formulation of general adiabatic pumping \citep{Buttiker94,Brouwer98} to the spin-polarized case. 
The spin pumping effect was described in Ref. \citep{Tserkovnyak02} in terms of spin-dependent transmission and reflection coefficients at the FN interface,
and  it was demonstrated that the two parameters, $A_{\rm r} $ and $A_{\rm i}$, are the real and the imaginary part of a complex parameter called the spin mixing conductance.
The spin mixing conductance, which is represented by transmission and reflection coefficients, turned out to be a convenient parameter for discussing spin current generation and other effects like the inverse spin-Hall effect. 
At the same time, scattering approach hides microscopic physical pictures of what is going on, as the scattering coefficients are not fundamental material parameters but are composite quantities of Fermi wave vector,  electron effective mass and the interface properties.
Formulation of spin pumping effect based on the Green's function method were presented in Refs. \citep{Chen_spinpump09,Mahfouzi12,Chen15,TataraSP16,TataraSP17}.
In this section, we describe the effect from a standard microscopic view point, following the approach of Ref. \citep{TataraSP17}.

Spin pumping effect is experimentally observed for both metallic and insulating ferromagnets.
From physical viewpoints, these two cases appear very different.
In the metallic case, conduction electron in the ferromagnet is excited by spin gauge field arising from spin dynamics, leading to a spin accumulation at FN interface and spin current generation in the normal metal. In contrast, in the case of insulator ferromagnet, the coupling between the magnetization and the conduction electron in normal metal occurs due to a magnetic proximity effect at the interface \citep{Kang17} and the pumping effect is a locally-induced perturbative effect.
In this paper, we consider the metallic case. The insulator case is discussed in Ref. \citep{TataraSP17}.

The model we consider is a junction of metallic ferromagnet (F) and a normal metal (N). 
The magnetization (or localized spins) in the ferromagnet is treated as spatially uniform but changing with time.
The frequency of magnetization precession is of the order of 10GHz, and is far low frequency compared to conduction electron spin's frequency determined by the $sd$ exchange interaction; For $\Jsd=0.1\ef$, the frequency is $\Jsd/\hbar\sim 2\times 10^{4}$ GHz if $\ef=1$ eV.  As a result, the conduction electron's spin follows instantaneous directions of localized spins, i.e., the system is in the adiabatic limit.
Adiabatic limit is described straightforwardly by introducing a unitary transformation that represents the time-dependence.
For the ferromagnet, we consider a simple quantum mechanical Hamiltonian, 
\begin{align}
 H_{\rm F} 
  &=  -\frac{\hbar^2\nabla^2}{2m} -\ef-\spol \nv(t)\cdot\sigmav ,
\end{align}
where $m$ is the electron's mass, $\sigmav$ is a vector of Pauli matrices, $\spol$ represents the energy splitting due to the $sd$ exchange interaction and $\nv(t)$ is a time-dependent unit vector  denoting the localized spin direction. 
The energy is measured from the Fermi energy $\ef$.
For simplicity, we consider the case $0<\spol<\ef$.

As a result of the $sd$ exchange interaction, the electron's spin wave function is given by \citep{Sakurai94}
\begin{align}
 \ket{\nv}\equiv \cos\frac{\theta}{2}|\!\uparrow\rangle+\sin\frac{\theta}{2}e^{i\phi}|\!\downarrow\rangle                                                                                                    
\end{align}
where $\ket{\uparrow}$ and $\ket{\downarrow}$ represent the spin up and down states, respectively, and $(\theta,\phi)$ are polar coordinates for $\nv$.
To treat slowly varying localized spin, we switch to a rotating frame where the spin direction is defined with respect to instantaneous direction $\nv$. 
This corresponds to diagonalizing the Hamiltonian at each time by  introducing a unitary matrix $U(t)$ as 
\begin{align}
 \ket{\nv(t)}\equiv U(t)|\!\uparrow\rangle,                                                                                             
\end{align}
where $ U(t)$ is a unitary matrix determined by polar angles $\theta$ and $\phi$ as in Eq. (\ref{Udef}), but angles are time-dependent.
The Hamiltonian in the rotated frame is diagonalized as (in the momentum representation)   
\begin{align}
 \widetilde{H}_{\rm F}\equiv U^{-1}H_{\rm F} U= \ekv-\spol \sigma_z ,
\end{align}
where $\ekv\equiv \frac{\hbar^2k^2}{2m}-\ef$ is the kinetic energy.
%
\begin{figure}[tbh]
\begin{minipage}{0.5\hsize}
  \begin{center}
  \includegraphics[height=8\baselineskip]{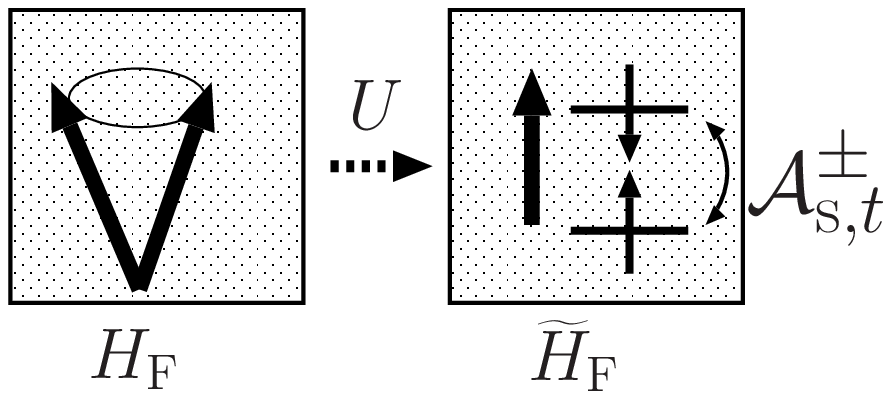}
  \end{center}
\caption{ Unitary transformation $U$ for conduction electron in ferromagnet converts the original Hamiltonian ${H}_{\rm F}$ into a diagonalized uniformly spin-polarized Hamiltonian $\widetilde{H}_{\rm F}$ and an interaction with spin gauge field, $\Ascalv{t}\cdot\sigmav$.
\label{FIGFNQM}}
\end{minipage}
\begin{minipage}{0.5\hsize}
  \begin{center}
    \includegraphics[height=8\baselineskip]{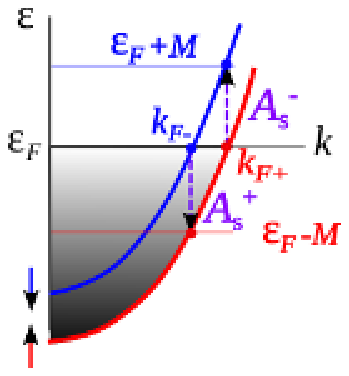}
  \end{center}
\caption{For uniform magnetization, the non-adiabatic components of the gauge field, $\Ascal{t}{\pm}$, induce a spin flip conserving the momentum. Excitation thus has an energy of $\spol$.
\label{FIGelectronband_sp}}
\end{minipage}
\end{figure}
As a result of unitary transformation, there arises in a rotated frame a time-component of a gauge field with three spin components, 
 $\Ascal{t}{} \equiv -i U^{-1}\delpo{t}U$  (Fig. \ref{FIGFNQM}).
Including the gauge field in the Hamiltonian, the effective Hamiltonian in the rotated frame reads 
\begin{align}
 \widetilde{H}_{\rm F}^{\rm eff}\equiv\widetilde{H}_{\rm F} +  {\bf {\cal A}}_{{\rm s},t} \cdot \sigmav
 = \lt( \begin{array}{cc} 
         \epsilon_{k}-\spol -\Ascal{t}{z}  & \Ascal{t}{-} \\
                    \Ascal{t}{+}  & \epsilon_{k}+\spol +\Ascal{t}{z} 
        \end{array} \rt)
        \label{HF}
\end{align}
where 
$\Ascal{t}{\pm}\equiv \Ascal{t}{x}\pm i \Ascal{t}{y}$.
We see that the adiabatic ($z$) component of the gauge field, $\Ascal{t}{z}$, acts as a spin-dependent chemical potential (spin chemical potential) generated by dynamic magnetization, while non-adiabatic ($x$ and $y$) components causes spin mixing.

The Hamiltonian Eq. (\ref{HF}) is diagonalized to obtain energy eigenvalues of  
$\tilde{\epsilon}_{k\spinindex}=\ekv-\spinindex\sqrt{(\spol+\Ascal{t}{z})^2+|\Ascal{t}{\perp}|^2}$, where 
$|\Ascal{t}{\perp}|^2\equiv \Ascal{t}{+}\Ascal{t}{-}$ and $\sigma=\pm$ represents spin ($\uparrow$ and $\downarrow$ correspond to $+$ and $-$, respectively).
We are interested in the adiabatic limit, and so the contribution lowest-order, namely, the first order, in the perpendicular component, 
$\Ascal{t}{\perp}$, is sufficient. 
In the present rotating-frame approach, the gauge field is treated as a static potential, since it already include time-derivative to the linear order (Eq. (\ref{Aexpression})). 
Moreover, the adiabatic component of the gauge field, $\Ascal{t}{z}$,  is neglected, as it modifies the spin pumping only at the second-order of time-derivative. 
The energy eigenvalues, $\ekvs\simeq \ekv-\spinindex\spol$,  are thus unaffected by the gauge field.

In the case of uniform magnetization we consider, the mixing due to the gauge field is between the electrons with different spin $\uparrow$ and $\downarrow$ but having the same wave vector $\kv$, because the gauge field  $\Ascal{t}{\pm} $ carries no momentum.
This leads to a mixing of states having an excitation energy of $\spol$ as shown in Fig. \ref{FIGelectronband_sp}.
In low energy transport effects, what concern are the electrons at the Fermi energy; The wave vector $\kv$ should be chosen as $\kfu$ and $\kfd$, the Fermi wave vectors for $\uparrow$ and $\downarrow$ electrons, respectively. 
The eigenstates we consider therefore read 
\begin{align}
\ket{{k_{{\rm F}\uparrow}\uparrow}}_{\rm F} &= \ket{k_{{\rm F}\uparrow}\uparrow} -\frac{\Ascal{t}{+}}{\spol}\ket{k_{{\rm F}\uparrow}\downarrow} \nnr
\ket{{k_{{\rm F}\downarrow}\downarrow}}_{\rm F} &= \ket{k_{{\rm F}\downarrow}\downarrow} +\frac{\Ascal{t}{-}}{\spol}\ket{k_{{\rm F}\downarrow}\uparrow} .
\label{FSstates}
\end{align}

\subsection{Generated spin current  \label{SEC:QMN}}
Spin pumping effect is now studied by taking account of the interface hopping effects on states in Eq. (\ref{FSstates}). 
The interface hopping amplitude of electron in F to N with spin $\spinindex$  is denoted by $\ttil_\spinindex$ and the amplitude from N to F is  $\ttil_\spinindex^*$. 
We assume that the spin-dependence of electron state in F is governed by the relative angle to the magnetization vector, and hence the spin $\spinindex$ is the one in the rotated frame.
Assuming moreover that there is no spin flip scattering at the interface, the amplitude 
$\ttil_\spinindex$ is diagonal in spin.  
Taking account of spin-dependent interface hopping, the non-adiabatic spin density (in the rotated frame) generated  in the N region at the interface was calculated field-theoretically by \cite{TataraSP17}.
The result is
\begin{align}
 \widetilde{\sv}^{\rm (N)}=
 (\pi\dos_{\rm N})^2 \chi_{\rm F} 
 \lt(\Re[T_{\uparrow\downarrow}]\Ascal{t}{\perp}+\Im[T_{\uparrow\downarrow}](\zvhat\times \Ascal{t}{\perp}) \rt) \label{sevrot}
 \end{align}
where $\Ascalv{t}^{\perp}=(\Ascal{t}{x},\Ascal{t}{y},0)=\Ascalv{t}-\zvhat \Ascal{t}{z}$
 is the transverse (non-adiabatic) components of spin gauge field, 
 \begin{align}
T_{\spinindex\spinindex'}\equiv {\ttil}^*_\spinindex {\ttil}_{\spinindex'},\label{Tdef}
\end{align}
$\dos_{\rm N}$ is electron density of states in N 
and $\chi_{\rm F}\equiv \frac{n_\uparrow -n_\downarrow}{2M}$ is susceptibility  ($n_{\sigma}$ is spin-resolved electron density in F).

The spin polarization in the laboratory frame is obtained by a rotation matrix
${\cal R}_{ij}$, defined by 
\begin{align}
U^{-1} \sigma_i U \equiv {\cal R}_{ij}\sigma_j,                  \label{Rdef}                                            
\end{align}
 as 
 \begin{align}
{s}_i^{\rm (N)}={\cal R}_{ij} \widetilde{s}_j^{\rm (N)}.    
\end{align}

Explicitly, 
\begin{align}
{\cal R}_{ij}=2m_im_j-\delta_{ij}   ,  \label{calRdef}                                        
\end{align}
where $\mv$ is in Eq. (\ref{mvdef}).
Using identities 
\begin{align}
{\cal R}_{iz}& = n_i \nnr
{\cal R}_{ij}(\Ascalv{t}^{\perp})_j &= \frac{\hbar}{2}(\nv\times \dot{\nv})_i \nnr
 {\cal R}_{ij}(\zvhat\times \Ascalv{t}^{\perp})_j &= \frac{\hbar}{2} \dot{\nv}_i ,
\end{align}
the induced interface spin density is finally obtained as
\begin{align}
 \sv^{\rm (N)}=\Re[\zeta^{\rm s}](\nv\times\dot{\nv})+\Im[\zeta^{\rm s}] \dot{\nv}
 \end{align}
where 
\begin{align}
    \zeta^{\rm s}&\equiv  \hbar(\pi\dos_{\rm N})^2\frac{n_\uparrow -n_\downarrow}{2M} T_{\uparrow\downarrow}.
    \label{zetadef}
\end{align}

Since the N electrons contributing to induced spin density is those at the Fermi energy, the spin current is simply proportional to the induced spin density as 
${\jsv}^{\rm N}=\frac{\hbar\kf}{m}{\sv}^{\rm (N)}$, resulting in  
\begin{align}
 \jsv^{\rm (N)}= \frac{\hbar\kf}{m}\lt[  \Re[\zeta^{\rm s}](\nv\times\dot{\nv})+ \Im[\zeta^{\rm s}]\dot{\nv} \rt].
 \label{jsresultQM}
 \end{align}
This is the result of spin current at the interface. The pumping efficiency is determined by the product of hopping amplitudes $t_\uparrow$ and  $t_\downarrow^*$.
The spin mixing conductance defined in Ref. \citep{Tserkovnyak02} corresponds to $T_{\uparrow\downarrow}$. 
In the scattering approach\citep{Tserkovnyak02} based on adiabatic pumping theory \citep{Buttiker94,Brouwer98,Moskalets12}, the expression for the spin mixing conductance in terms of scattering matrix element is exact as for the adiabatic contribution.
Our result (\ref{jsresultQM}), in contrast, is a perturbative one valid to the second order in the hopping amplitude.
To take full account of the hopping in the self energy is possible numerically in a field-theoretical approach.

In bulk systems without spin-orbit interaction and magnetic field, the hopping amplitudes $t_\spinindex$ are chosen as real, while at interfaces, this is not the case because inversion symmetry is broken.
Nevertheless, in metallic junctions such as Cu/Co, Cr/Fe and Au/Fe, first principles calculations indicate that imaginary part of spin mixing conductance (our $\zeta^{\rm s}$) is smaller than the real part by 1-2 orders of magnitude \citep{Xia02,Zwierzycki05}.
Large spin current proportional to $\dot{\nv}$ would therefore suggest existence of strong interface spin-orbit interaction, which gives rise to the imaginary part of  $\zeta^{\rm s}$.

\subsection{Adiabatic or nonadiabatic? \label{SECadornonad}}

\begin{figure}[tbh]
  \begin{center}
    \includegraphics[width=0.2\hsize]{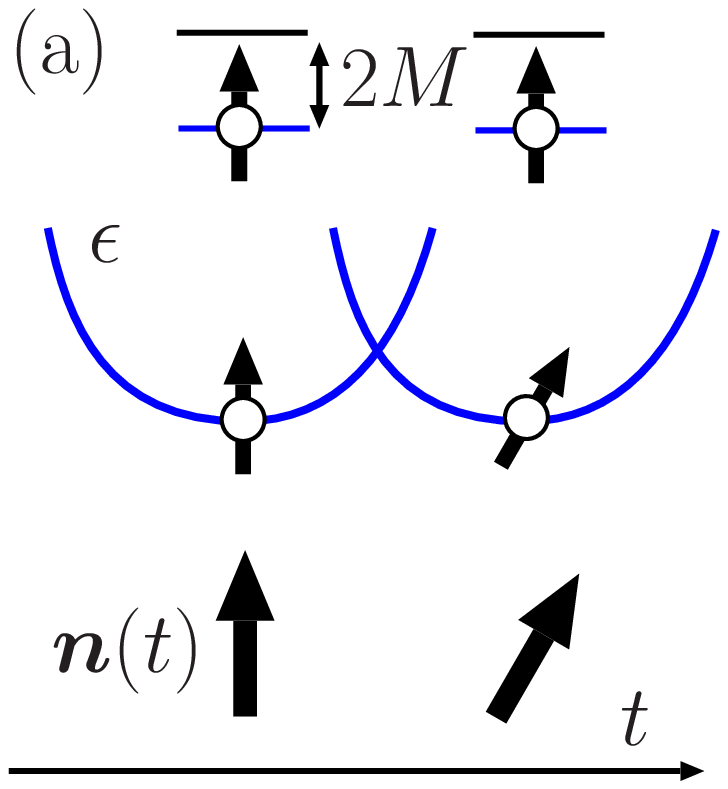}
    \includegraphics[width=0.2\hsize]{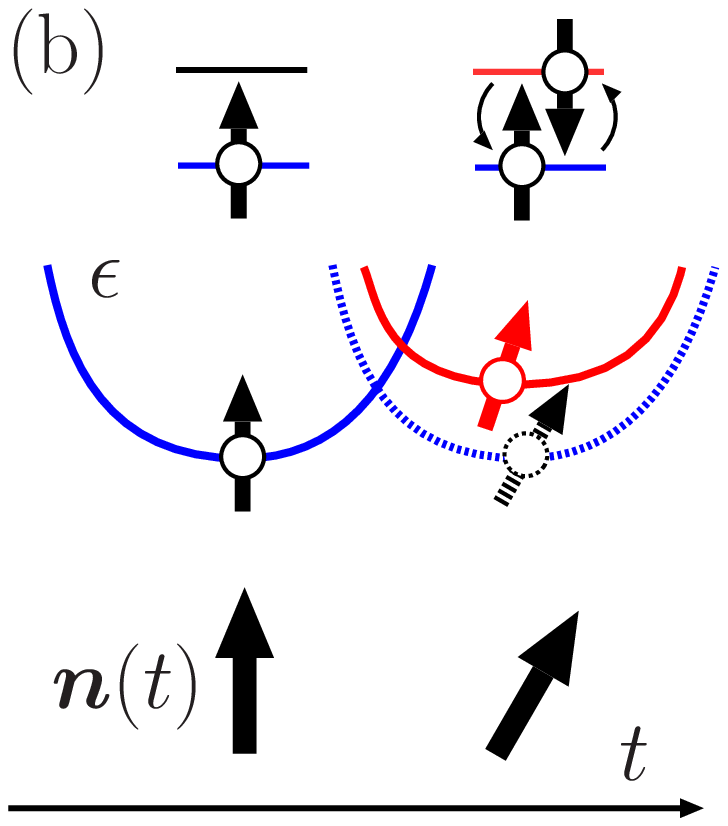}
  \end{center}
\caption{ Schematic figures of electron energy $\epsilon$ under precessing localized spin, $\nv(t)$, in the adiabatic limit (a) and with nonadiabaticity (b). 
Top figures represent energy levels with separation of $2M$  in the rotated frame. In the perfectly adiabatic case (a), the electron state keep the minimum energy state as $\nv(t)$ changes. Spin pumping does not occur in this limit.
Case (b) is with nonadiabaticity taken into account.
A perpendicular spin polarization along $\nv\times\dot{\nv}$ is induced by a 
temporal change of localized spin $\dot{\nv}$, resulting in a high energy state (shown in red). This nonadiabatic effect is essential for spin current generation.
\label{FIGadiabatic}}
\end{figure}

In our approach, spin pumping effect at the linear order in time-derivative is mapped to a static problem of spin polarization formed by a static spin-mixing potential in the rotated frame.
The rotated frame approach employed here provides clear physical picture, as it grasps the low energy dynamics in a mathematically proper manner. 
In this approach, it is clearly seen that pumping of spin current arises as a result of off-diagonal components of the spin gauge field that causes electron spin flip (Fig. \ref{FIGadiabatic}).

If so, is spin pumping an adiabatic effect or nonadiabatic one? 
Conventional adiabatic processes are those where the system under time-dependent external field remains to be the lowest energy state at each time (Fig. \ref{FIGadiabatic}(a)).
In the spintronics context, electron passing through a thick domain wall seems to be in the adiabatic limit in this sense; The electron spin keeps the lowest energy state by rotating it according to the magnetization profile at each spatial point as was argued in Sec. \ref{SECDWtransmit}. 
In contrast, as is seen from the above analysis, spin pumping effect does not arise in the same adiabatic limit; It is induced by the nonadiabatic (off-diagonal) spin gauge field, ${\cal A}_{{\rm s},t}^\pm$, which changes electron spin state in the local rotated frame with a cost of $sd$ exchange energy (Fig. \ref{FIGadiabatic}(b)). 
For spin pumping effect, therefore, nonadiabaticity is essential, as indicated also in a recent full counting statistics analysis \citep{Hashimoto17}.

\begin{figure}[tbh]
\begin{center}
\begin{tabular}{ccc}
& Spin-transfer effect for a domain wall & Spin pumping due to magnetization precession  \\ 
& \includegraphics[width=0.45\hsize]{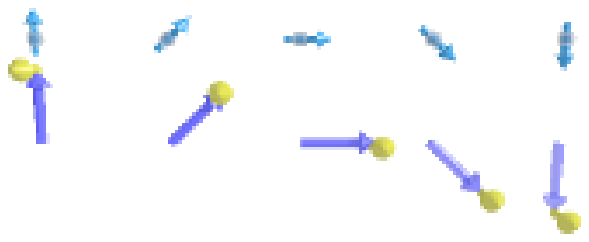}
&
\includegraphics[width=0.45\hsize]{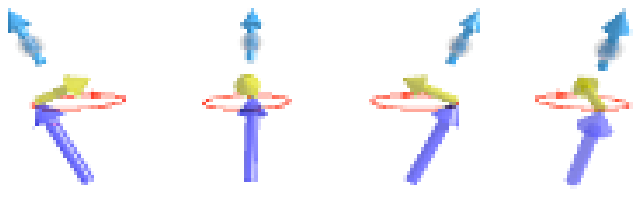}\\
& $ \delta\sv\propto \nv\times\nabla_x \nv \propto  {\cal R} {\cal A}_{{\rm s},x}^\pm$  
 & $ \delta\sv\propto \nv\times\dot{\nv} \propto {\cal R} {\cal A}_{{\rm s},t}^\pm$ 
\end{tabular}
\caption{ Comparison of electron transmission through a domain wall and spin pumping effect. Large arrows represent the localized spins, $\nv$, as function of position $x$ (left figure) or time $t$ (right figure), and electron spin is denoted by a small arrow with a circle. 
A nonadiabatic spin polarizations $\delta \sv$ induced by the nonadiabatic gauge field ${\cal A}_{{\rm s},\mu}^\pm$ are represented by yellow arrows. For the domain wall it is always perpendicular to the wall plane.
\label{FIGSPDW}} 
\end{center}
\end{figure}

A careful microscopic description indicates that a nonadiabaticity is essential even in spin-transfer effect. In fact, electron spin injected into a domain wall along $x$ direction is polarized along $\nv\times\nabla_x\nv$ as a result of nonadiabatic gauge field \citep{TKSLL07,TKS_PR08}, as shown in Eq. (\ref{sevres}). 
This non equilibrium spin polarization is perpendicular to the wall plane, and thus induces  translational motion of the wall. 
This is the physical mechanism of spin-transfer effect.
At the same time, spin-transfer effect can be discussed phenomenologically using conservation law of angular momentum. One should not forget, however, that nonadiabaticity is implicitly assumed  because spin rotation is caused only by a perpendicular component. 
Physically, the spin pumping effect is essentially the same as electron transmission through domain wall if we replace a spatial coordinate $x$ and the time, as summarized in Fig.  \ref{FIGSPDW}. 
In the case of domain wall, including the nonadiabatic gauge field to the next order leads to consideration of domain wall resistance and nonadiabatic $\beta$ torque \citep{GT00,GT01,TK04}.

\subsection{Spin accumulation in ferromagnet \label{SEC:damping}}
The spin current pumping is equivalent to the increase of spin damping due to magnetization precession, as was discussed in Refs. \citep{Berger96,Tserkovnyak02}.
The damping effect is discussed by calculating the torque by evaluating the spin polarization of the conduction electron spin in F region. 
To do this, a field theoretic method is convenient, as it enables a direct estimate of position-dependent spin density.
Details are shown in Ref. \citep{TataraSP17}, and we here present only the result.
The induced spin density in the ferromagnet is obtained as 
\begin{align}
{\sev}^{\rm (F)}(\rv,t)
 &= \frac{ m^2\dos_{\rm N}a^2}{2\kfu\kfd}  \sum_{\spinindex}
 \biggl[ (\nv\times\dot{\nv}) \overline{T_{\spinindex,-\spinindex}} e^{-i\spinindex(\kfu-\kfd)x}
   + \dot{\nv}(-i\spinindex)  \overline{T_{\spinindex,-\spinindex}} e^{-i\spinindex(\kfu-\kfd)x} \biggr],
\label{spinresult1}
\end{align}
where $\dos_{\rm N}$ is electron density of states in the normal metal and $\kf_{\sigma}$($\sigma=\pm$) denotes the Fermi wave vector for spin $\pm$ in the ferromagnet.
The induced spin accumulation density in the whole ferromagnet is
\begin{align}
 \overline{{\sev}^{\rm (F)}} & \equiv \frac{1}{d}\int_{-d}^0 dx {\sev}^{\rm (F)}(x) \nnr
 &= \frac{ m^2\dos_{\rm N}a^2}{\kfu\kfd(\kfu-\kfd)d} \lt[ (\nv\times \dot{\nv}) 
 \lt(-\Im[\overline{T_{\uparrow\downarrow}}](1-\cos\tilde{d}) +\Re[\overline{T_{\uparrow\downarrow}}]\sin\tilde{d} \rt) 
 + \dot{\nv} \lt( \Re[ \overline{T_{\uparrow\downarrow}} ](1-\cos \tilde{d}) +\Im[\overline{T_{\uparrow\downarrow}}]\sin \tilde{d} \rt)\rt],
 \label{totalaccumulation}
\end{align}
where $\tilde{d}\equiv (\kfu-\kfd)d$, and 
$d$ is the thickness of ferromagnet.
As a result of this induced electron spin density, $\overline{{\sev}^{\rm (F)}} $, 
the equation of motion for the averaged magnetization is modified to be \citep{Berger96}
\begin{align}
\dot{\nv}=-\alpha \nv\times\dot{\nv}-\gamma\Bv\times\nv - \spol \nv\times  \overline{{\sev}^{\rm (F)}} ,                                                                      
\label{LLGeq}
\end{align}
where $\Bv$ is the external magnetic field.

Let us first discuss thick ferromagnet case, $d \gg |\kfu-\kfd|^{-1}$, where oscillating part with respect to $\tilde{d}$ is neglected in Eq. (\ref{totalaccumulation}).
The equation of motion then reads   
\begin{align}
 (1+\delta ) \dot{\nv}= -(\alpha+\delta\alpha) \nv\times\dot{\nv} - \gamma\Bv\times\nv,
\end{align}
where 
\begin{align}
\delta\alpha&=\frac{ m^2\dos_{\rm N}a^2 M}{\kfu\kfd(\kfu-\kfd)d} \Re [\overline{T_{\uparrow\downarrow}}], \label{alphaenhance}
\end{align}
is the Gilbert damping enhancement by the effect of normal metal and 
\begin{align}
\delta&=\frac{ m^2\dos_{\rm N}a^2 M}{\kfu\kfd(\kfu-\kfd)d} \Im [\overline{T_{\uparrow\downarrow}}] , 
\end{align}
represents the shift of the precession angular frequency $\omega_B$ as 
\begin{align}
 \omega_B=\frac{\gamma B}{1+ \delta}.\label{resonancefreq}
\end{align}
This is equivalent to the modification of the gyromagnetic ratio, $\gamma$, or the $g$-factor.

For most 3d ferromagnets, we may approximate $\frac{ m^2\dos_{\rm N}a\spol\ef^2}{2\kfu\kfd(\kfu-\kfd)} \simeq O(1)$ (as $\kfu-\kfd\propto \spol$), resulting in $\delta\alpha\propto \frac{a}{d} \Re[\overline{T_{\uparrow\downarrow}}]$.
When interface spin-orbit interaction is taken into account, 
we have $T_{\uparrow,\downarrow}=\ttil^0_{\uparrow} \ttil^0_{\downarrow} 
 +i\widetilde{\gamma}_{xz}  (\ttil^0_\uparrow+\ttil^0_{\downarrow})+O((\widetilde{\gamma})^2)$, where $\ttil^0_\spinindex$ and $\widetilde{\gamma}_{xz}$ have usually small imaginary part compared to the real part \citep{Xia02,Zwierzycki05}.
Moreover, $\Re[\ttil^0_\spinindex]$ can be chosen as positive in most cases and thus $T_{\uparrow,\downarrow}>0$.
Equations (\ref{alphaenhance}) and (\ref{resonancefreq}) indicate that the strength of the hopping amplitude $\ttil^0_\spinindex$ and interface spin-orbit interaction $\widetilde{\gamma}_{xz}$ are experimentally accessible by measuring Gilbert damping and shift of resonance frequency as has been known \citep{Tserkovnyak02}.
A significant consequence of Eq. (\ref{alphaenhance}) is that the enhancement of the Gilbert damping, 
\begin{align}
\delta \alpha\sim \frac{a}{d}  \frac{1}{\ef^2} \ttil^0_{\uparrow} \ttil^0_{\downarrow} ,
\end{align}
can exceed in thin ferromagnets the intrinsic damping parameter $\alpha$, as the two contributions are governed by different material parameters.
In contrast to the positive enhancement of damping, the shift of the resonant frequency or $g$-factor can be positive or negative, as it is linear in the interface spin-orbit parameter $\widetilde{\gamma}_{xz}$.

Experimentally, enhancement of the Gilbert damping and frequency shift has been observed in many systems \citep{Mizukami01}.
In the case of Py/Pt junction, enhancement of damping is observed to be proportional to $1/d$ in the range of 2nm$<d<10$nm, and the enhancement was large, $\delta \alpha/\alpha\simeq 4$ at $d=2$ nm \citep{Mizukami01}.
These results appear to be consistent with our analysis.
Same $1/d$ dependence was observed in the shift of $g$-factor.
The shift was positive and magnitude was about 2\% for Py/Pt and Py/Pd with $d=2$nm, while it was negative for Py/Ta \citep{Mizukami01}.
The existence of both signs suggests that the shift is due to the linear effect of spin-orbit interaction, and the interface spin-orbit interaction we discuss is one of possible mechanisms.

For thin ferromagnet, $\tilde{d}(=(k_{{\rm F}+}-k_{{\rm F}-})d)\lesssim1$,   the spin accumulation of Eq. (\ref{totalaccumulation}) leads to 
\begin{align}
\delta{\alpha} & =  \frac{ m^2\dos_{\rm N}a^2\spol}{2\kfu\kfd} \Im[\overline{T_{\uparrow\downarrow}}]  \nnr
 \delta   & = - \frac{ m^2\dos_{\rm N}a^2\spol}{2\kfu\kfd} \Re[\overline{T_{\uparrow\downarrow}}] . \label{thin}
\end{align}
Thus, for weak interface spin-orbit interaction,  positive shift of resonance frequency is expected (if $\Re[\overline{T_{\uparrow\downarrow}}]>0$).
Significant feature is that the damping can be reduced or even be negative if  strong interface spin-orbit interaction exists with negative $\Im[\overline{T_{\uparrow\downarrow}}]$.
Our result indicates that 'spin mixing conductance' description of Ref. \citep{Tserkovnyak02} breaks down in thin metallic ferromagnet.

\subsection{Historical background and adiabatic pumping}
Spin current generation due to magnetization precession was pointed out  before Tserkovnyak theory by R. H. Silsbee {\it et al.} \citep{Silsbee79}, where the effect of interface spin accumulation  on the electron spin resonance in FN junction was focused on.
Enhancement of Gilbert damping constant in FN junction was  theoretically studied by L. Berger \citep{Berger96}, and developed by other authors \citep{SimanekHeinrich03,Simanek03}.
Experimental studies were also carried out and results were in agreement with thoeries \citep{Mizukami01,Urban01}.

In 2002, Tserkovnyak {\it et al.} presented a novel interpretation to those effects in terms of spin current generation, which they called the spin pumping effect \citep{Tserkovnyak02}.
Based on the adiabatic pumping theory, they showed that 
the spin current generated are determined by so-called the spin mixing conductance, which is written by use of scattering amplitudes.

Adiabatic pumping theory started by the seminal paper by Thouless, where he discussed that a current is induced in quantum system by applying a periodic modulation of a potential \citep{Thouless83}.
The study was described by use of scattering theory.
Current dynamically generated in electron system is generally written as \citep{Moskalets12}
\begin{align}
 I= \frac{e}{h}\int dE\lt(f_{\rm out}(E)-f(E)\rt),
\end{align}
where $f(E)$ is the equilibrium distribution with energy $E$ and $f_{\rm out}(E)$ is a nonequilibrium distribution function for the outgoing electron in the presence of external perturbation. 
Functions $f_{\rm out}(E)$ and $f(E)$ are related by scattering matrix element,  $S_{\alpha\beta}$, that also  connects the outgoing and incoming electron operator as $a_{{\rm out},\alpha}=S_{\alpha\beta}a_{{\rm in},\beta}$, where $\alpha,\beta$ are indeces of leads. 
$S_{\alpha\beta}$ is therefore reflection or transmission amplitudes between leads $\alpha$ and $\beta$.
When the perturbation is periodic with period ${\cal T}$, the current 
in the slow variation (adiabatic) limit is given by
\begin{align}
 I &= \frac{ie}{2\pi}\int dE　\lt(-\frac{\partial f}{\partial E}\rt) \int_0^{\cal T} \frac{dt}{{\cal T}} 
 \tr \lt[ S^\dagger (E,t)\frac{\partial}{\partial t} S(E,t) \rt] \nnr
 &= \frac{ie}{2\pi} \int_0^{\cal T} \frac{dt}{{\cal T}} 
 \tr \lt[ S^\dagger (\ef,t)\frac{\partial}{\partial t} S(\ef,t) \rt] .
\end{align}
The expression is written using the integral over the scattering matrix as
\begin{align}
 I 
 &= \frac{ie}{2\pi} \oint 
 \tr \lt[ S^\dagger (\ef,t)dS(\ef,t) \rt]. \label{SdS}
\end{align}
In the case of a single time-dependent parameter, the integral is trivial and vanishes, while for two parameters $\pv(t)=(p_1(t),p_2(t))$, it reads using Stokes theorem 
\begin{align}
 I 
 &= \frac{ie}{2\pi} \oint 
 \tr \lt[  \nabla_\pv \times \vv(\pv) \rt], \label{parea}
\end{align}
where $\nabla_\pv\equiv \frac{d}{d \pv} $ is a derivative in the parameter space. 
The pumped current is thus determined by the flux $\nabla_\pv \times \vv(\pv)$ in the parameter space \citep{Brouwer98,Moskalets12}.

\section{Brief remark on thermal transport}
Let us briefly mention transport driven by temperature gradient.
In metals, a temperature gradient gives rise to a force on electrons in the same manner as external electric field, and thus thermal transport effects appear to be discussed in parallel to the electrically-induced effects phenomenologically speaking.  
Strictly speaking, however, there is no rigorous formalism to incorporate temperature gradient in quantum systems, as the system is non-equilibrium and also because temperature is a concept defined in a macroscopic scale.
Nevertheless, as thermal transport effects are important for applications like Peltier effect, some theoretical approaches were proposed in the 1960's.

Of those, Luttinger's method \citep{Luttinger64} is commonly used nowadays.
He introduced a scalar potential to describe temperature gradient. 
The potential couples to the energy density of the system and was called the gravitational potential, perhaps because gravitational field couples to the energy density of the system in theory of general relativity.
Emergence of such a potential may be understood as follows.
A quantum system with Hamiltonian $H$ at equilibrium is described by the partition function,   
$\tr[e^{-\beta H}]$, where $\beta=1/(\kb T)$ is the inverse temperature.
When temperature is inhomogeneous, 
$T(\rv)=T_0+\delta T(\rv)$, 
we may expand (without justification) the partition function to the lowest order of  $\delta T$ to obtain (${\cal H}$ is the Hamiltonian density)\citep{Matsumoto14}
\begin{align}
\tr\lt[\exp\lt({-\frac{1}{\kb}\intr\frac{{\cal H}}{T}}\rt)\rt]
\simeq \tr\lt[\exp\lt({-\frac{1}{\kb}\intr\frac{{\cal H}}{T_0}\lt(1-\frac{\delta T}{T_0}\rt)}\rt)\rt].
\end{align}
We see that temperature inhomogeneity looks like a scalar potential  
\begin{align}
\psi_T \equiv -\frac{\delta T}{T_0},
\end{align}
which couples to the Hamiltonian density. 
Based on this 'gravitational' potential, Luttinger gave a prescription to calculate thermal transport coefficients in the framework of linear response theory.

For such treatment, temperature needs to be defined locally. In other words, the system needs to be in local equilibrium, satisfying energy conservation law of 
\begin{align}
\dot{{\cal E}}+\nabla\cdot\jv_{\cal E}=0,                                                                            \end{align}
where ${\cal E}$ and $\jv_{\cal E}$ are energy density and energy current density, respectively.
This indicates that there is approximately a U(1) gauge invariance for energy, similarly to that for charge.
(Correctly speaking, the energy conservation arises from translational invariance in time, and the corresponding symmetry is not the U(1) symmetry. For small variation, however, it is approximated as U(1) gauge invariance.)
Then the temperature gradient can be expressed in therms of an effective vector potential (thermal vector potential) \citep{Moreno96,Shitade14,Tatara15,TataraDW15}, which satisfies   
\begin{align}
 \dot{\Av}_{T}= \frac{\nabla T}{T_0}.
\end{align}

Based on the above approaches, thermal transport can be described in parallel to the case of electric cases by formally replacing the electric charge by energy density.
This feature, however, requires careful calculation of physical quantities because of enhancement at high energy, as noticed in some cases \citep{Qin11,Kohno16}.

Luttinger's approach has been employed to study thermally-induced electron transports \citep{Smrcka77,Oji85,Qin11,Eich14}, magnon transport \citep{Matsumoto11a} and thermally-induced torque \citep{Kohno14}. Vector potential form was applied in Refs. \citep{Shitade14,TataraDW15}.
Thermal transport is particularly important for magnon spintronics in insulator ferromagnets (magnonics) \citep{Murakami17}, as temperature gradient is most convenient driving field for magnons with no electric charge.

\section{Field theoretical approach}
So far we discussed in a quantum mechanical picture.
Such description may, however, lack transparency because we  always have to think in terms of wave functions.
In contrast, in  field theoretic formalisms, physical  observables are represented by fields, i.e., operators defined at each point in space and time, which have their own dynamics.
For instance, the Berry's phase is represented in quantum mechanics as an amplitude of a state change (Eq. (\ref{BerryphaseQM})), while it has a clear physical meaning of an effective gauge field in field theory.
Most importantly, field theory enables us to evaluate directly physical observables and provides clear theoretical scenario.
In this section, we introduce field theoretical description and discuss spintronics effects in the following sections.
It turns out that physics becomes clear and consistent in the field-theoretic formulation.

\subsection{Field operators \label{SEC:fieldop}}

Quantum particles can be created and annihilated by quantum fluctuation and accordingly their numbers fluctuate.
This fluctuation is neglected in quantum mechanics where a condition that the particle number in the whole space is always unity is imposed.
This constraint is removed by introducing creation and annihilation operators for the particle, which we denote here by $\ahat^\dagger$ and $\ahat$, respectively
\footnote{
In this section, field operators are denoted with $\hat{\ }$, although it shall be suppressed in the later sections.
}.
Creation and annihilation may occur at any position and any time, and so the operators are functions of space and time coordinates, i.e., fields.
The field operators acts on states which specifies how many particles exist at each space time point.
Any states are therefore constructed by applying necessary particle creation operators on a vacuum state $\ket{0}$.
The creation and annihilation operators are, by definition, 
not commutative with particle number operator, $\nhat$, because  particle numbers before and after creation have a difference of 1.
To put in equation, we need impose 
\begin{align}
\nhat\ahat^\dagger -\ahat^\dagger \nhat= \ahat^\dagger \label{nadagcom}
\end{align}
and
\begin{align}
\nhat \ahat -\ahat\nhat= -\ahat
\label{nacom}
\end{align}
These conditions are satisfied if we choose
$\nhat$ as 
\begin{align}
\nhat=\ahat^\dagger \ahat
\label{nadef}
\end{align}
and impose either 
\begin{align}
[\ahat,\ahat^\dagger]= 1, \;\;\; [\ahat,\ahat]=[\ahat^\dagger,\ahat^\dagger]=0,
\label{acom}
\end{align}
or
\begin{align}
\{\ahat^\dagger, \ahat\}= 1, \;\;\;  \{\ahat,\ahat\}=\{\ahat^\dagger,\ahat^\dagger\}=0,
\label{aacom}
\end{align}
for the operators.
Here $[A,B]\equiv AB-BA$ is a commutator and 
$\{A,B\}\equiv AB+BA$ is an anti commutator.
We have therefore either bosons described by Eq. (\ref{acom}) or fermions satisfying Eq. (\ref{aacom}).
For spintronics, the field of most interest is fermionic conduction electron, which we denote by $c^\dagger_{\sigma}(\rv,t)$ and $c_{\sigma}(\rv,t)$, where $\sigma=\pm$ denotes spin degrees of freedom. 
For field operators, the commutator and anticommutator become $\delta$-function in space and time as

\begin{align}
[\ahat(\rv,t),\ahat^\dagger(\rv',t')]= \delta(\rv-\rv')\delta(t-t')
\;\;\;
\mbox{\rm or} \;\;\;
\{\ahat^\dagger(\rv,t), \ahat(\rv',t')\}= \delta(\rv-\rv')\delta(t-t'),
\end{align}
because operators at different space time coordinates simply commute or anticommute.

In the absence of correlation effects between fields, the total  many particle Hamiltonian is simply a single-body Hamiltonian of quantum mechanics multiplied by particle number density, $\nhat=\chat^\dagger\chat$ for the case of electron.
We use a vector representation for two spin components of electron operator, i.e., 
$\chat=(\chat_+,\chat_-)$.
For the 1-particle quantum mechanical Hamiltonian with mass $m$ and potential $V$, $H_{\rm qm}=-\frac{\hbar^2\nabla^2}{2m}+V(\rv)$, 
the field version is 
\begin{align}
H &= \intr \chat^\dagger(\rv,t)\lt( -\frac{\hbar^2\nabla^2}{2m}+V(\rv)\rt) \chat(\rv,t) .
\label{Hpotfield}
\end{align}
Here electron density $\chat^\dagger\chat$ is split to make the Hamiltonian hermitian.
Correlation effects are straightforwardly included by replacing particle density by $\chat^\dagger\chat$.

The time-dependence of field operators are governed by the Hamiltonian by the Heisenberg equation,
\begin{align}
 \partial_t \chat &=\frac{i}{\hbar}[H,\chat], & \partial_t \chat^\dagger =\frac{i}{\hbar}[H,\chat^\dagger].
 \label{Heqfora}
\end{align}
The equation motions are derived from a field Lagrangian
\begin{align}
 L=\intr i\hbar \chat^\dagger \partial_t \chat -H. \label{Lfield}
\end{align}
The first time-derivative term represents the canonical relation between creation and annihilation operators.
In fact, Eq. (\ref{Lfield}) indicates that the canonical variable for $\chat$ is $\hbar \chat^\dagger$, and canonical commutation relation of $\{\chat,\chat^\dagger\}=1$ is derived.

\subsection{Field Lagrangian for $sd$ model \label{SEC:fieldlagrangian}}
The field representation of the Lagrangian for conduction electron interacting with localized spin is
\begin{align}
 \Lhat &= \intr \left[
i\hbar \chat^\dagger \dot{\chat} -  \chat^\dagger(\rv,t)\lt( -\frac{\hbar^2\nabla^2}{2m} - M (\nv\cdot \sigmav)\rt) \chat(\rv,t) 
 \right].
\label{L0}
\end{align}
Considering general case of inhomogeneous localized spin structure, we carry out a unitary transformation to choose electron spin's quantization axis along $z$-axis.
This assumes that the $sd$ exchange coupling is strong and certain adiabatic condition is satisfied.
For the spatial variation, the condition turns out to be Eq. (\ref{adiabaticcondition}) in the clean case, while for time-dependent localized spin with angular frequency of $\omega$, it would be $\omega\tau \ll 1$, where $\tau$ is the electron elastic lifetime. 
In the field representation, the unitary transformation corresponds to define a new electron operators, $a$ and $a^\dagger$ as 
\begin{align}
\ahat(\rv,t)=U(\rv,t)\chat(\rv,t).
\end{align}
A $2\times2$ matrix $U(\rv,t)$ is chosen to satisfy
\begin{align}
 U^{-1}(\nv\cdot\sigmav) U=\sigma_z,
\end{align}
at each point, and it is thus as given in Eq. (\ref{Udef}). 
Now the $sd$ interaction is diagonalized for the new electron, $a$ and $a^\dagger$, as 
\begin{equation}
 \Hhat_{sd}=  -\spol \intr \ahat^\dagger \sigma_z \ahat ,
  \label{Hsddiag}
\end{equation}
and thus this electron in the rotated frame is a good variable for describing low energy behavior. 
The unitary transformation affects, however, the kinetic term is modified as 
\begin{align}
\partial_\mu \chat 
=U\lt(\partial_\mu \pm\frac{i}{\hbar} {\cal A}_{{\rm s},\mu} \rt) \ahat,
\end{align}
where ${\cal A}_{{\rm s},\mu}$ is defined in Eq. (\ref{Aexpression}) and positive and negative signs correspond to $\mu=t$ and $\mu=x,y,z$, respectively.
The Lagrangian for $a$-electron is therefore the one with minimal coupling to the gauge field  (using integral parts, 
$\intr\chat^\dagger(\nabla^2 \chat)=-\int (\nabla\chat^\dagger)(\nabla \chat)$) 
\begin{align}
 \Lhat &= \intr \left[
i\hbar \ahat^\dagger \lt(\partial_t+ \frac{i}{\hbar}{\cal A}_{{\rm s},t}\rt) {\ahat} 
-\frac{\hbar^2}{2m}  \lt[\ahat^\dagger\lt(\nablal+\frac{i}{\hbar}\Acalsv\rt) \rt] \lt[\lt(\nablar- \frac{i}{\hbar} \Acalsv\rt) \ahat\rt]
+\eF \ahat^\dagger \ahat
 + \spol \ahat^\dagger \sigma_z \ahat
 \right],
\label{Lezexpression1}
\end{align}
where $\nablal$ and $\nablar(=\nabla)$ act on the field on the left and right side, respectively.
Defining spin density and spin current density operators in the rotated frame (without spin magnitude of $\frac{1}{2}$) as 
\begin{align}
 \hat{\se}_\alpha & \equiv \ahat^\dagger \sigma_\alpha \ahat \nnr
 \hat{j}_{{\rm s},i}^{\alpha} & \equiv \frac{-i\hbar}{2m} \ahat^\dagger \vvec{\nabla}_i \sigma_\alpha \ahat, 
\end{align}
it reads
\begin{align}
 \hat{L} &= \intr \left[
i\hbar \ahat^\dagger \dot{\ahat} -\frac{\hbar^2}{2m}|\nabla \ahat|^2 +\eF \ahat^\dagger \ahat
 +\spol \ahat^\dagger \sigma_z \ahat
  + \hat{j}_{{\rm s},i}^{\alpha} {\cal A}_{{\rm s},i}^\alpha
  -\frac{\hat{n} }{2m}{\cal A}_{\rm s}^2
   -  \hat{\se}_{\alpha} {\cal A}_{{\rm s},t}^\alpha
 \right].
\label{Lezexpression2}
\end{align}
Here it is clear that the spatial and time components of the gauge field, ${\cal A}_{{\rm s},i}$ ($i=x,y,z$) and ${\cal A}_{{\rm s},t}$, couples to spin current density and spin density, respectively.

The gauge field  ${\cal A}_{{\rm s},\mu}^\alpha$ is a SU(2) gauge field that have three spin components ($\alpha=x,y,z$) and four space-time components ($\mu=x,y,z,t$).
In the adiabatic limit, it reduces to a single spin component ${\cal A}_{{\rm s},\mu}^z\equiv A_{{\rm s},\mu}$
, i.e., to a U(1) gauge field we discussed in Sec. \ref{SEC:SEMF}. In fact, in this limit, the minority spin electron can be neglected due to a large electron spin polarization energy $\spol$. 
Thus electron field reduces to 
$a\ra\lt(\begin{array}{c} a_\uparrow \\ 0 \end{array}\rt)$ and we end up with the Lagrangian equivalent to the one with electromagnetic gauge field,
\begin{equation}
  L = \intr \lt[ 
i\hbar \ahat_\uparrow^\dagger \lt(\partial_t+ \frac{i}{\hbar}{A}_{{\rm s},t}\rt) {\ahat_\uparrow}
-\frac{\hbar^2}{2m} \lt[\lt(\nabla+\frac{i}{\hbar}\Asv\rt) \ahat_\uparrow^\dagger\rt] \lt[\lt(\nabla-\frac{i}{\hbar}\Asv\rt) \ahat_\uparrow\rt]
     +\spol\ahat_\uparrow^\dagger \ahat_\uparrow \rt].
     \label{gaugeH}
\end{equation}

\section{Effective Lagrangian for localized spin \label{SECeffectiveL}}
Once we know the field Lagrangian, field theory provides in principle any information on the system we want.
We  discuss  in term of Lagrangian, as Lagrangian contains information about canonical relations in the time-derivative term as we saw in Sec. \ref{SEC:fieldop}, while in the Hamiltonian approach, canonical relations need to be imposed, although both approaches lead to the same result if calculated correctly.

We first show how effective Lagrangian for localized spin is derived from the $sd$ model within equilibrium quantum statistical physics.
Effective Lagrangian is the one obtained by integrating out, in other words, evaluating quantum trace of, other quantum degrees of freedom \citep{Sakita85}, which is conduction electron in our case. 
All the effects of electrons are formally contained in the effective Lagrangian.
In Sec. \ref{SECminimamcoupling},  we discussed that spin current induces  Dzyaloshinskii-Moriya interaction (Eq. (\ref{Djs})).
In the effective Lagrangian study, this fact is discussed systematically on the equal footing as ferromagnetic exchange interaction induced by conduction electron.

\begin{figure}[tb]
\begin{center}
\includegraphics[width=0.2\hsize]{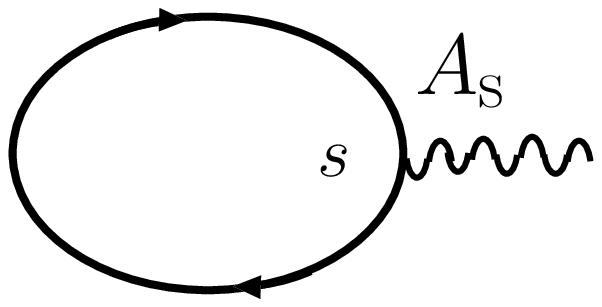}
\includegraphics[width=0.2\hsize]{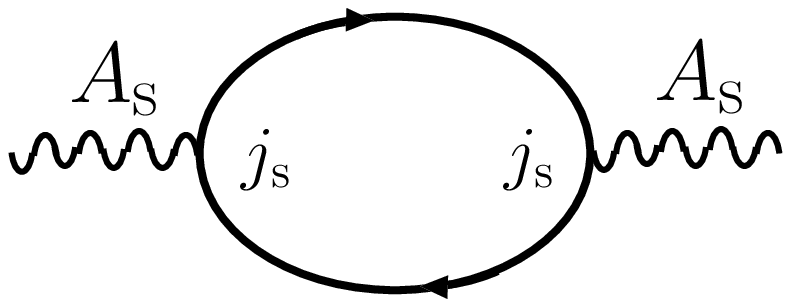}
\includegraphics[width=0.2\hsize]{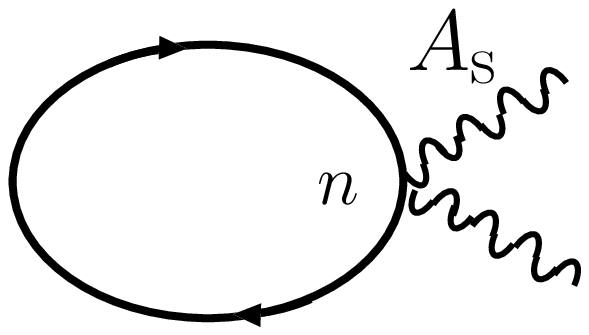}
\end{center}
\caption{ Feynman diagram representaion of contributions to the effective  Lagrangian to the second order in spatial derivative and linear order in time derivative. 
The spin gauge field ${\cal A}_{\rm s}$ represented by wavy lines is linear in the derivative of localized spin. 
Solid lines represent electron Green's functions.
}
\label{FIGAsecond}
\end{figure}

The effective Lagrangian for localized spin is obtained by evaluating the expectation value for the electron treating the spin gauge field perturbatively.
The spatial component of the spin gauge field is taken account to the second order, while only the first order of temporal component needs to be retained.
By use of path-integral method, the effective action, time-integral of effective Lagrangian, is obtained as 
\begin{align}
 \Delta S & =  \int d\tau \intr \lt[  
 -{\se}^{\alpha} {\cal A}_{{\rm s},0}^\alpha 
 +\jspin{,i}{\alpha}{\cal A}_{{\rm s},i}^\alpha
 -\frac{\hbar^2}{2m}{\cal A}_{\rm s}^2 {n} 
     \rt] 
   -\hf \int d\tau \int d\tau' \intr \intr' 
 {\cal A}_{{\rm s},i}^\alpha(\tau,\rv) {\cal A}_{{\rm s},j}^\beta(\tau',\rv') 
  \chi_{ij}^{\alpha\beta}(\tau,\rv,\tau',\rv'),
   \label{DeltaSdef}
\end{align}
where  ${\se}^{\alpha}\equiv \average{\hat{\se}}$ and  
$\jspin{,i}{\alpha}\equiv \average{\hat{j}_{{\rm s},i}^\alpha}$ are expectation values of spin density and spin current, respectively ($\average{\ }$ is quantum statistical average).
The second term on the right hand side contains spin current correlation functions
\begin{align}
 \chi_{ij}^{\alpha\beta}(\tau,\rv,\tau',\rv') 
 & \equiv \average{ \hat{j}_{{\rm s},i}^{\alpha}(\tau,\rv) \hat{j}_{{\rm s},j}^{\beta}(\tau',\rv')   }
 \nnr
 &= \frac{1}{\beta V}\sum_{\qv \omega_\ell}e^{-i\omega_\ell(\tau-\tau')}e^{i\qv\cdot(\rv-\rv')} 
 \chi_{ij}^{\alpha\beta}(i\omega_\ell,\qv),
\end{align}
where 
\begin{align}
 \chi_{ij}^{\alpha\beta}(i\omega_\ell,\qv)
 & = \frac{\hbar^2}{\beta V}\sum_{\kv \omega_n}\frac{k_ik_j}{m^2} 
 \tr[\sigma_\alpha G_{\kvmq,\omega_n}\sigma_\beta G_{\kvpq,\omega_n+\omega_\ell} ] ,\label{chissdef}
\end{align}
in terms of the imaginary time free Green's function ($\omega_n\equiv (2n-1)\pi/\beta$ is fermionic thermal frequency, with $n$ an integer), 
\begin{align}
G_{\kv,\omega_n} &\equiv \frac{1}{i\omega_n -\ek+\spol\sigma_z}.
\end{align}
The effective action is diagrammatically represented in Fig. \ref{FIGAsecond}.

The summation over thermal frequency in Eq. (\ref{chissdef}) is evaluated using contour integration as 
\begin{align}
 \frac{1}{\beta} \sum_{\omega_n}  G_{\kv,\omega_n,\sigma}\sigma_\beta G_{\kv',\omega_n+\omega_\ell,\sigma'}
 &=
-\int_C \frac{dz}{2\pi i}f(z) \frac{1}{z-\epsilon_{\kv\sigma}}\frac{1}{z+i\omega_\ell-\epsilon_{\kv'\sigma'}} \nnr
=\frac{f(\epsilon_{\kv\sigma})-f(\epsilon_{\kv'\sigma'})}{\epsilon_{\kv\sigma}-\epsilon_{\kv'\sigma'}+i\omega_\ell},
\end{align}
where $z\equiv i\omega_n$ is a complex thermal frequency and $C$ is a contour surrounding the imaginary axis.
As the electrons connects localized spins at different position and time, the contribution of the effective action containing correlation functions are nonlocal in general.
For our purpose of looking into the second-order derivatives, however, it is sufficient to consider the local components, because the spin gauge field contains a first-order derivative. 
The correlation functions are therefore approximated as 
\begin{align}
 \chi_{ij}^{\alpha\beta}(\tau,\rv,\tau',\rv') 
 &= \delta(\rv-\rv')\delta(\tau-\tau') 
 \chi_{ij}^{\alpha\beta}(i\omega_\ell=0,\qv\ra0),
\end{align}
where $\chi_{ij}^{\alpha\beta}(i\omega_\ell=0,\qv\ra0)$ denotes that the limit of $\qv\ra0$ is taken after setting $\omega_\ell=0$.
The effective action is then represented by a local effective Lagrangian as 
$\Delta S=\int d\tau L_{\rm eff}(\tau)$, where 
\begin{align}
 L_{\rm eff} & = \intr \lt[
 - {\se}_{\alpha} {\cal A}_{{\rm s},t}^\alpha 
 +\jspin{,i}{\alpha}{\cal A}_{{\rm s},i}^\alpha
 -\frac{\hbar^2}{2m}{\cal A}_{\rm s}^2 {n} 
   -\hf \chi_{ij}^{\alpha\beta}(0,0){\cal A}_{{\rm s},i}^\alpha  {\cal A}_{{\rm s},j}^\beta \rt] .
   \label{Leff}
\end{align}
We here see that spin accumulation of electron contributes to an additional spin Berry's phase term (${\cal A}_{{\rm s},0}^\alpha $), and linear term in spatial component ${\cal A}_{{\rm s},i}^\alpha$ arises if expectation value of spin current is finite.
The terms second order in the gauge field describes  exchange interaction arising from electron conduction. 
They are in general anisotropic in space and spin, such as, 
$J_{ij}^{\alpha\beta}\nabla_i n_\alpha \nabla_j n_\beta$ ($J_{ij}^{\alpha\beta}$ is a coefficient), if symmetry of the system is broken.

Below, we look into the exchange interaction focusing on the isotropic electron dispersion and neglecting spin-orbit interaction.
Spin current vanishes in the  absence of external current, and 
the spin density is diagonal, $\se_\alpha\equiv s\delta_{\alpha z}$, where 
($s$ is defined without the factor of $\frac{1}{2}$ of spin)
\begin{align}
 s\equiv \frac{1}{V}\sumkv(f_{\kv+}-f_{\kv-}), \label{seldef}
\end{align}
is the electron spin polarization density.
The correlation function is written as 
\begin{align}
 \chi_{ij}^{\alpha\beta}(0,0)
 & = \delta_{ij} [(\delta_{\alpha\beta} -\delta_{\alpha z} \delta_{\beta z})\chi^{xx}(0,0)
 +\delta_{\alpha z} \delta_{\beta z}\chi^{zz}(0,0)
+ \epsilon_{\alpha\beta z} \chi^{xy}(0,0)],
\end{align}
and each component is evaluated as 
\begin{align}
 \chi_{ij}^{xx}(\omega_\ell=0,\qv\ra0)
 &=
 -\frac{\hbar^2(\kfu^5-\kfd^5)}{30\pi^2m^2\spol}\delta_{ij}\nnr
 \chi_{ij}^{xy}(\omega_\ell=0,\qv\ra0)&=0 \nnr
 \chi_{ij}^{zz}(0,0)&=-\frac{n}{m}\delta_{ij},
\end{align}
where $k_{F\pm}\equiv \sqrt{2m(\ef\pm\spol)}/\hbar$ is the Fermi wave vector of spin $\pm$ electron.
The effective Lagrangian is therefore obtained as  
\begin{align}
 L_{\rm eff} & = \intr \lt[  
   -{s} {\cal A}_{{\rm s},0}^z 
   -\frac{n}{2m}  \lt(1- \frac{\hbar^2(\kfu^5-\kfd^5)}{30\pi^2mn\spol} \rt)  
 \lt[ ({\cal A}_{{\rm s},i}^x)^2 +({\cal A}_{{\rm s},i}^y)^2 \rt] \rt] \nnr
 &=\intr \lt[  
   \frac{s\hbar}{2} \dot{\phi}(\cos\theta-1) 
   -\frac{J_{\rm e}}{2} (\nabla\nv)^2 \rt],
   \label{Heff2}
\end{align}
where 
\begin{align}
 J_{\rm e} &= \frac{n\hbar^2}{4m}  \lt[1- \frac{\hbar^2(\kfu^5-\kfd^5)}{30\pi^2mn\spol} \rt]  ,
\end{align}
is the exchange interaction induced by electron, which is positive in the present model.

In the presence of applied current, spin current along localized spin ($z$ direction in the rotated frame) is finite. 
In this case we retain the spin current term to obtain 
\begin{align}
 L_{\rm eff}  
 &=
 \intr \lt[  
   \frac{s\hbar}{2} [\lt(\partial_t+\vsv\cdot\nabla\rt){\phi}](\cos\theta-1) 
   -\frac{J_{\rm e}}{2} (\nabla\nv)^2 \rt]
   ,
   \label{Heff3}
\end{align}
where $\vsv$ is given by Eq. (\ref{vstt}) with spin magnitude $S$ replaced by $s/2$.

In the presence of spin-orbit interaction in systems with broken inversion symmetry, perpendicular spin current, $\jsv^\perp$, arises in general.
Then we have  a Lagrangian with Dzyaloshinskii-Moriya interaction interaction,
\begin{align}
 L_{\rm eff} 
 &=
 \intr \lt[  
   \frac{s\hbar}{2} \lt(\partial_t+\vsv\cdot\nabla\rt){\phi}(\cos\theta-1) 
   -\frac{J_{\rm e}}{2} (\nabla\nv)^2 
   -D_i^\alpha(\nv\times\nabla_i\nv) \rt]	
   ,
   \label{Heff4}
\end{align}
with DM constant given by Eq. (\ref{Djs}).

Here we presented a simple case of single band electron. Contributions from many bands need to be included for quantitative estimates for real materials.
Evaluation of strengths of exchange interaction and DM interaction is important in studies of magnetic structures \citep{Katsnelson10,Freimuth14,Mikhaylovskiy15,Belabbes16}. 
Effective Hamiltonian approach gives a straightforward method to evaluate interaction strengths when combined with the first principles calculations \citep{Kikuchi16}.

\section{Landau-Lifshitz-Gilbert (LLG) equation with electron effects
\label{SECLLGwithelectron}}
In this section we consider effects of conduction electron on localized spin dynamics by directly calculating the electron spin polarization and torque.
The Hamiltonian we consider is 
\begin{align}
 H&=H_S - \spol\intr \nv\cdot\hat{\sev} +H_{\rm e},  
\end{align}
where $\hat{\sev}\equiv c^\dagger\sigmav c$ is the electron spin field operator and $H_{\rm e}$ represents the conduction electron Hamiltonian. 
Effects other than electrons are included as an effective magnetic field in $H_S$.
To describe the effect of applied electric current, we include an electromagnetic gauge field (vector potential), $\Av$. The electron Hamiltonian is (see Eq. (\ref{AphicouplingQM}))
\begin{align}
H_{\rm e}^0\equiv 
  \intr \frac{\hbar^2}{2m} \lt[ \chat^\dagger \lt(\nablal+i\frac{e}{\hbar}\Av\rt) \rt]\lt[\lt(\nabla-i\frac{e}{\hbar}\Av\rt)\chat\rt]
  =\intr \lt[ \frac{\hbar^2}{2m} |\nabla\chat|^2 
  +\frac{ie\hbar}{2m} \Av\cdot\chat^\dagger \vvec{\nabla}_i \chat
  +\frac{e^2A^2}{2m}\nhat \rt]. \label{He0}
\end{align}
The total current density operator including the 'diamagnetic part' due to the electromagnetic vector potential is 
\begin{align}
\hat{\jv} \equiv - \frac{\delta H_{\rm e}}{\delta \Av} 
=\frac{-ie\hbar}{2m} \chat^\dagger \vvec{\nabla} \chat-\frac{e^2}{m}\nhat \Av.                                                            \end{align}
We do not consider electron spin relaxations like due to spin-orbit interaction.
Relaxation effects are only briefly mentioned later (See. Ref. \citep{KTS06}).

Including the effect of the electron, the equation of motion for localized spin (LLG equation)  reads
\begin{align}
 \dot{\nv}&= -\gamma \Bv_{S}\times\nv +\frac{\spol a^3}{\hbar S} \nv\times \sev,\label{LLGwithse}
\end{align}
where $\gamma\Bv_{S}\equiv -\frac{1}{\hbar S}\frac{\delta H_S}{\delta \nv}$, $a$ is the lattice constant, and 
\begin{align}
\sev\equiv \average{\hat{\sev}},
\end{align}
is the conduction electron spin polarization density, which contains all the electron effects  such as spin-transfer torque. 

\begin{figure}[tb]
\begin{center}
\includegraphics[width=0.12\hsize]{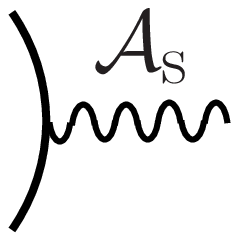}
\includegraphics[width=0.12\hsize]{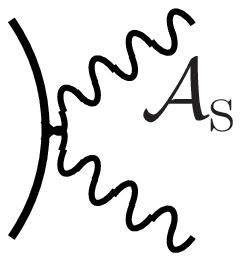}
\includegraphics[width=0.12\hsize]{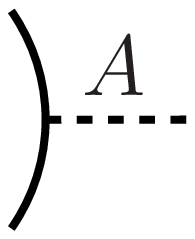}
\includegraphics[width=0.12\hsize]{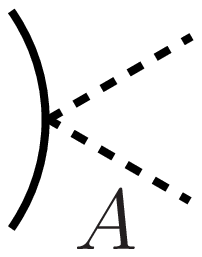}
\includegraphics[width=0.12\hsize]{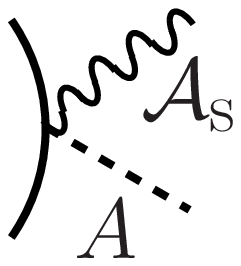}
\end{center}
\caption{ Diagrammatic representation of  interaction verteces including spin gauge field and  electromagnetic gauge field.  
Wavy, dotted and solid lines represent the spin gauge field ${\cal A}_{\rm s}$, electromagnetic gauge field, $\Av$, and  electron, respectively.
}
\label{FIGEMint}
\end{figure}

Calculation is carried out in the rotating frame that diagonalizes the $sd$ exchange interaction, where low energy behavior of electron spin is correctly described.
The electromagnetic interaction in Eq. (\ref{He0}) with spatial derivative is modified by the  transformation, resulting in 
\begin{align}
 \hat{H}_{\rm e} &\equiv H_{\rm e}^0+H_{sd} \\
 &= \intr \left[ \frac{\hbar^2}{2m}|\nabla \ahat|^2 -\eF \ahat^\dagger \ahat
 -\spol \ahat^\dagger \sigma_z \ahat
  - \hat{j}_{{\rm s},i}^{\alpha} {\cal A}_{{\rm s},i}^\alpha
  +\frac{1}{2m}{\cal A}_{\rm s}^2 \hat{n} 
  -\hat{\jv}\cdot\Av+\frac{e^2}{2m}{A}^2 \hat{n} 
  +\frac{e}{m} \ahat^\dagger A_i {\cal A}_{{\rm s},i} \ahat
 \right],
\label{He2}
\end{align}
where $\spol\equiv \Jsd S$, and 
\begin{align}
\hat{j}^0_i \equiv \frac{-ie\hbar}{2m} \ahat^\dagger \vvec{\nabla}_i \ahat,                                                             \end{align}
is the bare (paramagnetic) current density of the rotated electron (superscript $^0$ is for bare part).
The interaction verteces in Eq. (\ref{He2}) are shown diagrammatically in Fig. \ref{FIGEMint}.

The electron spin density in the rotated frame, 
$\svtil\equiv \average{\ahat^\dagger \sigmav\ahat}$, is related to the laboratory frame one by rotation matrix ${\cal R}$ as 
\begin{align}
 \se_i={\cal R}_{ij}\stil_j. 
\end{align}
It is represented by lesser Green function, defined in the rotated frame by  
\begin{align}
G_{\sigma\sigma'}^<(\rv,t,\rv',t')\equiv \frac{i}{\hbar}\average{\ahat^\dagger_{\sigma'}(\rv',t')\ahat_{\sigma}(\rv,t)}, 
\end{align}
as 
\begin{align}
 \svtil(\rv,t) &\equiv -i\hbar\tr[\sigmav G^<(\rv,t,\rv,t)].
\end{align}

Our objective is to evaluate the Green's function including the effect of localized spin structure, which is expressed by the spin gauge field, $\Ascal{}{}$, and electromagnetic gauge field, $\Av$, represented in Fig. \ref{FIGEMint}.
The applied electric field is written as $\Ev=-\dot{\Av}$. We consider spatially uniform $\Ev$ and $\Av$.
We consider DC case (static electric field) by treating the angular frequency $\Omega$ of $\Av$ as finite during the calculation and taking the limit of $\Omega\ra0$ at the end \citep{Rammer86}.
(The electric field is expressed also by use of a scalar potential $\Phi$ as $\Ev=-\nabla\Phi$, but the calculation is easier if we use vector potential.) 

The lesser Green's function is calculated by solving for the path-ordered Green's function defined by
\begin{equation}
G_{\sigma,\sigma'}(\rv,t,\rv',t') \equiv
 -\frac{i}{\hbar} \average{T_C a_{\sigma}(\rv,t) \adag_{\sigma'}(\rv',t')},
\end{equation}
where $t, t'$ are defined on a contour $C$ which goes from $-\infty$ to $\infty$ on the upper plane and comes back from  $\infty$ to $-\infty$ on the lower plane in a complex time plane.
This Green's function satisfies 
\begin{eqnarray}
i\hbar \partial_t G_{\kv\kv'}(t,t')
  &=& \delta(t-t')\average{ \{ a_{\kv}(t),\adag_{\kv'}(t') \} }
  + \frac{i}{\hbar} \average{ T_C [H,a_{\kv}(t)]\adag_{\kv'}(t') },
\end{eqnarray}
where $G_{\kv\kv'}$ is the Fourier transform of $G(\rv,\rv')$ and $H$ is the total Hamiltonian.
We are interested in the adiabatic limit, and treat the spin gauge field perturbatively to the linear order. The effect of the applied current is also discussed to the linear order in $\Av$.
Choosing the initial and final wave vectors as $\kvpq$ and $\kvmq$, respectively, the Dyson equation on time contour $C$ reads 
\begin{align}
G_{\kvmq,\kvpq}(t,t')
&=  g_{\kv}(t-t')\delta_{\qv,0}   
   + \hbar \sum_{\alpha} \int_{C}\! dt_1 
   \lt[ \frac{\hbar}{m} k_i {\cal A}_{{\rm s},i}^{\alpha}(\qv,t_1)+{\cal A}_{{\rm s},t}^{\alpha}(\qv,t_1) \rt]
g_{\kvmq}(t-t_1) \sigma_{\alpha} g_{\kvpq}(t_1-t') \nonumber\\
&
-\frac{e\hbar}{m}k_i 
\sum_{\alpha i} \int_{C}\! dt_1  A_i(t_1)
 g_{\kv}(t-t_1) g_{\kv}(t_1-t')
+\frac{e}{m} 
\sum_{\alpha i} \int_{C}\! dt_1  A_i(t_1)
{\cal A}_{{\rm s},i}^{\alpha}(\qv,t_1) 
 g_{\kvmq}(t-t_1) \sigma_{\alpha} g_{\kvpq}(t_1-t')
\nonumber\\
&+\frac{e\hbar^2}{m^2} 
\sum_{\alpha i} k_j 
\int_{C}\! dt_1 \! \int_{C}\! dt_2 
\lt[\lt(k+\frac{q}{2}\rt)_i  A_i(t_2) {\cal A}_{{\rm s},j}^{\alpha}(\qv,t_1)
   g_{\kvmq}(t-t_1) \sigma_{\alpha} g_{\kvpq}(t_1-t_2) g_{\kvpq}(t_2-t') 
\rt. \nnr & \lt.
+ \lt(k-\frac{q}{2}\rt)_i  A_i(t_1)  A_{j}^{\alpha}(\qv,t_2) 
g_{\kvmq}(t-t_1)  g_{\kvmq}(t_1-t_2) 
   \sigma_{\alpha} g_{\kvpq}(t_2-t') \rt]
\nonumber\\
& +O(({\cal A}_{\rm s})^2,A^2)
\label{dysonfull},
\end{align}
where $g_{\kv}(t)$ is the free path-ordered Green's function.

\subsection{Induced electron spin density}
Electron spin density, written in terms of the lesser Green's function at equal time and position, 
$G^<(\rv,t,\rv,t)=\sum_{\kv\kv'}e^{i(\kv-\kv')\cdot\rv}G^<_{\kv\kv'}(t,t)$, 
is represented by bubble diagrams starting from and ending at $\rv,t$ as 
\begin{align}
 \svtil(\rv,t) &= 
\raisebox{-\baselineskip}{\includegraphics[height=2\baselineskip]{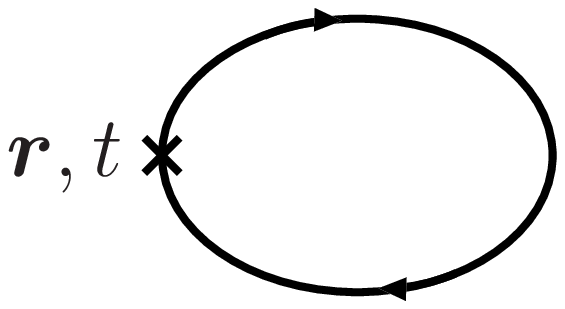}}
+
\raisebox{-\baselineskip}{\includegraphics[height=2\baselineskip]{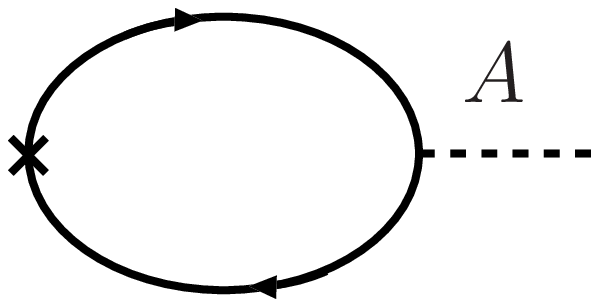}}
+
\raisebox{-\baselineskip}{\includegraphics[height=2\baselineskip]{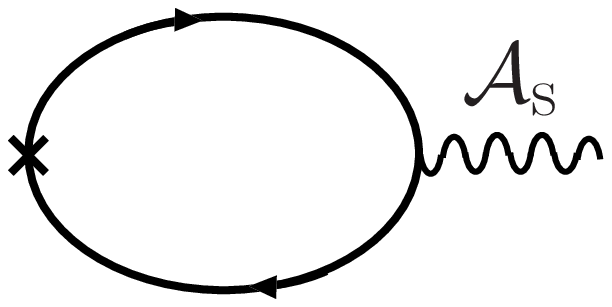}}
+
\raisebox{-\baselineskip}{\includegraphics[height=2\baselineskip]{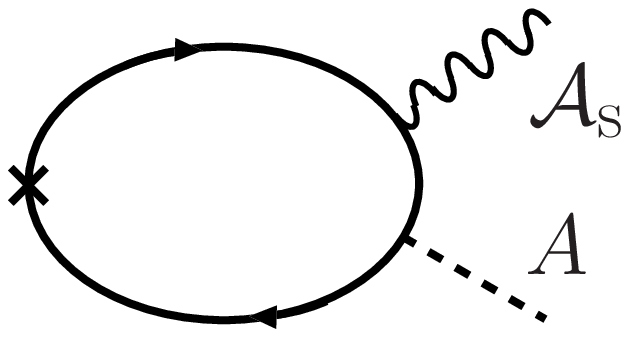}}
+
\raisebox{-\baselineskip}{\includegraphics[height=2\baselineskip]{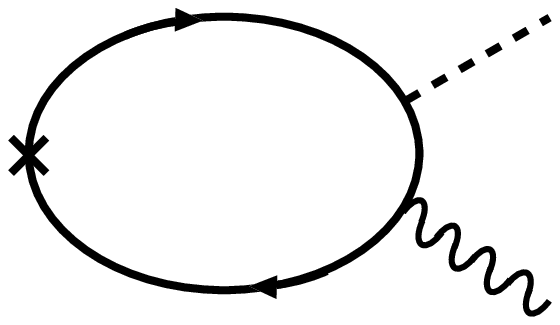}}
+
\raisebox{-\baselineskip}{\includegraphics[height=2\baselineskip]{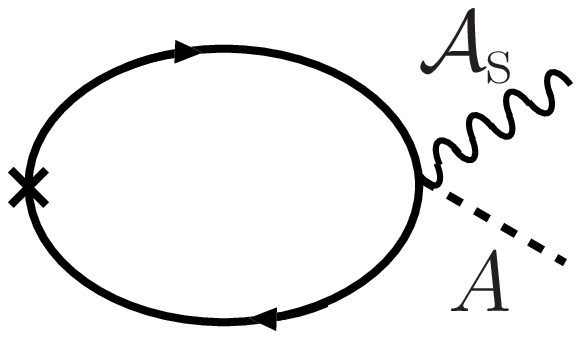}}
+\cdots.
\end{align}
The first and second diagrams on the right-hand side are trivial one without effects of spin structure, and are neglected.
The third term containing ${\cal A}_{\rm s}$ (represented by wavy line) only represents the equilibrium contribution in the presence of spin structure.
It was  already  discussed  in Sec. \ref{SECeffectiveL} in the effective Lagrangian approach 
(the term of $s$ in Eq. (\ref{Heff2})), but we here argue in a different approach.
The fourth, fifth and sixth contributions including both  ${\cal A}_{\rm s}$ and electromagnetic gauge field $A$ (represented by dotted line) represent the current-induced contribution, which is of our most interest.

Evaluation of lesser Green's function from the path-ordered one is carried out using projection formula (called the Langreth formula),
\begin{align}
\lt[\int_C dt_1 A(t,t_1) B (t_1,t')\rt]^<
&=\intinf dt_1 (A^{\rm r}(t,t_1) B^<(t_1,t')+A^{<}(t,t_1) B^{\rm a}(t_1,t')) \nonumber\\
\lt[\int_C dt_1 A(t,t_1) B (t_1,t')\rt]^{\rm r}
&=\intinf dt_1 A^{\rm r}(t,t_1) B^{\rm r}(t_1,t'),\label{decomposition}
\end{align}
where $^\ret$ and $^\adv$ denote the retarded and advanced components, respectively.

The perpendicular components of the spin density is defined as 
$ \stil^{\pm}\equiv \frac{1}{2}(\stil^{x}\pm i\stil^{y})$.
Its equilibrium contribution  (denoted by $\stil^{(0)}$) 
as function of wave vector $\qv$ is given as 
($\sigma^{\pm}\equiv \frac{1}{2}(\sigma_{x}\pm i\sigma_{y})$ and 
${\cal A}_{\rm s}^{\pm}\equiv \frac{1}{2}({\cal A}_{\rm s}^{x}\pm i{\cal A}_{\rm s}^{y})$)
\begin{align}
\stil^{\pm,(0)}(\qv) & = \raisebox{-\baselineskip}{\includegraphics[height=2.5\baselineskip]{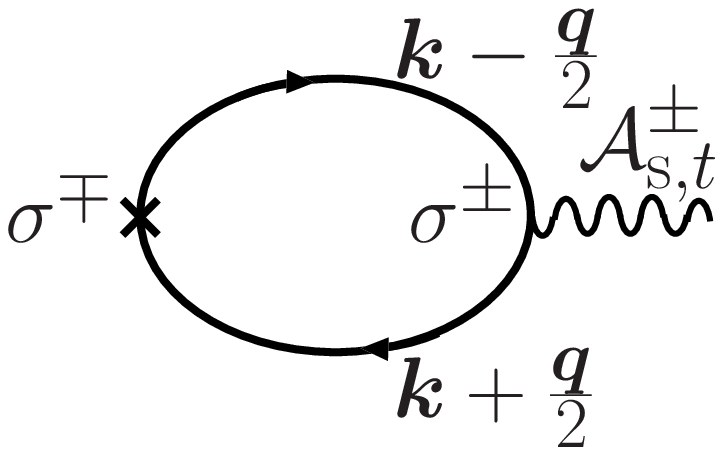}}  \nnr
& = 
-i \frac{\hbar}{V}  \sumom \sum_{\kv}  A_{{\rm s},t}^\pm (\qv) 
\lt[g_{\kvmq,\mp,\omega} g_{\kvpq,\pm,\omega}\rt]^<.
\end{align}
As the spin gauge field is static, all the electron Green's functions carries the same angular frequency, $\omega$, while the wave vector is shifted by that for the gauge field, $\qv$. 
Using Eq. (\ref{decomposition}) and 
\begin{equation}
g^<_{\kv\sigma}(\omega)
 = f_{\kv\sigma}\delta(\hbar\omega-\ekvs),
\label{freelesser2}
\end{equation} 
where $f_{\kv\sigma}\equiv \frac{1}{e^{\beta\ekvs}+1}$ is the Fermi distribution function, 
it reads 
\begin{align}
\stil^{\pm,(0)}(\qv) 
&=
-i \frac{\hbar}{V}  \sumom \sum_{\kv} 
\sum_{\mu} A_{{\rm s},t}^\pm (\qv) 
\lt[g^\ret_{\kvmq,\mp,\omega} g^<_{\kvpq,\pm,\omega}+g^<_{\kvmq,\mp,\omega} g^\adv_{\kvpq,\pm,\omega}\rt]\nnr
&
= \frac{2}{V}\Ascal{t}{\pm}(\qv)\sumkv 
     \frac{f_{\kvpq,\pm}-f_{\kvmq,\mp}}{\epsilon_{\kvpq,\pm}-\epsilon_{\kvmq,\mp}}.
\end{align}
The wave vector $\qv$  for the localized spin structure  is neglected in the adiabatic limit for evaluating the electron energy, resulting in
\begin{align}
\stil^{\pm,(0)}(\qv)
&\simeq 
 -\frac{s}{2\spol}\Ascal{t}{\pm}(\qv),\label{selinear0}
\end{align}
where $s$ is the equilibrium conduction electron spin density defined in Eq. (\ref{seldef}).

Current-induced contributions are calculated as 
\begin{align}
\stil^{\pm,(1{\rm a})}(\qv) 
& = \raisebox{-\baselineskip}{\includegraphics[height=2.5\baselineskip]{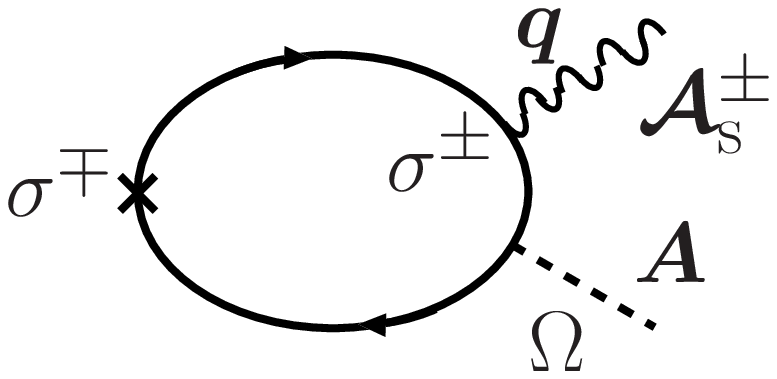}} 
+\raisebox{-\baselineskip}{\includegraphics[height=2.5\baselineskip]{selinear_c2_e}}  \nnr
& = -i \frac{e\hbar^2}{m^2 V}
\lim_{\Omega\ra0}
\sumom \sum_{\kv} \sum_{ij} 
A_i(\Omega){\cal A}_{{\rm s},j}^\pm (\qv) k_j 
\lt[
\lt( k+\frac{q}{2} \rt)_i 
\lt[ g_{\kvmq,\mp,\ommOm} g_{\kvpq,\pm,\ommOm} g_{\kvpq,\pm,\ompOm}\rt]^<
\right.\nonumber\\
& \left.
+
\lt(k-\frac{q}{2}\right)_i 
\lt[g_{\kvmq,\mp,\ommOm} g_{\kvmq,\mp,\ompOm} g_{\kvpq,\pm,\ompOm}\rt]^<
\rt]
\nnr
\stil^{\pm,(1{\rm b})}(\qv) & \equiv 
 \raisebox{-\baselineskip}{\includegraphics[height=2.5\baselineskip]{selinear_d}}
\nnr
&= -i\frac{e}{m V}\lim_{\Omega\ra0} \sumom\sum_{\kv} A_i(\Omega) {\cal A}_{{\rm s},i}^\pm (\qv) 
\lt[
g_{\kvmq,\mp,\ommOm}g_{\kvpq,\pm,\ompOm} 
\rt]^<.
\label{sV}
\end{align}
Evaluating the lesser component, the first contribution reads 
\begin{align}
\stil^{\pm,(1a)} & (\qv)  
= 
-i \frac{e\hbar^2}{m^2 V} \lim_{\Omega\ra0}\sumom\sum_{\kv}  \sum_{ij} A_i(\Omega) {\cal A}_{{\rm s},j}^\pm (\qv)k_j 
\nnr
&\times 
\lt[
-f\lt(\ompOm\rt) \lt[
\lt( k+\frac{q}{2} \rt)_i \gr_{\kvmq,\mp,\ommOm}
{\gr_{\kvpq,\pm,\ommOm} \gr_{\kvpq,\pm,\ompOm} }
+\lt( k-\frac{q}{2} \rt)_i 
{ \gr_{\kvmq,\mp,\ommOm}\gr_{\kvmq,\mp,\ompOm} } \gr_{\kvpq,\pm,\ompOm}\rt]
\rt.\nonumber\\
&  \lt.
+f\lt(\ommOm\rt) \lt[
\lt( k+\frac{q}{2} \rt)_i \ga_{\kvmq,\mp,\ommOm}
{ \ga_{\kvpq,\pm,\ommOm} \ga_{\kvpq,\pm,\ompOm}}
+\lt( k-\frac{q}{2} \rt)_i 
{\ga_{\kvmq,\mp,\ommOm}\ga_{\kvmq,\mp,\ompOm}} \ga_{\kvpq,\pm,\ompOm}\rt]
\rt.\nonumber\\
& 
\lt.
+\lt[f\lt(\ompOm\rt) -f\lt(\ommOm\rt) \rt] \rt. \nnr
& \lt. \times 
\lt[ \lt( k+\frac{q}{2} \rt)_i\gr_{\kvmq,\mp,\ommOm}\gr_{\kvpq,\pm,\ommOm} \ga_{\kvpq,\pm,\ompOm}
+\lt( k-\frac{q}{2} \rt)_i \gr_{\kvmq,\mp,\ommOm}\ga_{\kvmq,\mp,\ompOm}\ga_{\kvpq,\pm,\ompOm}\rt]
\rt]
.
\end{align}
We expand with respect to $\Omega$ and take the limit of $\Omega\ra0$ using identities 
\begin{align}
\gr_{\kvpq,\pm,\ommOm} \gr_{\kvpq,\pm,\ompOm}
& = (\gr_{\kvpq,\pm,\ompOm})^2 -(\gr_{\kvpq,\pm,\ompOm}-\gr_{\kvpq,\pm,\ommOm})\gr_{\kvpq,\pm,\ompOm} \nnr
& =(\gr_{\kvpq,\pm,\ompOm})^2 + \Omega (\gr_{\kvpq,\pm,\ompOm})^3
+O(\Omega^2)
\end{align}
and
\begin{align}
 \frac{\hbar^2\lt( k+\frac{q}{2} \rt)_i}{m}(\gr_{\kvpq,\sigma,\omega})^2 
 =\frac{\partial}{\partial k_i} \gr_{\kvpq,\sigma,\omega}.
\end{align}
After integral by parts with respect to $\kv$, we obtain
\begin{align}
\stil^{\pm,(1{\rm a})} & (\qv) 
=
i \frac{e}{m V}\lim_{\Omega\ra0} \sumom\sum_{\kv}  \sum_i A_i(\Omega)
\nnr
& \times \biggl[
{\cal A}_{{\rm s},i}s^\pm (\qv)
\lt\{
-f\lt(\ompOm\rt) 
 \gr_{\kvmq,\mp,\ommOm} \gr_{\kvpq,\pm,\ompOm}
+f\lt(\ommOm\rt) 
 \ga_{\kvmq,\mp,\ommOm} \ga_{\kvpq,\pm,\ompOm}
\rt\} \nnr
& +\frac{\Omega}{2}{\cal A}_{{\rm s},i}^\pm (\qv) f\lt(\omega\rt) 
\lt[\gr_{\kvmq,\mp,\omega} (\gr_{\kvpq,\pm,\omega})^2
   - (\gr_{\kvmq,\mp,\omega})^2 \gr_{\kvpq,\pm,\omega} 
   -(\ga_{\kvmq,\mp,\omega} (\ga_{\kvpq,\pm,\omega})^2
   - (\ga_{\kvmq,\mp,\omega})^2 \ga_{\kvpq,\pm,\omega} )
   \rt] 
\nnr
& 
+\sum_{j} \frac{\hbar^2\Omega}{2m}{\cal A}_{{\rm s},j}^\pm (\qv) f\lt(\omega\rt)
k_j q_i [(\gr_{\kvmq,\mp,\omega})^2 (\gr_{\kvpq,\pm,\omega})^2 -(\ga_{\kvmq,\mp,\omega})^2 (\ga_{\kvpq,\pm,\omega})^2 ]
\nnr
& 
-\sum_{j} \frac{\hbar^2\Omega}{m} {\cal A}_{{\rm s},j}^\pm (\qv) f'\lt(\omega\rt) k_j 
\lt[ \lt( k+\frac{q}{2} \rt)_i\gr_{\kvmq,\mp,\omega}\gr_{\kvpq,\pm,\omega} \ga_{\kvpq,\pm,\omega}
+\lt( k-\frac{q}{2} \rt)_i \gr_{\kvmq,\mp,\omega}\ga_{\kvmq,\mp,\omega}\ga_{\kvpq,\pm,\omega}
\rt]\biggr]\nnr
&+O(\Omega^2)
.
\end{align}
Similarly, we have 
\begin{align}
\stil^{\pm,(1{\rm b})}(\qv) 
&= -i \frac{e}{m V}\lim_{\Omega\ra0}\sumom\sum_{\kv}  A_i(\Omega) {\cal A}_{{\rm s},i}^\pm (\qv) 
\lt[
-f\lt(\ompOm\rt) \gr_{\kvmq,\mp,\ommOm}\gr_{\kvpq,\pm,\ompOm} 
+f\lt(\ommOm\rt)\ga_{\kvmq,\mp,\ommOm}\ga_{\kvpq,\pm,\ompOm} 
\rt. \nnr &\lt.
+\Omega f'(\omega) \gr_{\kvmq,\mp,\ommOm}\ga_{\kvpq,\pm,\ompOm} \rt]
+O(\Omega^2).
\label{s2til1b}
\end{align}
The sum of the two contributions thus is (using $-i\Omega A_i=E_i$)
\begin{align}
 \stil^{\pm,(1)}(\qv)& \equiv  
\stil^{\pm,(1{\rm a})}(\qv) + \stil^{\pm,(1{\rm b})}(\qv) \nnr
&= 
\frac{e}{m V}\sum_{i}E_i 
\sumom\sum_{\kv}  \nnr
&\times \biggl[ \sum_{j}  \frac{\hbar^2}{m}{\cal A}_{{\rm s},j}^\pm (\qv) f'\lt(\omega\rt) k_j
\lt[ \lt( k+\frac{q}{2} \rt)_i\gr_{\kvmq,\mp,\omega}\gr_{\kvpq,\pm,\omega} \ga_{\kvpq,\pm,\omega}
+\lt( k-\frac{q}{2} \rt)_i \gr_{\kvmq,\mp,\omega}\ga_{\kvmq,\mp,\omega}\ga_{\kvpq,\pm,\omega}
\rt] 
\nnr
&+{\cal A}_{{\rm s},i}^\pm (\qv) 
 f'(\omega) \gr_{\kvmq,\mp,\omega}\ga_{\kvpq,\pm,\omega} 
 \nnr
& +{\cal A}_{{\rm s},i}^\pm (\qv)
f\lt(\omega\rt) 
{
\lt[\gr_{\kvmq,\mp,\omega} (\gr_{\kvpq,\pm,\omega})^2
   - (\gr_{\kvmq,\mp,\omega})^2 \gr_{\kvpq,\pm,\omega} -({\rm c.c.})\rt] 
   } \nnr
& 
+\sum_{j}  {\cal A}_{{\rm s},j}^\pm (\qv) f\lt(\omega\rt)
k_j 
q_i [(\gr_{\kvmq,\mp,\omega})^2 (\gr_{\kvpq,\pm,\omega})^2 -(\ga_{\kvmq,\mp,\omega})^2 (\ga_{\kvpq,\pm,\omega})^2 ] \biggr]
.
\label{s2til1res}
\end{align}.

Considering the adiabatic limit, electron Green's functions are now evaluated at $q=0$.
Using $f'(\omega)=-\delta(\omega)$ valid at low temperatures,  we obtain
\begin{align}
 \stil^{\pm,(1)}(\qv) 
&=
- \frac{e}{2\pi m V}  \sum_{ij}  E_i{\cal A}_{{\rm s},j}^\pm (\qv)\sum_{\kv} \nnr
&\times \biggl[ 
\frac{\hbar^2}{m}k_i k_j
\lt[ \gr_{\kv,\mp}\gr_{\kv,\pm} \ga_{\kv,\pm}
+\gr_{\kv,\mp}\ga_{\kv,\mp}\ga_{\kv,\pm}
\rt] 
+\delta_{ij} \gr_{\kv,\mp}\ga_{\kv,\pm} 
 \nnr
& -\delta_{ij} \int d\omega f\lt(\omega\rt) 
\lt[\gr_{\kv,\mp,\omega} (\gr_{\kv,\pm,\omega})^2
   - (\gr_{\kv,\mp,\omega})^2 \gr_{\kv,\pm,\omega} -({\rm c.c.})\rt] \biggr],
\label{s2til1resad}
\end{align}
where $\ga_{\kv,\sigma}\equiv \ga_{\kv,\sigma,\omega=0}$ and  c.c. denotes complex conjugate.
The summation over $\kv$ in the first term on the right-hand side is carried out as follows.
Using 
$\gr_{\kv,\sigma}\ga_{\kv,\sigma}=i\tau_{\rm e}(\gr_{\kv,\sigma}-\ga_{\kv,\sigma})$ 
(we assume that elastic lifetime $\taue$ for the two spins are equal),
the first term of the right-hand side of Eq. (\ref{s2til1resad}) is evaluated as  
\begin{align}
\sum_{\kv} k_i k_j 
[ \gr_{\kv,\mp}\gr_{\kv,\pm} \ga_{\kv,\pm}+\gr_{\kv,\mp}\ga_{\kv,\mp}\ga_{\kv,\pm} ]
&=
i\tau_{\rm e} \frac{\delta_{ij}}{3}\sum_{\kv} k^2 
[ \gr_{\kv,\mp}\gr_{\kv,\pm} -\ga_{\kv,\mp}\ga_{\kv,\pm} ].
\end{align}
Retarded contribution is evaluated by use of contour integral as 
\begin{align}
\frac{1}{V}  \sum_{\kv} k^2 \gr_{\kv,\mp}\gr_{\kv,\pm} 
&= \int_{-\ef}^\infty d\epsilon \dos(\epsilon)(k(\epsilon))^2 
\frac{1}{\epsilon\pm\spol-i\eta_{\rm e}}\frac{1}{\epsilon\mp\spol-i\eta_{\rm e}} \nnr
&=\frac{i\pi}{2\spol}[\dos_+(\kfu)^2 -\dos_-(\kfd)^2],
\end{align}
where $\dos(\epsilon)\equiv \frac{mk(\epsilon)}{2\pi^2}$, 
$k(\epsilon)\equiv \sqrt{2m(\epsilon+\ef)}/\hbar$ and $\eta_{\rm e}\equiv \frac{\hbar}{2\taue}$.
Other terms of Eq. (\ref{s2til1resad}) are smaller by order of $\frac{\hbar}{\ef\taue}$ and are neglected. 
The result of induced spin density is therefore 
\begin{align}
 \stil^{\pm,(1)}(\qv) 
 &= 
-\frac{1}{2\spol} \jsv\cdot \Ascalv{}^{\pm}, \label{selinerad}
\end{align}
where 
\begin{align}
\jsv\equiv \frac{1}{e}(\sigma_+-\sigma_-)\Ev\equiv \frac{P}{e}\jv,                                          
\end{align}
and $P\equiv \frac{\sigma_+-\sigma_-}{\sigma_++\sigma_-}$ is the spin polarization of the current,  $\sigma_{\pm}\equiv\frac{e^2 n_\pm \tau_{\rm e}}{m}$ being spin-resolved conductivity ($n_\pm$ is spin-resolved electron density).

The non-adiabatic contribution (finite $q$ contribution) has a form like
$\stil^{\pm,{\rm na}}(\qv) = \chi_{ij}^\pm(\qv) E_i \Ascal{j}{\pm}(\qv)$, where $\chi_{ij}^{\pm}$ denotes a correlation function, and has a non-local form in the real space as \citep{TKSLL07,TKS_PR08}
\begin{align}
\stil^{\pm,{\rm na}}(\rv) 
& = E_i \intr' \chi_{ij}^\pm(\rv-\rv') \Ascal{j}{\pm}(\rv').\label{stilna}
\end{align}
This nonlocal  spin polarization represents the force due to electron reflection by localized spin structure like in the case of domain wall \citep{TK04}, as will be discussed later.

From Eqs.(\ref{selinear0})(\ref{selinerad})(\ref{stilna}), the spin polarization density is obtained as
\begin{align}
 \svtil = -\frac{1}{2\spol}\biggl[ \se  \Ascalv{t}+ \jspin{,i}{} \Ascalv{i} \biggr]^\perp +  \svtil^{\rm na}.
\end{align}
The laboratory frame spin density is finally obtained as 
\begin{align}
 \sev&\equiv {\cal R}\svtil 
 =\frac{1}{2\spol}\biggl[ \se(\nv\times\dot{\nv})
 + [ \nv\times(\jsv\cdot\nabla)\nv] \biggr]+\sv^{\rm na},\label{sevres}
\end{align}
where $\sv^{\rm na}\equiv {\cal R}\svtil^{\rm na}$.

When spin relaxation is included in the electron Hamiltonian $H_{\rm e}$, spin polarization perpendicular to the adiabatic case is induced \citep{Zhang04}.
The effect of spin relaxation (sr) leads to spin polarization of \citep{KTS06,Kohno07,TE08}
\begin{align}
 \sev^{\rm sr} & =
 -\frac{1}{\Jsd a^3}\biggl[ \alpha_{\rm sr}\dot{\nv}
 +\frac{a^3}{2S}\beta_{\rm sr}(\jsv\cdot\nabla)\nv\biggr],
\end{align}
where $\alpha_{\rm sr}$ and $\beta_{\rm sr}$ are dimensionless parameters proportional to spin relaxation rate.

The LLG equation including the effects of conduction electron, Eq. (\ref{LLGwithse}), is explicitly given by 
\begin{align}
  (1+\delta) \dot{\nv}& = -\gamma \Bv\times\nv -{\alpha_{\rm sr}} (\nv\times\dot{\nv})
   - \frac{a^3}{2S}(\jsv\cdot\nabla)\nv - \beta_{\rm sr} \frac{a^3}{2S}[\nv\times(\jsv\cdot\nabla)\nv]
   +\torquev_{\rm na}
  \label{LLGJ},
\end{align}
where $\delta\equiv \frac{sa^3}{2S}$ is the renormalization of localized spin due to the conduction electron spin polarization.

\section{Current-driven  domain wall motion}
We discuss here dynamics of a domain wall based on the LLG equation including the current-induced torques, Eq. (\ref{LLGJ}). 
Applied magnetic field and pinning, represented by including a local magnetic field, are not considered.
Similarly to the field-driven case discussed in Sec. \ref{SECDWB}, 
the equation of motion for current-driven wall is obtained by putting the wall profile (\ref{DWsol}) in Eq. (\ref{LLGJ}) and integrating over spatial coordinate as 
\begin{align}
 \dot{\phi}+\alpha\frac{\dot{X}}{\lambda} =&
  \frac{\beta_{\rm w}}{\lambda}  \jtil \nnr
 \dot{X}-\alpha\lambda \dot{\phi} =& 
 -\vc\sin2\phi + \jtil ,
 \label{eqs}
\end{align}
where both $\vc\equiv \frac{K_{\perp}\lambda S}{2\hbar}$ and $\jtil \equiv \frac{a^3P}{2eS}j$ have dimension of velocity,  $P\equiv e\js/j$ being spin polarization of the current.
Here
\begin{align}
\beta_{\rm w}\equiv \beta_{\rm sr}+\beta_{\rm na},
\end{align}
represents the total current-induced force as a result of spin relaxation and electron reflection ($\beta_{\rm na}$).
The reflection force term $\beta_{\rm na}$ arises from the nonadiabatic (nonlocal) torque 
$\torquev_{\rm na}$, and  is written  in terms of electric registance due to the wall, $R_{\rm w}$ \citep{TK04,TKS_PR08} as
\begin{align}
\beta_{\rm na}\equiv \frac{e^2}{h}\overline{N}R_{\rm w},
\end{align}
where $\overline{N}=nA\lambda$ is the total number of spins in wall with thickness $\lambda$, where $n$ and $A$ are electron density and cross sectional area of the system, respectively.

\begin{figure}[t]
\begin{center}
\includegraphics[width=0.5\hsize]{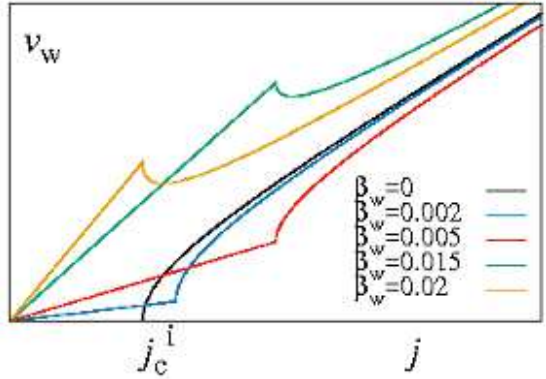}
\caption{ Time averaged wall velocity $v_{w}$ as function of applied spin-polarized current $j$ for $\alpha=0.01$.
Intrinsic pinning threshold $j_{\rm c}^{\rm i}$ exists only for $\betaw=0$. 
The current density where derivative of $v_w$ is discontinuous corresponds to $\jatil$.
\label{FIGvj} }
\end{center}
\end{figure}

When $\beta_{\rm w}=0$ and without magnetic field, the wall velocity when a constant $\jtil$ is applied is easily obtained as \citep{TK04} 
\begin{equation}
\overline{\dot{X}} =
\left\{ \begin{array}{lrr} 
  0  &  \;\;\;\;\; & (\jtil < \jcitil) \\
 \frac{1}{1+\alpha^2}\sqrt{\jtil^2-(\jcitil)^2} 
  & \;\;\;\;\; & (\jtil \geq \jcitil )
\end{array}\right.
\end{equation}
and $\jcitil \equiv {\vc}$ is the intrinsic threshold current density \citep{TK04}.
Namely, the wall cannot move if the applied current is lower than the threshold value as shown in black line in Fig. \ref{FIGvj}.
This is because the torque supplied by the current is totally absorbed by the wall by tilting the out of plane angle to be 
$\sin 2\phiz = \jtil /\vc$ when the current is weak ($| \jtil /\vc|\leq 1$)  and thus the wall cannot move.  This effect is called the intrinsic pinning effect \citep{TKS_PR08}.
For larger current density, the torque carried by the current induces an oscillation of the angle similar to the Walker's breakdown in an applied magnetic field, and the wall speed also becomes an oscillating function of time. 
%

When nonadiabaticity parameter $\betaw$ is finite, the behavior changes greatly and intrinsic pinning effect is removed and the wall can move with infinitesimal applied current if extrinsic pinning is neglected. 
In fact, when the applied current density is $ |\jtil|>|\jatil|$, where 
\begin{equation}
\jatil \equiv \frac{\vc}{1-{\betaw}/{\alpha}},
\label{jatildef}
\end{equation}
the solution of Eq. (\ref{eqs}) is an oscillating function  given by \citep{TKS_PR08} 
\begin{eqnarray}
{\dot{X}}
&=&
\frac{\betaw}{\alpha} {\jtil} 
+\frac{\vc}{1+\alpha^2}\frac{\lt({\jtil}/{\jatil}\rt)^2-1}
  {{\jtil}/{\jatil}-\sin(2\om t-\vartheta)},
\label{dwvelocity1}
\end{eqnarray}
where 
\begin{align}
\om & \equiv \frac{\vc}{\lambda}\frac{\alpha}{1+\alpha^2}
\sqrt{\lt(\frac{\jtil}{\jatil}\rt)^2-1} , &
\sin \vartheta \equiv \frac{\vc}{({\betaw}/{\alpha}-1)\jtil}.
\end{align} 
The time-average of the wall speed is 
\begin{eqnarray}
{\overline{\dot{X}}}
&=&
\frac{\betaw}{\alpha} \jtil
 +\frac{\vc}{1+\alpha^2}\frac{1}{\jatil}
 \sqrt{\jtil^2-\jatil^2}
.\label{averagewallvelocity}
\end{eqnarray}
For current density satisfying $\jtil<\jatil$, $\om$ becomes imaginary and the oscillation in Eq. (\ref{dwvelocity1})
is replaced by an exponential decay in time. The wall velocity then reaches a terminal value of 
 \begin{eqnarray}
\dot{X}
&\ra&
\frac{\betaw}{\alpha} {\jtil} .
\label{dwvelocity2}
\end{eqnarray}
The angle of the wall also reaches a terminal value determined by  
\begin{equation}
\sin 2\phiz \rightarrow  
 \left(\frac{\betaw}{\alpha}-1\right) \frac{\jtil}{\vc}. \label{phizterm}
\end{equation}
The averaged wall speed (Eq. (\ref{averagewallvelocity})) is plotted in Fig. \ref{FIGvj}.

\subsection{Threshold current for domain wall motion}
The intrinsic pinning is a unique feature of current-driven domain wall, as the wall cannot move even in the absence of pinning center. 
In the unit of A/m$^2$, the intrinsic pinning threshold is
\begin{align}
 j_{\rm c}^{\rm i}= \frac{eS^2}{Pa^3\hbar}K_\perp \lambda.
\end{align}
For device applications, this threshold needs to be lowered by reducing the hard-axis anisotropy and wall width \citep{Fukami08}. 
The intrinsic pinning regime is promising for stable device operations, because   the threshold current and dynamics is insensitive to extrinsic pinning and external magnetic field \citep{TK04}, as was confirmed experimentally \citep{Koyama11}.
This is due to the fact that the wall dynamics in the intrinsic pinning regime is governed by a torque, which governs the wall velocity $\dot{X}$, while extrinsic pinning and magnetic field induce force, which governs $\dot{\phi}$; The forces due to sample irregularity therefore does not modify the  motion induced by a torque in the intrinsic pinning regime. 
The insensitivity is therefore a consequence of the fact that the wall carries both linear and angular momenta, and thus has two different mechanisms for driving.

Experimentally, intrinsic pinning is observed in perpendicularly magnetized materials \citep{Koyama11}, while materials with in-plane magnetization mostly are in the extrinsic pinning regime governed by the nonadiabatic parameter $\betaw$ and extrinsic pinning. 
In this regime, the threshold current of the wall motion is given by \citep{TTKSNF06} 
\begin{align}
 j_{\rm c}^{\rm e} \propto \frac{V_{\rm e}}{\betaw},
\end{align}
where $V_{\rm e}$ represents strength of extrinsic pinning potential like those generated by geometrical notches and defects.
Control of nonadiabaticity parameter is therefore expected to be useful for driving domain walls at low current density.

Close to the threshold current density, thermal assist \citep{TVF05} and creep motion \citep{Lemerle98} becomes important. 

Of recent interest from the viewpoint of low current operation  is to use multilayer structures. 
For instance, heavy metal layers turned out to lower the threshold current by exerting a torque as a result of spin Hall effect \citep{Emori13}, and synthetic antiferromagnets turned out to be suitable for fast domain wall motion at low current \citep{Saarikoski14,YangParkin15,Lepadatu17}.

It was recently shown theoretically that strong Rashba-induced magnetic field works as a strong pinning center when introduced locally, and that this Rashba pinning effect is useful for highly reliable control of domain walls in racetrack memories \citep{TataraDW16}. 

\newcommand{\boson}{b}
\newcommand{\AU}{A_U}
\newcommand{\AUv}{\Av_U}
\section{Magnon gauge field}
Let us discuss that effective gauge field exist for magnons. 
Magnon (spin wave) is an excitation representing fluctuation of localized spin.
For localized spin configuration polarized uniformly along $z$ direction, magnon is  introduced by use of the Holstein-Primakov boson, represented by field operators $b$ and $b^\dagger$.
For treating localized spins with spatial and temporal variations, we introduce unitary transformation, in the same manner as the case of electron.
The localized spin is represented as
\begin{align}
  \Sv &= U_3(\rv,t) \widetilde{\Sv},
  \label{Sdef}
\end{align}
where $\widetilde{\Sv}$ is spin vector with average along the $z$ direction and  
 $U_3$ is a $3\times3$ unitary matrix describing a rotation of a vector $\hat{\zv}$ to the direction $(\theta,\phi)$.
The spin vector $\widetilde{\Sv}$ is repressented by magnon field as
\begin{align}
\widetilde{\Sv} &\equiv  S\hat{\zv}+\delta \sv \nnr
  \delta \sv &= \lt( \begin{array}{c}
         \gamma (\boson^\dagger +\boson) \\
         i\gamma(\boson^\dagger-\boson) \\
         - \boson^\dagger \boson 
       \end{array}  \rt),\label{HPboson}
\end{align}
where $\gamma\equiv \sqrt{\frac{S}{2}}$, and the terms third- and higher-order in boson operators are neglected.
The unitary matrix is chosen as \cite{Thiele73}
\begin{align}
  U &= \lt( \begin{array}{ccc}
         \cos\theta\cos\phi & -\sin\phi & \sin\theta\cos\phi  \\
         \cos\theta\sin\phi &  \cos\phi & \sin\theta\sin\phi  \\
         -\sin\theta & 0 & \cos \theta 
       \end{array}  \rt).
\end{align}
The unitary transformation modifies derivatives of spin as
\begin{align}
  \nabla_i\Sv=U_3\lt(\nabla_i-\frac{i}{\hbar}A_{U,i}\rt	) \widetilde{\Sv},
\end{align}
where 
\begin{align}
 A_{U,i}   & \equiv i U_3^{-1} \nabla_i U_3,
\end{align}
is a spin gauge field represented by a $3\times3$ matrix.
Explicitly, the spin gauge field reads 
\begin{align}
 A_{U,i}   & = i\hbar \lt[
   \nabla_\mu\theta 
        \lt( \begin{array}{ccc}
         0 & 0 & 1 \\
         0 & 0 & 0  \\
         -1 & 0 & 0
       \end{array}  \rt)
          +\nabla_\mu\phi 
           \lt( \begin{array}{ccc}
         0 & -\cos\theta & 0 \\
         \cos\theta & 0 & \sin\theta  \\
         0 & -\sin\theta & 0
       \end{array}  \rt)\rt].
\end{align}

Here we consider the simple system with ferromagnetic exchange ineteraction, represented by the Lagrangian 
\begin{align}
 L_{\rm F} &= \sumr \lt[ - \hbar S(1-\cos\theta)\dot{\phi}-\frac{J}{2}  (\nabla\Sv)^2 \rt].
\label{LFerro}
\end{align}
The exchange interaction is written in the rotated frame as
\begin{align}
  (\nabla \Sv)^2 &= (\nabla \widetilde{\Sv})^2 
  + \frac{i}{\hbar} \widetilde{\Sv}^\dagger (\AUv \cdot \nablalr) \widetilde{\Sv}
+O((A_U)^2).
\end{align}
The second term contains terms linear in boson operators. They describe the interaction of localized spin structure and magnons, and are neglected.
We thus obtain to the linear order in the derivative of localized spin structure 
\begin{align}
  (\nabla \Sv)^2 &= 2S\biggl[|\nabla b|^2 -\frac{i}{\hbar}\Asv(b^\dagger\nablalr b)\biggr],
\end{align}
where $\Av_{{\rm s}}=\frac{\hbar}{2}(\nabla{\phi})\cos\theta$ agrees with the spin  gauge field for conduction electron.
(Although magnon spin is $-1$, we keep the prefactor of $\frac{1}{2}$ for electron spin in $\Av_{{\rm s}}$ to avoid confusion.) 
The time-derivative term of Eq. (\ref{LFerro}), $L_{\rm B}$,  is written in terms of magnon opearators as
\begin{align}
 L_{\rm B}=-i\frac{\hbar}{2} \lt[ b^\dagger \partial_t b+\frac{i}{\hbar}A_{{\rm s},t}b^\dagger b \rt],
\end{align}
where $A_{{\rm s},t}$ is the time component of spin gauge field.
The Lagrangian (\ref{LFerro}) is therefore written as a Lagrangian for a boson interacting with an effective U(1) gauge field \citep{TataraDW15},
\begin{align}
 L_{\rm F} &= \sumr \lt[ -i\frac{\hbar}{2} b^\dagger \partial_t b+ A_{{\rm s},t}b^\dagger b  -JS|\nabla b|^2  -\Asv\cdot\jv_{\rm m}
 \rt],
\label{LFerro2}
\end{align}
where
\begin{align}
 \jv_{\rm m}\equiv -i\frac{JS}{\hbar}b^\dagger\nablalr b,
\end{align}
is magnon current.
As magnon carry negative spin, the sign of the gauge coupling term, $\Asv\cdot\jv_{\rm m}$,   of Eq. (\ref{LFerro2}) is negative.
Equation (\ref{LFerro2}) indicates an interesting fact that magnon feels an effective U(1) gauge field that is  the same as the one $A_{{\rm s},\mu}$ for conduction electron spin. This fact might be natural from symmetry point of view, but is not obvious. 
At equilibrium, magnon chemical potential is zero, and expectation value of magnon number is determined by temperature.  
Equation (\ref{LFerro2}) indicates that an effective chemical potential for magnon, $A_{{\rm s},t}$, emerges from dynamics of localized spin. 
Experimental observation of dynamically-induced magnon chemical potential was reported recently \citep{Du17}.
Spin structure with finite effective magnetic field $\Bvs$ like magnetic skyrmion induces magnon Hall effect \citep{Dugaev05}.

\section{Interface Rashba spin-orbit effects}
Although theoretical physics has been focusing on infinite systems for exploring beautiful general law supported by symmetries, studying such 'beautiful' systems seems to becoming insufficient in condensed matter physics.
This is because demands to understand interfaces and surfaces has been increasing rapidly as devices are becoming smaller and smaller to meet the needs for fast processing of huge data.
Systems with lower symmetry are therefore important subjects of material science today.

Surfaces and interfaces have no inversion symmetry, and this leads to emergence of an antisymmetric exchange interaction (Dzyaloshinskii-Moriya interaction) \citep{Dzyaloshinsky58,Moriya60} in magnetism .
As for electrons, broken inversion symmetry leads to a peculiar spin-orbit interaction, called the Rashba interaction \citep{Rashba60}, whose quantum mechanical Hamiltonian is
\begin{align}
{H}_{\rm R} &= i \alphaRv \cdot (\bm{\nabla} \times \bm{\sigma}), \label{RashbaH}
\end{align}
where $\bm{\sigma}$ is the vector of Pauli matrices and $\alphaRv$ is a vector representing the strength and direction of the interaction.
The form of the interaction is the one derived directly from the Dirac equation as a relativistic interaction, but the magnitude can be strongly enhanced in solids having heavy elements compared to the vacuum case.

As is obvious from the form of the Hamiltonian, the Rashba interaction induces electromagnetic cross correlation effects where a magnetization and an electric current are induced by external electric and magnetic field, $\Ev$ and $\Bv$, respectively, like represented at finite frequency $\omega$ as 
\begin{align}
\Mv&={\kappa_{ME}}({\bm{{\alpha}}_{\rm R}}\times\Ev), 
&\jv=i\hbar{\omega}\gamma\kappa_{ME}({\bm{{\alpha}}}_{\rm R}\times\Bv), \label{spinchargemixing}
\end{align}
where $\kappa_{ME}$ is a  coefficient  depending on frequency \citep{Shibata16}.
The emergence of spin accumulation  from the applied electric field, mentioned in Refs. \citep{Rashba60,Dyakonov71}, was studied by Edelstein \citep{Edelstein90} in detail, and the effect is sometimes  called Edelstein effect. 
The generation of electric current by magnetic field or magnetization, called the inverse Edelstein effect \citep{Shen14}, was recently observed in multilayer of Ag, Bi and a ferromagnet \citep{Sanchez13}.

\subsection{Effective magnetic field}
When a current density $\jv$ is applied, the conduction electron has average momentum of $\pv=\frac{m}{en}\jv$ ($n$ is electron density), and Eq. (\ref{RashbaH}) indicates that an effective magnetic field of 
$
\Bv_{\rm e}=\frac{ma^3}{e\hbar^2\gamma}\alphaRv\times\jv,
$
acts on the conduction electron spin ($\gamma(=\frac{e}{m})$ is  the gyromagnetic ratio).
When the $sd$ exchange interaction between the conduction electron and localized spin is strong, this field multiplied by the the spin polarization, $P$, 
is the field acting on the localized spin. 
Namely, the localized spin feels a current-induced effective magnetic field of 
\begin{align}
\Bv_{\rm R}=\frac{Pma^3}{e\hbar^2\gamma}\alphaRv\times\jv . \label{BR}                                                                                                                                               \end{align}

The strength of the Rashba-induced magnetic field is estimated (choosing $a=2$\AA) as  
$B_{\rm R}=2\times 10^{16} \times \alphaR$(Jm)$\js$(A/m$^2$); 
For a strong Rashba interaction $\alphaR=1$ eV\AA\ like at surfaces \citep{Ast07}, $B_{\rm R}=4\times 10^{-2}$ T at $\js=10^{11}$ A/m$^2$.
This field appears not very strong, but is sufficient at modify the magnetization dynamics. 
In fact, for the domain wall motion, when the Rashba-induced magnetic field is along the magnetic easy axis, the field  is equivalent to that of  an effective $\beta$ parameter of
\begin{align}
\beta_{\rm R}=\frac{2m \lambda}{\hbar^2}\alphaR                                        ,
\end{align}
 where $\lambda$ is the wall thickness.
If  $\alphaR=1$ eV\AA, $\beta_{\rm R}$ becomes extremely large like $\beta_{\rm R}\simeq 250$ for $\lambda=50$ nm. Note that $\beta$ arising from spin relaxation is the same order as Gilbert damping constant, namely of the order of $10^{-2}$.
Such a large effective $\beta_{\rm R}$ from Rashba effect is expected to leads to an extremely fast domain wall motion under current \citep{Obata08,Manchon09}.

Experimentally, it was argued that fast domain wall motion observed in Pt/Co/AlO was due to the Rashba interaction \citep{Miron10}, but the result is later associated with the torque generated by spin Hall effect in Pt layer \citep{Emori13}.

\begin{figure}[tb]
  \begin{center}
   \includegraphics[width=0.4\hsize]{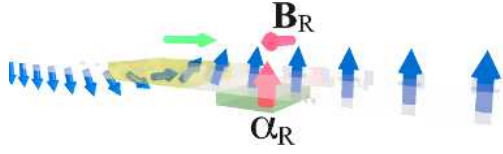}
  \end{center}
\caption{ 
Domain wall (DW) transported with current $j$ in a magnetic wire with a local Rashba spin-orbit pinning potential $\alphaR$ of finite width used to capture and pin a moving wall.
\label{FIGRashbaDWstopping}}
\end{figure}

Domain walls in ferromagnetic nano wires are potential building-blocks of future technologies such as racetrack memories \citep{YangParkin15}.
For such memories, efficient mechanisms to initiate and stop domain-wall motion are necessary. 
It was pointed out theoretically that a locally embedded spin-orbit interaction of Rashba type (Fig. \ref{FIGRashbaDWstopping}) acts as a strong pinning center for current-driven domain walls and that efficient capturing and depinning of the wall is realized even using a weak spin-orbit interaction \citep{TataraDW16}. 

\subsection{Rashba-induced spin gauge field}
Since the interaction (\ref{RashbaH}) is the one coupling to the spin current, the Rashba interaction 
is regarded as a gauge field acting on electron spin as far as the linear order concerns.
In the field representation, the interaction is 
\begin{align}
{H}_{\rm R} &= i \intr c^\dagger \alphaRv \cdot (\bm{\nablalr} \times \bm{\sigma}) c.
\label{RashbaHF}
\end{align}
Considering the case of strong $sd$ exchange interaction, the interaction is expressed in terms of the rotated frame electron, $a\equiv U^{-1}c$, where $U$ is defined in Eq. (\ref{Udef}), as 
\begin{align}
H_{\rm R} &= -\frac{e}{2m} \intr a^\dagger (-i\nablalr) \cdot {{\cal A}}_{\rm R} a ,
\label{RashbaHF2}
\end{align}
where 
\begin{align}
 { {\cal A}}_{{\rm R},\alpha} &\equiv -\frac{m}{e\hbar}\epsilon_{\alpha\beta\gamma} \alpha_{{\rm R},\beta} {\cal R}_{\gamma\delta}\sigma_{\delta}
 \equiv  { {\cal A}}_{{\rm R},\alpha}^\delta \sigma_{\delta},
\end{align}
and ${\cal R}$ is defined in (\ref{calRdef}).
We neglect contributions including derivatives of localized spin structure, namely, spin gauge field ${\cal A}_{\rm s}$.
In the strong $sd$ exchange interaction case, a gauge field ${\cal A}_{\rm R}$ is projected to the diagonal component, 
${\cal A}_{{\rm R},\alpha}\ra {\cal A}_{{\rm R},\alpha}^z\equiv A_{{\rm R},\alpha}$, giving rise to a U(1) effective gauge field of  
\begin{align}
\Av_{\rm R}\equiv -\frac{m}{e\hbar}(\alphaRv\times\nv).
\label{AvRdef}
\end{align}
Existence of a gauge field naturally leads to an effective electric and magnetic field \citep{Kim12,Nakabayashi14} 
\begin{eqnarray}
  \Ev_{\rm R} = -\dot{\Av}_{\rm R} = \frac{m}{e\hbar}(\alphaRv\times\dot{\nv})  \nnr
  \Bv_{\rm R} = \nablav \times \Av_{\rm R} = - \frac{m}{e\hbar}\nablav\times(\alphaRv\times\nv).
  \label{ERBR}
\end{eqnarray}
In the presence of electron spin relaxation, the electric field has a perpendicular component \citep{Tatara_smf13}
\begin{equation}
  \Ev_{\rm R}' = \frac{m}{e\hbar}\beta_{\rm R}[\alphaRv\times(\nv\times\dot{\nv})] , \label{ERp}
\end{equation}
where $\beta_{\rm R}$ is a coefficient representing the strength of spin relaxation.
For the case of strong Rashba interaction of $\alpha_{\rm R}=3$ eV\AA, as realized in Bi/Ag,  
the magnitude of the electric field is $|E_{\rm R}|=\frac{m}{e\hbar}\alpha_{\rm R}\omega=26$kV/m if the angular frequency $\omega$ of magnetization dynamics is 10 GHz. 
The magnitude of relaxation contribution is $|E_{\rm R}'|\sim260$V/m if $\beta_{\rm R}=0.01$.
The effective magnetic field in the case of spatial length scale of 10 nm is high as well; $B_{\rm R}\sim 260$T. 

\begin{figure}[tb]
  \begin{center}
   \includegraphics[width=0.35\hsize]{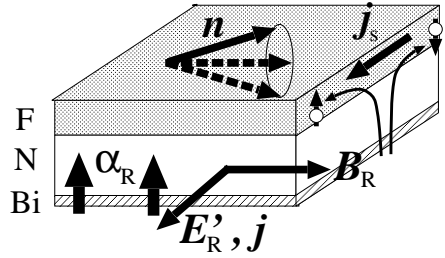}
  \end{center}
\caption{ 
Schematic figure depicting (spin relaxation contribution of) Rashba-induced spin electric field $\Ev_{\rm R}'$ generated by magnetization precession in a junction  of a ferromagnet (F), a nonmagnetic spacer (N) and a heavy atom layer (Bi), where the Rashba interaction is induced.
Electric current $\jv$ is induced as a result of motive force $\Ev_{\rm R}'$ in the direction perpendicular to both $\nv\times\dot{\nv}$ and Rashba field $\alphaRv$.
The magnetic field component $\Bv_{\rm R}$ lies in-plane and is expected to induce 'giant' spin Hall effect when an electric field is applied perpendicular to the plane.
\label{FIGEs_alpha}}
\end{figure}

The Rashba-induced electric fields, $\Ev_{\rm R}$ and $\Ev_{\rm R}'$, are important from the viewpoint of spin-charge conversion.
In fact, results (\ref{ERBR})(\ref{ERp}) indicates that a voltage is generated by a dynamics magnetization if the Rashba interaction is present.
Importantly, this effect emerges even from a spatially uniform  magnetization precession, in sharp contrast to the conventional adiabatic effective electric field 
(spin motive force) (Eq. (\ref{EsBsdef})).
In the case of a think film with Rashba interaction perpendicular to the plane and with a precessing  magnetization,
the component $\Ev_{\rm R}\propto \dot{\nv}$ has no DC component, while 
the relaxation contribution  $\Ev_{\rm R}'$ has a DC component perpendicular to $\overline{\nv\times\dot{\nv}}\parallel \overline{\nv}$, where $\overline{\nv\times\dot{\nv}}$ and $\overline{\nv}$ denote time-averages.
The geometry of this current pumping effect, 
$\jv\propto  \Ev_{\rm R}' \propto  \alphaRv\times \overline{\nv} $ (Fig. \ref{FIGEs_alpha}),
is therefore the same as the one expected in the case of inverse Edelstein effect, conventionally discussed in terms of spin current \cite{Shen14}. 

In the present form, Eqs. (\ref{ERBR})(\ref{ERp}), the Rashba-induced electric field is a local quantity; a voltage is generated by a direct contact between the Rashba interaction and magnetization. 
It is expected, however,  to become long-ranged if electron diffusion is taken into account.   
The inverse Edelstein effect reported in Ref. \citep{Sanchez13} in a system with a Ag spacer, would therefore be explained by the long-ranged Rashba-induced voltage.
For this scenario to be justified, it is crucial to confirm the existence of magnetic component,  $ \Bv_{\rm R} $, which can be of the order of 100T.
In the setup of Fig. \ref{FIGEs_alpha},  $ \Bv_{\rm R} $ is along $\overline{\nv}$. 
The field can therefore be detected by measuring ``giant'' in-plane spin Hall effect when a current is injected perpendicular to the plane.

Electric current generation by spin dynamics with Rashba spin-orbit interaction was theoretically discussed in the case of a dot \citep{Levitov03} and two-dimensional electron gas \citep{Tokatly10}

\section{Anomalous optical properties of Rashba conductor}

The idea of effective gauge field  is useful for extending the discussion to include other degrees of freedom, like optical properties.
In fact, the fact that the Rashba interaction coupled with magnetization leads to an effective vector potential $\Av_{\rm R}$ (Eq. (\ref{AvRdef})) for electron spin indicates that the existence of intrinsic spin flow.
Such intrinsic flow affects the optical properties, as  incident electromagnetic waves get Doppler shift when interacting with flowing electrons, resulting in a transmission depending on the direction (directional dichroism), as was theoretically demonstrated in Refs. \citep{Shibata16,Kawaguchi16}.
The magnitude of the directional dichroism for the case of wave vector $\qv$ is determined by $\qv\cdot(\alphaRv\times\nv)$. 
The vector $(\alphaRv\times\nv)$, which breaks both time-reversal and spatial inversion symmetries, is called in the context of multiferroics the toroidal moment, and it was argued to acts as an effective vector potential for light \citep{Sawada05}.

\subsection{Electromagnetic metamaterial property}
\begin{figure}[tb]
\begin{minipage}{0.3\hsize}
\centering
\includegraphics[height=5\baselineskip]{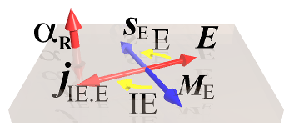}
\caption{ 
Schematic figure showing the cross-correlation effects in the plane perpendicular to the Rashba field $\alphaRv$.
Edelstein effect (E) generates spin density, $\sv_{\rm E}$, from the applied electric field, and inverse Edelstein effect (IE) generates current $\jv_{{\rm IE}\cdot{\rm E}}$ from magnetization $\Mv_{\rm E}$.
\label{FIGEIE}}
\end{minipage}
\hfill
\begin{minipage}{0.3\hsize}
\centering
\includegraphics[height=8\baselineskip]{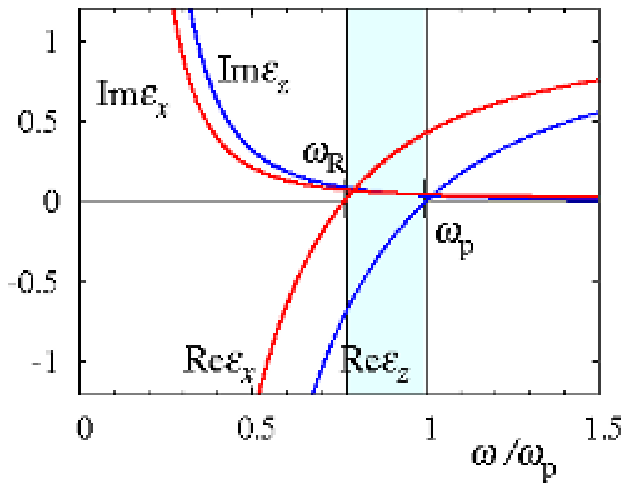}
\caption{ Electric permittivity of Rashba conductor with $\tilde{\alpha}\equiv \frac{m\alpha}{\hbar^2\kf}=1.0$ and
$\eta/(\hbar\omegaP) = 0.01$.
The shaded region between $\omegaR$ and $\omegaP$ is the hyperbolic region showing metamaterial behavior.
\label{FIGepsilon}}
 \end{minipage}
\hfill
\begin{minipage}{0.3\hsize}
\centering
\includegraphics[height=7\baselineskip]{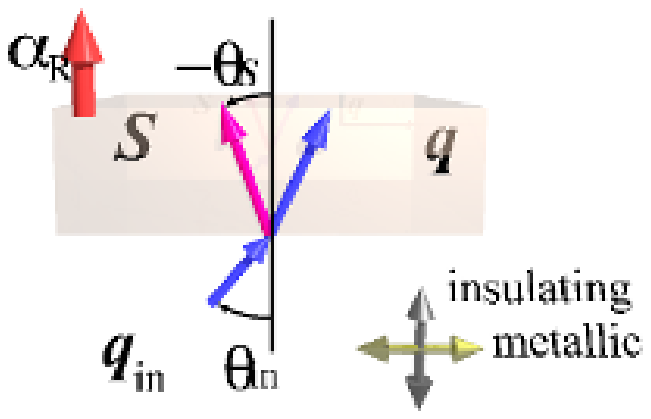}
\caption{ Schematic picture showing negative refraction of Rashba conductor. 
The physical flow of light (the Poyinting vector $\Sv$) incident with an angle $\theta_{\rm in}$ with respect to $\alphaRv$ get refracted in the negative direction. The wave vector $\qv$ of the refracted wave are in the normal direction, but it does not describe physical flow.
\label{FIGnegative}}
 \end{minipage}
\end{figure}

It was shown that Rashba conductor itself, without magnetization, shows peculiar optical properties such as negative refraction as a result of spin-charge mixing effects \citep{Shibata16,Shibata18}.
In fact, spin-charge mixing effects of Eq. (\ref{spinchargemixing})  leads to a current generated by applied electric field, $\Ev$,  given by (schematically shown in Fig. \ref{FIGEIE}) 
\begin{align}
 \jv_{{\rm IE}\cdot{\rm E}} = -i\hbar\omega\gamma{(\kappa_{ME})^2} [\alphaRv\times(\alphaRv\times\Ev)]. 
\end{align}
As it is opposite to the applied field, the mixing effect results in a softening of the plasma frequency as for the $\Ev$ having components perpendicular to $\alphaRv$.
The electric permittivity of the system is therefore anisotropic; Choosing $\alphaRv$ along the $z$ axis, we have 
\begin{align}
\varepsilon_{z}&=1-\frac{\omegaP^2}{\omega(\omega+i\eta)}, 
& 
\varepsilon_{x}=\varepsilon_{y}
=1-\frac{\omegaR^2}{\omega(\omega + i\eta)}, \label{e-perp}
\end{align}
where $\omegaP = \sqrt{{e^2n_{\rm e}}/{\varepsilon_{0}m}}$ 
is the bare plasma frequency ($\nel$ is the electron density), and 
\begin{align}
  \omegaR \equiv \omegaP\sqrt{1+\Re C(\omegaR)}<\omegaP,
\end{align}
 is the plasma frequency reduced by the spin mixing effect.
Here 
\begin{align}
C(\omega) \equiv -\frac{{\alphaR}^2\kf^2}{\ef n}\int\frac{d^3k}{(2\pi)^3} \frac{\gamma_k s_k}{(\hbar\omega+i\eta)^2-4\gamma_k^2},         
\end{align}
with  
$s_\kv\equiv\sum_{\sigma=\pm}\sigma f_{\kv\sigma}$ is the electron spin polarization and  $\gamma_k\equiv |\kv\times\alphaRv|$, 
represents the correlation function representing the Rashba-Edelstein effect \citep{Shibata16}.
The real part of $C(\omega)$ is negative near $\omega\sim\omegaP$ and thus  
$\omegaR<\omegaP$.
The frequency region $\omegaR<\omega<\omegaP$ is of interest, as 
 the system is insulating ($\varepsilon_z>0$) in the direction of the Rashba field but 
metallic in the perpendicular direction  ($\varepsilon_x<0$).
The dispersion in this case becomes hyperbolic, and the group velocity and phase velocity along $\qv$ can have opposite direction, resulting in negative refraction.
Rashba system is, therefore a natural hyperbolic metamaterial \citep{Narimanov15}.
A great advantage of Rashba conductors are that the metamaterial behavior arises in the infrared or visible light region, which is not easily accessible in artificially fabricated systems. 
In the case of BiTeI with Rashba splitting of $\alpha=3.85$ eV\AA \citep{Ishizaka11}, 
the plasma frequency is $\omegaP=2.5\times10^{14}$ Hz (corresponding to a wavelength of $7.5\mu$m) for 
$\nel=8\times 10^{25}$ m$^{-3}$ and $\ef=0.2$ eV \citep{Demko12}.
We then have  $\omegaR/\omegaP=0.77$ ($\omegaR=1.9\times 10^{14}$ Hz, corresponding to the wavelength of $9.8\mu$m), and  hyperbolic behavior arises in the infrared regime. 

Spintronics devices have potential applications for electromagnetic metamaterials as was argued for spin torque oscillators \citep{Tatara_meta13}.

\subsection{Directional dichroism of magnetic Rashba conductor}
Rashba interaction breaking spatial inversion symmetry leads to various interesting spin transport and optical responses as we have seen.
When the time reversal symmetry is broken in addition, other anomalous properties are expected. 
Here we discuss such a optical property.
The system we consider is a Rashba conductor with magnetization $\Mv$ (or in a magnetic field), which breaks the time-reversal symmetry.
Unique feature in this system is that we have a vector 
\begin{align}
\uv\equiv \alphaRv\times\Mv,                                                                                  \end{align}
 that has the same symmetry as a velocity; Namely, $\uv$ represents a sort of intrinsic flow in magnetic Rashba conductor.
Such a vector can induce intriguing cross correlation effect, because it allows a coupling between the electric and the magnetic fields like \citep{Kawaguchi16}
\begin{align}
{\cal{H}}_{u}&=g\uv\cdot(\Ev\times\Bv), \label{ExB}
\end{align}
where $g$ is a coefficient.
The vector $\frac{1}{\mu}\Ev\times\Bv$ (where $\mu$ is the magnetic permeability of solids) is the Poynting vector representing the momentum of the electromagnetic wave. The vector coupling of Eq. (\ref{ExB}) is thus essentially $\uv\cdot\kv$ ($\kv$ is the wave vector of the electromagnetic wave), and represents the Doppler shift with respect to the intrinsic 'velocity', $\uv$.
In other words, $\uv$ acts as an effective vector potential for electromagnetic vector potential $\Av$. 
Quantity $\uv$ has the same symmetry as toroidal moment in multiferroics, and thus it is reasonable that it causes directional dichroism.  
As seen from Eq. (\ref{AvRdef}), $\uv$ is also an effective vector potential for conduction electron spin. It is interesting that the same quantity works as a vector potential for both electron spin and photon. 
Note that the intrinsic flow represented by $\uv$ is not charge current nor spin current; those currents are driven by effective electric field of Eq. (\ref{ERBR}) and are proportional to time derivative of $\uv$. 

The Maxwell's equations taking account of Eq. (\ref{ExB}) read 
\begin{align}
 \bm{\nabla}\cdot\Ev =& \frac{\rho}{\ez}- \frac{1}{\ez}\bm{\nabla}\cdot(\uv\times\Bv), \nnr
 \bm{\nabla}\times\Bv =& \muz\jv+\ez\muz\frac{\partial \Ev}{\partial t} +\muz\frac{\partial }{\partial t}(\uv\times\Bv)-\muz\bm{\nabla}\times(\uv\times\Ev).
\end{align}
Therefore the total electric and magnetic fields in the present system read 
\begin{align}
 \Ev_{\rm tot} =& \Ev+\frac{1}{\ez}(\uv\times\Bv), \nnr
 \Bv_{\rm tot}= & \Bv+\muz(\uv\times\Ev). \label{EBcross}
\end{align}
These relations are the same as what we obtain in a moving frame and they clearly represent a cross-correlation effect as a result of the Doppler shift.
The interaction (\ref{ExB}) modifies the electric permittivity as 
\begin{align}
\epsilon_{ij}&=\epsilon^{(0)}_{ij}
-\frac{1}{\ez\omega}(u_{i}\delta_{lj}+\delta_{il}u_{j})k_{l}
+\frac{2}{\ez\omega}\delta_{ij}(\bm{u}\cdot\kv) \label{epqu}.
\end{align}
The last diagonal term linear in $\kv$ leads to transmission and reflection depending on the direction $\kv$ with respect to a vector $\uv$, namely, an asymmetric light propagation (directional dichroism).

\section{Summary}
We have discussed spintronics effects from a unified viewpoint of effective gauge field (spin gauge field) coupling to electron spin current.
Effective gauge field is a general concept to describe low energy behavior of smooth background structure, which is localized spin (magnetization) structure in the present case. 
As we have seen, gauge field description has an advantage that driving field is clearly identified and provides solid physical picture of the effects. 
In fact, spin motive force, a voltage generated by dynamic magnetization, is induced by the adiabatic component of spin gauge field, while spin pumping effect is driven by the nonadiabatic component.
Spin-charge conversion and anomalous optical properties due to Rashba spin-orbit interaction were also discussed.
We have seen that localized spin structures generates the same effective adiabatic gauge field for both magnon and conduction electron spin. Moreover, a 'troidal' moment arising from the Rashba interaction and localized spin acts as an effective gauge field for electron spin and photons. 
Effective gauge field therefore describes transport and responses of various degrees of freedom in a unified viewpoint.

\vspace{\baselineskip}

\noindent \textbf{Acknowledgements}

The author thanks 
a Grant-in-Aid for Scientific Research (B) (No. 17H02929) from the Japan Society for the Promotion of Science 
and  
a Grant-in-Aid for Scientific Research on Innovative Areas (No.26103006) from The Ministry of Education, Culture, Sports, Science and Technology (MEXT), Japan for financial support.

\appendix
\section{Summary of path-ordered Green's function}
Here we present  a practival introduction for non-equilibrium Green's function.
\subsection{Obseervable and path-ordering}
\begin{figure}[t]
\centering
 \includegraphics[width=0.4\hsize]{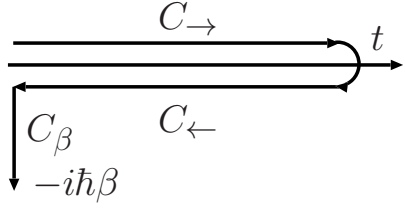}
 \caption{ The contour $C\equiv C_\ra+C_\la+C_\beta$ which arises to describe physical observables. Path-ordered Green's function is defined on this contour. 
 \label{FIGCO}}
\end{figure}

An expectation value of observable $\Ohat$ at time $t$ is
\begin{align}
 \overline{O}(t) &= \frac{1}{Z} \tr \lt[ e^{-\beta \Hhat_0} \OH(t) \rt],
\end{align}
where 
$Z\equiv \tr [e^{-\beta \Hhat_0}]$ is the partition function, 
\begin{align}
 \OH(t) &\equiv \Uhat(t-\tminf)^\dagger  \Ohat \Uhat(t-\tminf),
\end{align}
is the Heisenberg representation with unitary operator representing the time-development
\begin{align}
\Uhat(t-\tminf)=Te^{-\frac{i}{\hbar}\int_{\tminf}^t dt \Hhat(t)},
\end{align}
$T$ being the time-ordering operator.
Writing explicitly the time-dependence, we have
\begin{align}
 \overline{O}(t) &= \frac{1}{Z} \tr \lt[ e^{-\beta \Hhat_0} [\overline{T} e^{\frac{i}{\hbar}\int_{\tminf}^t dt \Hhat(t)} ] \Ohat  [T e^{-\frac{i}{\hbar}\int_{\tminf}^t dt \Hhat(t)}] \rt],
 \label{Ohatexpect}
\end{align}
where 
$\overline{T}$ is the anti time-ordering operator.
Physical observable is therefore written as a path-ordered expectation value along a path $C\equiv C_\ra+C_\la+C_\beta$, where $C_\ra$ is a path from   
$\tminf$ to $\tinf$, $C_\la$ is from $\tinf$ to $\tminf$ and imaginary direction, $C_\beta$
(Fig. \ref{FIGCO}), as
\begin{align}
 \overline{O}(t) &= \frac{1}{Z} \tr \lt[{T_C} e^{-\frac{i}{\hbar}\int_C dt \Hhat(t)} \Ohat  \rt],
\label{OhatexpectC}
\end{align}
where $T_C$ is the ordering operator along $C$.
\subsection{Interaction representation}
Time-development operator $\Uhat$ is separated into a free part $\Uhat_0$ described by $\Hhat_0$ and an interaction part as
\begin{align}
\Uhat(t-\tminf)=\Uhat_0(t-\tminf)T[e^{-\frac{i}{\hbar}\int_{\tminf}^t dt \Vtil(t)}]
\end{align}
Observable of Eq. (\ref{Ohatexpect}) then reads 
\begin{align}
 \overline{O}(t) &= \frac{1}{Z} \tr \lt[e^{-\beta \Hhat_0} \overline{T}[e^{\frac{i}{\hbar}\int_{\tminf}^{t} dt \Vtil(t)} ]\Otil(t) T[e^{-\frac{i}{\hbar}\int_{\tminf}^{t} dt \Vtil(t)} ]  \rt],
\label{OhatexpectInt}
\end{align}
where 
\begin{align}
\Otil(t) \equiv [\Uhat_0(t-\tminf)]^{-1}\Ohat \Uhat_0(t-\tminf).
\end{align}

If the observable is quadratic in the field operators as $\Ohat=\sum_{\alpha\beta}\chat_\alpha {\cal O}_{\alpha\beta} \chat_\beta$
($\alpha$ and $\beta$ are index specifying the field, like spin indices), physical ovservable is written as
\begin{align}
 \overline{O}(t) &= \tr \lt[ {\cal O}  G(\tc,\tc+0) \rt],
\label{OGf}
\end{align}
where 
\begin{align}
G^c(\tc,\tc')&\equiv -\frac{i}{Z}\tr \lt[ {T_C e^{-\frac{i}{\hbar}\int_{C} dt \Hhat(t)} \chat(\tc)\chat^\dagger(\tc')} \rt],
\label{noneqGf}
\end{align}
is path-ordered Green's function defined for complex time $\tc,\tc'$, $T_C$ being the path-ordering operator.
To calculate physical observable, we need to know how to calculate the path-ordered Green's function, and to evaluate real-time value of path-ordered Green's function, defined on a complex time contour.
By definition, the path-ordered Green's function agrees with the time-ordered Green's function if the two time are on the upper contour, $C_\ra$, and it is anti-time-ordered if two times are on $C_\la$;
\begin{align}
G^c(\tc\in C_\ra,\tc'\in C_\ra)&= -i \average{T \cH(t)\cH^\dagger(t')} \equiv \Gto(t,t') \nnr 
G^c(\tc\in C_\la,\tc'\in C_\la)&= -i \average{\overline{T} \cH(t)\cH^\dagger(t')} \equiv \Gat(t,t') .
\label{Gftoat}
\end{align}
When one of the two times is on $C_\ra$, the ordering of the operator is fixed irrespective of $\tc$ and $\tc'$.
When $\tc\in C_\ra$ and $\tc'\in C_\la$, we have the lesser Green's function,
\begin{align}
G^c(\tc\in C_\ra,\tc'\in C_\lar)&= i \average{\cH^\dagger(t')\cH(t)} \equiv \Gless(t,t'),
\label{Gfless}
\end{align}
while we have the greater Green's function for the opposite case, 
\begin{align}
G^c(\tc\in C_\la,\tc'\in C_\ra)&= -i \average{\cH(t)\cH^\dagger(t')} \equiv \Ggrt(t,t').
\label{Gfgrt}
\end{align}
Those four Green's functions are not independent, because 
\begin{align}
 \Gto(t,t') =\theta(t-t') \Ggrt(t,t') +\theta(t'-t)\Gless(t,t').
\end{align}
In fact,
\begin{align}
\Gto(t<t',t')&= i \average{\cH^\dagger(t')\cH(t)} =\Gless(t,t') \nnr
\Gto(t>t',t')&=- i \average{\cH(t)\cH^\dagger(t')} =\Ggrt(t,t') .
\end{align}

\subsection{Free electron Green's functions}
Let us derive explicit form of Green's functions for free electron described by the Hamiltonian \begin{align}
\Hhat= \intr \chat^\dagger \lt(\frac{-\hbar^2\nabla^2}{2m}-\ef\rt)\chat.                                                                                               \end{align}
The field operator in the momentum representation is
\begin{align}
 \ctil_\kv(t) &= e^{-\frac{i}{\hbar}\ekv t} \chat_{\kv},
\end{align}
where 
$\ekv\equiv \frac{\hbar^2k^2}{2m}-\ef$.
The time-ordered Green's function of free electron is therefore 
\begin{align}
 \gto_\kv(t-t') &= -i e^{-\frac{i}{\hbar}\ekv t}
    (\theta(t-t')\average{\chat_\kv \chat_{\kv}^\dagger }-\theta(t'-t)\average{\chat_{\kv}^\dagger\chat_\kv  }) \nnr
  &= -i e^{-\frac{i}{\hbar}\ekv t}    (\theta(t-t')(1-f_{\kv})-\theta(t'-t)f_{\kv}),
   \label{gtokresult}
\end{align}
where 
\begin{align}
 f_{\kv}\equiv \frac{1}{e^{\beta\ekv}+1},
\end{align}
is the Fermi distribution function.
In the frequency representation, we have 
\begin{align}
 \gto_{\kv,\omega}&=\frac{1-f_\kv}{\hbar\omega-\ekv+i0} -\frac{f_\kv}{\hbar\omega-\ekv-i0}, \label{gtokcomplete}
\end{align}
where $\pm i0$ denote infinitesimal positive imaginary part.
Lessor and greater Green's function are
\begin{align}
 \gless_\kv(t-t') &= i e^{-\frac{i}{\hbar}\ekv t}  f_{\kv} \nnr
 \ggrt_\kv(t-t') &= - i e^{-\frac{i}{\hbar}\ekv t} (1- f_{\kv}),
   \label{glgkresult}
\end{align}
and the Fourier representations are
\begin{align}
 \gless_{\kv,\omega}&= 2\pi i f_\kv\delta(\hbar\omega-\ekv) \nnr
 \ggrt_{\kv,\omega}&= -2\pi i (1-f_\kv)\delta(\hbar\omega-\ekv).
 \label{glgkomresult}
\end{align}

\subsection{Calculation of lesser component}
When physical quantities are calculated perturbatively, we need to calculate the lesser component of products of path-ordered Green's functions.
We consider here the case of a scattering by a potential ${\cal V}$.
The Dyson equation for the path-ordered Green's function reads 
\begin{align}
G^{c}{}_{\alpha,\beta}(\tc,\tc')
&= \sum_{n=0}^\infty \lt(-\frac{i}{\hbar}\rt)^n  \lt[\prod_{i=1}^n \int_{C} d\tc{}_{i} \rt]
\lt[ G^c(\tc,\tc{}_1){\cal V}(t_1) G^c(\tc{}_1,\tc{}_2)\cdots {\cal V}(t_n) G^c(\tc{}_n,\tc') \rt]_{\alpha\beta} .
\label{Gcperturbation}
\end{align}
The lesser component of the first order term is 
\begin{align}
\Gless{}^{(1)}(t,t')& =
G^{c}{}^{(1)}(\tc\in C_\ra,\tc'\in C_\la) \nnr
&= \int_{C} d\tc{}_{1} G^c(\tc,\tc{}_1){\cal V}(t_1) G^c(\tc{}_1,\tc').
\end{align}
Writing explicitly the contributions from the contour $ C_\ra$ and $C_\la$, we have 
\begin{align}
\Gless{}^{(1)}(t,t')
 &= \int_{C_\ra} d\tc{}_{1} \Gto(\tc,\tc{}_1){\cal V}(t_1) \Gless(\tc{}_1,\tc')  
 +\int_{C_\la} d\tc{}_{1} \Gless(\tc,\tc{}_1){\cal V}(t_1) \Gat(\tc{}_1,\tc') \nnr
 &= \int_{-\infty}^\infty d t_{1} \lt[ \Gto(t,t_1){\cal V}(t_1) \Gless(t_1,t')  
 - \Gless(t,t_1){\cal V}(t_1) \Gat(t_1,t') \rt].
\label{Gcperturbationless}
\end{align}
Using identities 
\begin{align}
 \Gto(t,t') &= \Gr(t,t')+\Gless(t,t') \label{Gtoless} \\
 \Gat(t,t') &= -\Ga(t,t')+\Gless(t,t') \label{Gatless},
\end{align}
we have a formula for evaluating the lesser component, 
\begin{align}
\lt[ \int_{C} d\tc{}_{1} G^c(\tc,\tc{}_1){\cal V}(t_1) G^c(\tc{}_1,\tc') \rt]^<
 &= \int_{-\infty}^\infty d t_{1} \lt[ \Gr(t,t_1) {\cal V}(t_1)\Gless(t_1,t')  
 + \Gless(t,t_1) {\cal V}(t_1)\Ga(t_1,t') \rt].
\label{Glesserdecompose}
\end{align}
Let us denote Eq. \ref{Glesserdecompose} simply as 
\begin{align}
 [G^c G^c]^< &= \Gr\Gless+ \Gless\Ga.   \label{Glesserdecompose2}
\end{align}
For retarded and advanced components, 
\begin{align}
[G^c G^c ] ^\adv &= \Ga \Ga \nonumber \\
[G^c G^c ] ^\ret &= \Gr \Gr . \label{Gardecompose}
\end{align}
This formula is called the Langreth theorem.
Terms with more Green's functions are evaluated by applying the formula multiply.
For the three Green's function case, we have
\begin{align}
 [G^c G^c G^c]^< &= \Gr[G^c G^c]^<+ \Gless[G^c G^c]^\adv =  \Gr\Gr\Gless+ \Gr\Gless\Ga+\Gless\Ga\Ga.
\end{align}

\section*{References}

\end{document}